\documentclass[sn-apa,oneside,a4paper,pdflatex,referee]{sn-jnl}

\newgeometry{left=1.55cm,right=1.55cm,top=1.75cm,bottom=1.75cm}
\usepackage{graphicx}%
\usepackage{booktabs, makecell, multicol, multirow, tabularx}   
\usepackage{amsmath,amssymb,amsfonts}%
\usepackage{amsthm}%
\usepackage{mathrsfs}%
\usepackage[title]{appendix}%
\usepackage{color}
\usepackage[table, xcdraw, x11names, usenames, dvipsnames, svgnames]{xcolor}
\usepackage{ragged2e}                       

\usepackage{tabulary}
\usepackage{textcomp}%
\usepackage{manyfoot}%
\usepackage{footnote}
\usepackage{verbatim}
\usepackage{enumitem}
\usepackage{algorithm}%
\usepackage{algorithmicx}%
\usepackage{algpseudocode}%
\usepackage{listings}%
\usepackage{caption}
\usepackage{enumitem}
\usepackage{courier}
\usepackage{lscape}
\usepackage{adjustbox}
\usepackage{setspace}
\usepackage{textgreek} 
\usepackage{bm}
\usepackage{nicefrac}  
\usepackage{microtype}  
\usepackage[utf8]{inputenc} 
\usepackage[T1]{fontenc}    
\usepackage{lipsum}
\usepackage{caption}
\usepackage{subcaption}
\usepackage{fancyhdr}       
\usepackage{url} 
\usepackage{hyperref}

\AtBeginDocument{%
\hypersetup{
    citecolor=Purple3,
    linkcolor=OliveGreen,
    urlcolor=NavyBlue,
    pdfstartview=FitH
    }
}

\theoremstyle{thmstyleone}%
\theoremstyle{thmstyletwo}%

\theoremstyle{thmstylethree}%

\begin{document}
\title[Supporting single responsibility through automated extract method refactoring]{Supporting single responsibility through automated extract method refactoring}


\author[1]{\fnm{Alireza } \sur{Ardalani}}\email{alirezaardalani1996@gmail.com}

\author*[1]{\fnm{Saeed} \sur{Parsa}}\email{parsa@iust.ac.ir}

\author[1]{\fnm{Morteza} \sur{Zakeri-Nasrabadi}}\email{morteza\_zakeri@comp.iust.ac.ir}

\author[2]{\fnm{Alexander} \sur{Chatzigeorgiou}}\email{achat@uom.gr}

\affil*[1]{\orgdiv{School of Computer Engineering}, \orgname{ Iran University of Science and Technology}, \orgaddress{\street{Resalat, Hengam St.}, \city{Tehran}, \postcode{16846-13114}, \state{Tehran}, \country{Iran}}}

\affil[2]{\orgdiv{Department of Applied Informatics}, \orgname{University of Macedonia}, \orgaddress{\street{156 Egnatia Str.}, \city{Thessaloniki}, \postcode{54636}, \state{Thessaloniki}, \country{Greece}}}


\abstract{
The responsibility of a method/function is to perform some desired computations and disseminate the results to its caller through various deliverables, including object fields and variables in output instructions. Based on this definition of responsibility, this paper offers a new algorithm to refactor long methods to those with a single responsibility. We propose a backward slicing algorithm to decompose a long method into slightly overlapping slices. The slices are computed for each output instruction, representing the outcome of a responsibility delegated to the method. The slices will be non-overlapping if the slicing criteria address the same output variable. The slices are further extracted as independent methods, invoked by the original method, if certain behavioral preservation are made. The proposed method has been evaluated on the GEMS extract method refactoring benchmark and three real-world projects. On average, our experiments demonstrate at least a 29.6\% improvement in precision and a 12.1\% improvement in the recall of uncovering refactoring opportunities compared to the state-of-the-art approaches. Furthermore, our tool improves method-level cohesion metrics by an average of 20\% after refactoring. Experimental results confirm the applicability of the proposed approach in extracting methods with a single responsibility.}


\maketitle


\section{Introduction}
\label{introduction}
Responsibility is a task or duty to be performed. According to the single responsibility principle, a module should have a single responsibility to avoid having more than one reason to change \citep{martin2003agile}. When a design element, such as a class, has many different responsibilities, it has many reasons to change, making it hard to test. On the other hand, if a class does not encapsulate its responsibility, changes will cascade through its code, and it will be harder to test. A reason to change can be the change in the requirements set forth by the actors using the class \citep{martin2003agile}. The single responsibility principle \citep{robert2017clean} states that a module or a class in a program should be responsible to one, and only one, actor" to have a single reason to change. In this way, each class will be modified due to changes to solely one actor’s requirement. Hence, the responsibility of a software product is to satisfy the requirements set forth by the stakeholders \citep{robert2017clean}.

Requirements are the leaves of the goal model. Therefore, using a goal-oriented architecture, the software architect may assign the responsibility for obtaining each goal to a different software component and the subgoals to its sub-components. This way, the software’s structural view is obtained by converting each goal and subgoal to a corresponding component \citep{cheng2009goal}. At the lowest level, each method, and at the higher levels, each class and package should primarily meet a single requirement, objective, or goal \citep{ampatzoglou2019applying}. Otherwise, any changes in requirements, objectives, or goals may entail changes to different components scattered across the code. According to Robert C. Martin \citep{martin2009clean}: “functions should do one thing; they should do it well; they should do it only.” Otherwise, as described by Robert Martin, the function will violate the SRP, and there will be more than one reason to change the function. The reason can be any change in the actor’s requirement. At the lowest level, a method's responsibility could be to satisfy a requirement or fulfill some tasks concerned with a requirement \citep{martin2009clean}.

The question is, what is responsibility? A class responsibility is anything that it knows or does. The responsibility is operationalized by defining one or more methods that may collaborate with other classes and methods to fulfill the responsibility. Class responsibility collaboration cards (CRC)  define what a class does as one or more actions or methods constituting its responsibility \citep{fayad2003pattern}. The responsibility of each method/function is delegated to a sequence of instructions that are executed whenever the function is invoked. Hence, if a method has more than one responsibility, the sequences of instructions, each fulfilling a distinct responsibility, will have minimal overlapping. Such sequences could be identified using backward slices for the instructions where the method terminates or exposes a result to the outside. Each of these slices includes output statements representing the method's results to the outside of the method body. The exposition could be either directly to return or any output stream (e.g., a file, console, and network port) or indirectly through the side effects on objects and global variables.

JDeodorant \citep{tsantalis2011identification} considers side effects on objects as an extract method refactoring opportunity for extracting the sequence of instructions as a distinct method. However, it does not consider output instructions, e.g., the return statements. The authors argue that the operation of a return statement is directly associated with the enclosing method. Thus, a return statement cannot be moved or copied to another method \citep{tsantalis2011identification}. We introduce a new output-based slicing algorithm to overcome the JDeodorant slicing algorithm limitations and identify parts of code fulfilling a distinct responsibility. Output-based slicing focuses on the variables used in output instructions, exposing the outcome of any responsibility a method holds. Such slicing-based extract method refactoring is applicable when method level SRP \citep{ampatzoglou2019applying, charalampidou2016identifying} is violated.  Extract Method refactoring is a common technique to fix Long Method, one of the most frequent code smells in software systems, as well as to remove code duplication and improve readability. \citep{brdar2022semi, lacerda2020code}. Long methods often violate the single responsibility design principle \citep{robert2017clean}. We call method-level SRP violations ‘Non-SRP Method’ smell. This paper proposes an automated Extract Method refactoring approach to achieve method-level single responsibility by fixing Non-SRP Method and Long Method code smells.

Our proposed approach consists of four main steps. First, seed criteria to apply slicing algorithms are selected. To this end, a distinct list of slicing criteria \footnote{The slice is defined for a slicing criterion $C=(x,v)$ where $x$ is a statement (line number) in program $P$ and $v$ is a variable in $x$. A static slice includes all the statements that can affect the value of variable $v$ at statement $x$ for any possible input.}  is created for each variable, $x$, used in an output statement, ots. The list includes the slicing criterion (ots, $x$) and the criteria ($d_{i}$, $x$), where $d_{i}$ addresses any statement defining $x$ provided that the statement is located between ots and a previous output statement using $x$. If ots is the first statement using $x$, then all the statements before ots are considered. In the second step, static backward slices are computed for each criterion in each list. The slices calculated for each variable $x$ include only the statements between two consecutive output instructions using $x$. In the third step, the union of all slices for each list is computed in the regions determined by the block-based slicing technique \citep{maruyama2001automated} to form the initial set of extract method opportunities. In situations where there is only one output instruction, our approach uses the block-based slicing algorithm \citep{tsantalis2011identification}, \citep{maruyama2001automated} on all variables defined or modified in the method body to identify possible extract method refactoring opportunities. Finally, in the fourth step, the candidate methods are checked with several precondition rules to ensure their usefulness and legality. The primary contributions of this article can be summarized as follows:

\begin{enumerate}
    \item {We define the responsibility of a method as its impact on the outer scope and propose a novel slicing algorithm, Output-based Slicing, to identify extract method opportunities based on the method responsibilities.}
  
    \item {Five types of output instructions are introduced and considered by the proposed output-based slicing algorithm to achieve single responsibility while refactoring long methods.}
    
    \item {We propose a comprehensive set of rules used as refactoring precondition to verify the correctness and usefulness of the extract method candidates regarding the single responsibility and program semantics and behavior preservation.} 
    
    \item {Our approach is supported by an open-source tool, \textit{slice-based single responsibility extraction} (SBSRE), integrated as an Eclipse plugin to be effectively used by the developers. The proposed tool identifies and applies the extract method refactoring of long methods with multiple responsibilities automatically at the source code level. The source code of the developed tool is available on our GitHub repository} \textit{\href{https://github.com/Alireza-Ardalani/SBSRE}{https://github.com/Alireza-Ardalani/SBSRE}.}
\end{enumerate}

The remainder of this paper is structured as follows. Section \ref{RelatedWorks} provides background on single responsibility and discusses the related works on automated extract method refactoring of long methods. Section \ref{motivating} provides a motivating example to demonstrate the shortcomings of the existing extract method approaches and how our approach overcame them. Section \ref{sbsre} describes our extract method refactoring methodology. The evaluation and results obtained from our experiment are presented in Section \ref{evaluation}. Threats to validity are described in Section \ref{validity}. Section \ref{conclusion} ends this paper with concluding remarks and future work.

\section{Motivating Example}
\label{motivating}

Slicing based on output statements can be very efficient for extracting meaningful extract method opportunities because each slice is, in essence, a responsibility implemented as a sequence of activities leading to an output. The output can be represented by a return statement or writing in an output stream, such as a console, file, database table, I/O port, or global variable (discussed in \ref{sec:outputCriteria}). In this section, a motivating example is used to demonstrate the applicability of our definition of responsibility by comparing our extract method refactoring results with SEMI \citep{charalampidou2016identifying}, JDeodorant \citep{tsantalis2011identification}, JExtract \citep{silva2014recommending}, GEMS \citep{xu2017gems}, and Segmentation \citep{tiwari2022identifying}.  Moreover, more motivating examples are publicly available on our GitHub repository \textit{\href{https://github.com/Alireza-Ardalani/SBSRE/tree/main/motivatingExamples}{github.com/Alireza-Ardalani/SBSRE}}.

        \subsection{Clarification of responsibilities in the motivating example}

         As a motivating example, we present the \texttt{CVResultsString} method from the Weka project \citep{Weka2011}, shown in Figure \ref{fig:1}, with two distinguishable responsibilities. As shown in Figure \ref{fig:1}, the \texttt{CVResultsString} method is responsible for two tasks: outputting the result of statistical calculations in the global variable, \texttt{m\_rankResults}, and returning the summary of cross\-validations results stored in the \texttt{CvString} variable through the return statement. However, these two responsibilities are currently intertwined in the code and can be altered independently. It is generally considered bad programming practice to use global variables, so modifying the code by returning the computed standard deviation and mean value in an array would be advisable instead. Additionally, instead of summarizing the cross\-validation evaluation results in a string, modifying the code to save them in a file may be necessary. As a result, there are two reasons for changing the method, and this violates the Single Responsibility Principle (SRP).

        \subsection{The result of the state-of-the-art approaches}

         JDeodorant \citep{tsantalis2011identification} does not suggest any extract method opportunity for the method shown in \ref{fig:1}. JDeodorant \citep{tsantalis2011identification} does not handle return statements because it considers return statements as indispensable parts of its enclosing methods. This belief could be correct, provided that there is only one return statement in a method, and other statements in the method body cooperate to calculate the return value. However, a method could have more than one return statement. In addition, apart from the return statements, there are many other ways to represent the results of a method's activities, such as modifying global variables that JDeodorant does not consider too.

         \vspace{0.3cm}

         SEMI \citep{charalampidou2016identifying} is the only available tool suggesting the extract method opportunities based on the SRP. The SEMI tool takes a different approach than JDeodorant and our tool. Instead of slicing, SEMI seeks to extract statements that somehow cooperate to increase cohesion. SEMI tool also has no extract method suggestion for this method. JExtract \citep{silva2014recommending}, GEMS \citep{xu2017gems}, and Segmentation \citep{tiwari2022identifying} are other approaches that do not directly focus on the Single Responsibility Principle, but they can be examined in this section as they suggest the extract method opportunity. Extract method suggestions provided by each tool for method \texttt{CVResultsString}  are as follows:

         \begin{itemize}
            \item \textbf{Tool 1: JExtract.} Suggestion 1: (12-27), suggestion 2: (10-27), suggestion 3: (14-27). 
            \item 	\textbf{Tool 2: GEMS.} Suggestion 1: (9-46), Suggestion 2: (21- 27), Suggestion 3: (4-8).
            \item 	\textbf{Tool 3: Segmentation.} Suggestion 1: (10- 21) And (28-46), Suggestion 2: (4-8), Suggestion 3: (22, 27).
            \item 	\textbf{Tool 4: JDeodorant.} No suggestion.
            \item  	\textbf{Tool 5: SEMI.} No suggestion.
        \end{itemize}

         All three extract method suggestions of JExtract \citep{silva2014recommending} cover almost the same calculations corresponding to the \texttt{rankResults} variable. GEMS's \citep{xu2017gems} first suggestion includes nearly the entire body of the method, and the second suggestion covers only a portion of the \texttt{rankResults} variable's calculations, and the third proposal includes a small part of the method. Segmentation's \citep{tiwari2022identifying} first suggestion doesn't have a specific responsibility and includes part of the \texttt{rankResults} and \texttt{CvString} variables’ computation, and the other two suggestions include a small part of the \texttt{CVResultsString} method.

            \subsection{Resolving the challenges of SEMI and JDeodorant}

\begin{figure*}[ht]
             \centering
             \includegraphics[width=0.7\linewidth]{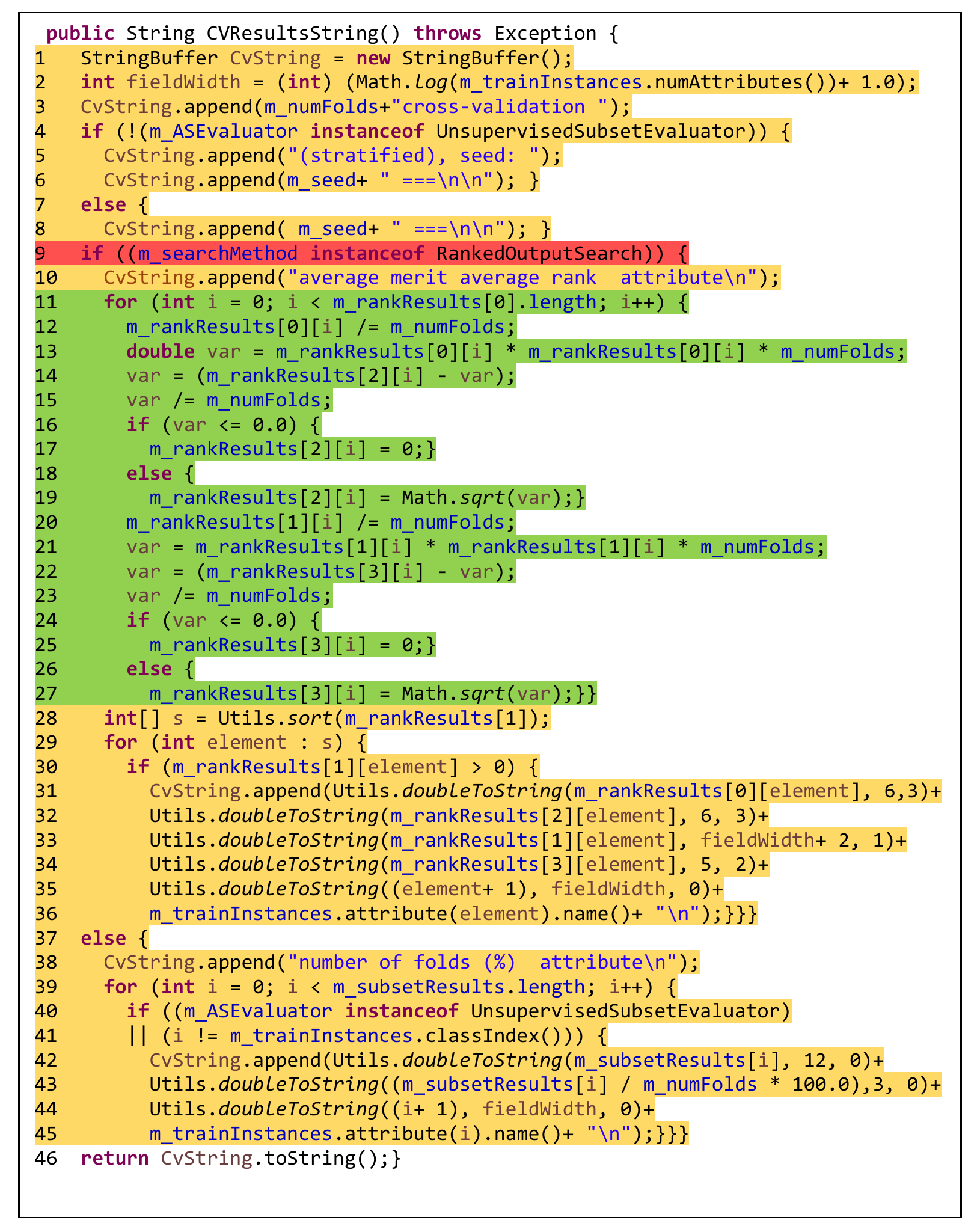}
                \caption{The \texttt{CVResultsString } method with two distinct slicing criteria, (27, \texttt{m\_rankResults}) and (46, \texttt{CvString}).}
                  \label{fig:1}
            \end{figure*}

            Our motivation for the design and implementation of SBSRE has been to resolve the problems we come across when using other extract-method tools to identify and apply the extract-method refactoring automatically. We focus on output statements because they represent the result of the method activities in compliance with its responsibility. The results are provided by the statements covered by backward slices of the output statements where the results are revealed. Since we consider backward slices for method extraction, we are not restricted to extracting only continuous sequences of statements. The result of applying our approach, SBSRE, to the \texttt{CVResultsString} method is the backward slices of the variables \texttt{rankResults} and \texttt{CvString}, which appeared in the output instructions. In Figure \ref{fig:1}, the extract method recommendations identified by our approach are highlighted in green and yellow colors, and common statements between the suggestions are highlighted in red. The statements of each extract method candidate are as follows:

            \begin{enumerate}
            \item Statements numbers {1-10} and {28-46}.
            \item Statements numbers {9}  and {11-27}.
            \end{enumerate}
            The SBSRE approach described in the next section aims to identify and recommend the above extract method candidates.

\section{Related Works}
\label{RelatedWorks}

    Various approaches have been presented to automate the extract method refactoring \citep{tsantalis2011identification, charalampidou2016identifying, maruyama2001automated, silva2014recommending, xu2017gems, hubert2019implementation}. None of them have considered single responsibility extraction, applicability, behavior preservation, and full automation simultaneously, to the best of our knowledge. The prominent techniques for identifying extract method opportunities are program slicing \citep{tsantalis2011identification} and cohesion metrics \citep{charalampidou2016identifying}. Approaches based on the former technique avoid identifying the responsibilities, while approaches based on the latter cannot extract non-consecutive statements.

    There has to be the detection of long method bad practices before refactoring recommendations can take place. Many studies have proposed approaches to detect long method code smell instances. However, they have not provided any refactoring approaches to fix the detected smells \citep{arcelli2016comparing, palomba2016textual,pecorelli2020large, hadj2018hybrid}. On the other hand, extract method refactoring tools, such as JDeodorant \citep{tsantalis2011identification}, SEMI \citep{charalampidou2016identifying}, JExtract \citep{silva2014recommending}, and GEMS \citep{xu2017gems} fix long method smells without directly identifying code long method instances. Indeed, there is not a separate long method detection step in these tools. However, it should be noted that methods with one or more extract method refactoring opportunities detected by any refactoring tools, mainly our proposed tool, can be considered long methods. In addition to identifying refactoring opportunities, the automatic application of extract method refactoring is a challenging and fault-prone task. According to Alcocer et al. \citep{alcocer2020improving}, only 51\% of the extract method refactoring attempts in integrated development environments (IDEs) are successful. Their analysis reveals that the most prominent failure issues are related to the selection of code snippets for extracting a method. Therefore, detecting long method instances is not enough to fix the long method code smells. Appropriate detection of extract method refactoring opportunities is required to fully automated the process of this refactoring and reduce the failure rate.

    Maruyama \citep{maruyama2001automated} has proposed block-based slicing to extract a complete slice of a given variable from a specific region in the method body. Block-based slicing does not extract the code fragments from the whole method body but the region consisting of some consecutive basic blocks. A basic block is a sequence of consecutive statements in which the flow of control enters at the beginning and leaves at the end without any halt or possibility of branching except for the last statement. Block-based slicing's idea is to avoid the remaining statements within a basic block, which are not detected by simple slicing but should be extracted to preserve the semantics and behavior of the program. For instance, non-executable or non-equivalent code might remain when a loop statement or a variable related to a loop condition is casually extracted based on the simple static slicing technique. To overcome this, block-based slicing introduces the concept of the region into the original slicing algorithm and defines the region of code from which a new method is extracted. The block-based slicing approach shows promising results regarding program behavior preservation. However, the programmer should manually determine the slicing criteria to receive a recommendation. As a result, finding all refactoring opportunities and selecting the appropriate one is error-prone, time-consuming, and requires an experienced developer.

    Tsantalis et al. \citep{tsantalis2011identification} have automated the selection of seed statements for block-based slicing to address the shortcomings of the Maruyama approach \citep{maruyama2001automated}. They have proposed two slicing algorithms, (i) complete computation slice, and (ii) object state slice. The former slicing algorithm considers variables having primitive data types or object references whose values are modified by assignment statements throughout the body of the original method. The later slicing algorithm takes into account object references, including local variables or fields of the method enclosing class, that refer to objects whose state is affected by method invocations throughout the body of the original method. Their approach is supported by a tool, JDeodorant \citep{tsantalis2018ten}, which does not require manual slicing criteria selection. This way, JDeodorant can also be used to detect long method code smells, which indeed are methods with at least one refactoring opportunity. However, we noticed that JDeodorant often comes up with a long list of candidate refactoring opportunities suggested to the developer to manually choose the best one. Another disadvantage of JDeodorant is that extracted candidates might have many duplications since block-based slicing requires some statements to be repeated in both the initial and extracted methods to preserve the program behavior \citep{hubert2019implementation}. JDeodorant extract method module also excludes the return statements and does not consider the usage of variables in output instructions, leading to the recommendation of extract methods that violate single responsibility. Compared with JDeodorant, our proposed approach could select more slicing criteria and, subsequently, more extract method refactoring opportunities. We focus on output variables and output instructions that expose the method's behavior to outer scopes to generate single responsible slices and avoid making useless suggestions.

    Extract method recommendations must be as independent as possible from the original method concerning the dependencies. Silva et al. \citep{silva2014recommending} have proposed a hierarchical model representing the method body as a block structure. A block is a sequence of continuous statements that follow a linear control flow. Here, the concept of the block is different from the basic block in the way that the proposed block structure can be nested and all block statements are not necessary in the same basic block. Every combination of blocks in this structure that simultaneously satisfies a set of quality, syntactic, and behavioral preconditions has been considered an extract method candidate. The feasible candidate blocks have been then ranked with a scoring function that computes the structural dependencies between each candidate and the remaining part of a method. The authors have not directly compared their tool with other automated extract method refactoring approaches. Their approach, supported by a tool called JExtract \citep{silva2014recommending}, heavily relies on a scoring function, which considers the structural dependencies present in the code to filter and rank the recommendations.
    
    In contrast to JExtract, JDeodorant relies on specific slicing criteria \citep{tsantalis2011identification}, which is not applicable to scoring candidates. However, generating valid code blocks by exhaustively checking numerous preconditions and ranking them is time-consuming and computationally intensive. In addition, the only quality metric employed by Silva et al. \citep{silva2014recommending}is the number of statements or lines of code of the candidate method to filter out the extreme cases in which a candidate would result in a poor-quality recommendation because of its small size. Focusing on the method's lines of code as the only quality metric may result in non-cohesive methods \citep{tsantalis2011identification}.
    
    The SEMI approach \citep{charalampidou2016identifying} uses cohesion between pairs of statements to recognize which code fragments work together and provide specific functionality. Two statements in the code are considered coherent if they are accessing the same variable, calling a method for the same object, or calling the same method for a different object of the same type. The statements are then clustered based on the agglomerative hierarchical clustering technique \citep{han2012data} to find larger sets of cohesive statements as refactoring opportunities. The lack of cohesion in methods (LCOM) \citep{al2010measuring} measure and the candidate size has been used to rank the generated candidates. The evaluations by experts have demonstrated a higher F-measure of SEMI \citep{charalampidou2016identifying} compared to the JDeodorant extract method module \citep{tsantalis2011identification}. However, it makes no distinction between different types of instructions within the method body when performing clustering. We hypothesize that variables in output instructions are more important in determining the given method's responsibility. Moreover, a non-slicing approach employed by SEMI prevents it from suggesting extract method candidates with non-consecutive statements. We consider a slice-based approach that preserves the cohesion of statements by focusing on the output variables in output instructions when searching for extract method opportunities. 

    JDeodorant \citep{tsantalis2011identification}, JExtract \citep{silva2014recommending}, and SEMI \citep{charalampidou2016identifying} generate many extract method candidates for the long method. The JDeodorant approach \citep{tsantalis2018ten} uses no scoring mechanism to rank the generated candidate. JExtract \citep{silva2014recommending} and SEMI \citep{charalampidou2016identifying} use a simple scoring function based on a single metric which is not precise in most situations. GEMS \citep{xu2017gems} uses a machine-learning approach that recommends the most useful extract method refactorings to address the problem of ranking refactoring candidates. A classifier is trained on a practical dataset containing the real-world extract method refactoring applied by developers in open-source projects. First, all possible extract method refactoring candidates are extracted by JExtract \citep{silva2014recommending}. Then, the feature vector containing structural source code metrics for each pair of the extracted method and the remaining method is computed to be fed into the model along with the result of positive or negative refactoring as labels. Finally, all candidate methods are extracted and fed to the learned model for a given method. The model gives a probability of usefulness to each input sample, which is used to rank existing candidates. Although a relatively higher F-measure compared to JDeodorant \citep{tsantalis2011identification}, JExtract \citep{silva2014recommending}, and SEMI \citep{charalampidou2016identifying}, GEMS \citep{xu2017gems} initial refactoring candidates generated using JExtract \citep{silva2014recommending} may not cover all single responsible opportunities. Our proposed approach can improve the initial extract method candidate generation to be used in similar learning-to-rank techniques.

    Long Method Remover (LMR) \citep{meananeatra2018refactoring} employs a set of four refactoring techniques including ‘replace temp with query’, ‘introduce parameter object’, ‘preserve whole object’, ‘decompose conditional’, in addition to extract method, to completely remove the long method code smell, while minimizing the number of statements impacted by the refactorings. However, LMR uses the JDeodorant plugin \citep{meananeatra2018refactoring} for identifying extract method refactoring opportunities without any significant improvement.

    Additional information in the method body, such as blank lines and comments, can be used to help the identification of the extract method opportunities. Hubert \citep{hubert2019implementation} has proposed an approach that suggests extract method refactoring based on quality analysis considering this supplementary information. Their approach receives a long method as input and generates an exhaustive list of refactoring candidates by transferring the source code into a block structure called the statement graph \citep{silva2014recommending}. Each node of the statement graph represents a block containing one line (statement) in the method body, and the root node is the method definition. The nesting depth of each statement in the code is assigned to its corresponding node in the graph. The statement graph also stores data edges between statements that contain one or more shared variables. The nodes of the statement graph are combined under several preconditions to determine an initial set of extract method candidates with continuous statements. The ranking algorithm is based on the number of lines of code, parameters, blank lines, and comments in the code. Blank lines and comments are used since programmers usually separate or mark the beginning and end of each responsibility to explain it with comments.

     The code fragments written between the comments most presumably perform a separate task that can be extracted. However, this assumption makes the ranking technique highly dependent on the coding style. Developers may not adhere to such coding styles and commenting principles. Thus, Hubert's approach \citep{hubert2019implementation} is ineffective in ranking candidates for methods with no comments and blank lines. Our extract method approach does not rely on any coding style or non-executing parts in the source code. Tiwari and Joshi \citep{tiwari2022identifying} have proposed the Segmentation approach, which clusters related statements with distinct functionalities as separate refactoring extract method refactoring opportunities. The count of suggestions can be used to detect the long method code smells and compute the smell severity level of the detected long method. However, Segmentation only finds extract method opportunities, including consecutive statements. Fernandes et al. \citep{fernandes2022live} have proposed a block-based extract method refactoring for JavaScript codes in which blocks of code are formed only by consecutive statements that have more than three statements. Blocks that contain more than 80\% of statements of the original method are considered refactoring candidates. However, the authors have not proposed any evaluation for their proposed method. Moreover, their approach only identifies extract method refactoring candidates with consecutive statements.
    
    Shahidi et al. \citep{shahidi2022automated} have proposed a graph-based approach to identify extract method opportunities. First, the adjacency matrix of the method dependency graph is checked to determine whether it is decomposable into modular subgraphs \citep{fortunato2010community} or not. If the dependency graph is decomposable, then the subset of nodes with the highest clustering coefficient with a starting point is chosen and considered an extract method candidate. The starting point is the first line that can be separated from a method that does not have any control or data dependency on other statements. In the method dependency graph, each line of code represents a node, and each edge of the graph indicates the data or control dependency \citep{aho2007compilers} between two lines of code. The approach performs well where the extracted statements appear in consecutive lines. Otherwise, it does not recommend any refactoring opportunity. In addition, the choice of the right starting points is challenging and error-prone. For example, output statements considered seeds for slicing in our approach cannot be selected as starting points in the graph-based approach proposed by Shahidi et al. \citep{shahidi2022automated}.

\section{SBSRE approach}
\label{sbsre}

This section describes our proposed SBSRE approach to identify extract method opportunities based on the single responsibility principle. The main innovation of SBSRE is to focus on output instruction since methods with more than one output instruction may violate the single responsibility principle. According to the number of output instructions, methods may violate the single responsibility principle (SRP) in the following situations:

        \begin{enumerate}[label=(\Alph*)]
            \item A long method with a single output instruction and a single object or variable to output.  
            \item A method with a single output instruction and output more than one variable or object, provided that the statements computing the outputs do not overlap much.
            \item A method with more than one output instruction presenting different values provided that the statement computing the outputs does not overlap much with each other. 
        \end{enumerate}

    The output-based slicing algorithm identifies all the output instructions and the variables or objects used in them. After that, the backward slicing regions associated with each output instruction are determined using enhanced block-based slicing (mentioned in section \ref{sec:block-based}). Thus, the output-base slicing algorithm suggests method extraction opportunities that violate the single responsibility principle for methods of types (B) and (C). The key point is that the statements supporting the calculations of the output variables should have little overlap; hence, each backward slice of output variables can perform a separate responsibility. According to Rule 6 mentioned in section \ref{sec:Rules}, our suggested threshold is less than 0.75 duplication statements between backward slices of output variables.

    Complete computation and object state slicing algorithms aim to find chunks of a method that perform specific functionality. By adding some behavior conservation rules to complete computation and object state slicing algorithms, they were prepared to use in the SBSRE tool. These two algorithms have been used to discover methods from type (A) that violate the single responsibility principle.  Neither of the mentioned algorithms considers the output instructions, therefore they cannot find SRP violations caused by multiple output instructions. However, if we only use output-based slicing in such cases ( types (B) or (C)), we may end up with a long method corresponding to an output (type (A)). To maximize the detection of SRP violations, the SBSRE tool applies the three mentioned algorithms to all types: (A), (B), and (C).

\begin{figure*}[]
    \centering
    \includegraphics[width=0.8\linewidth]{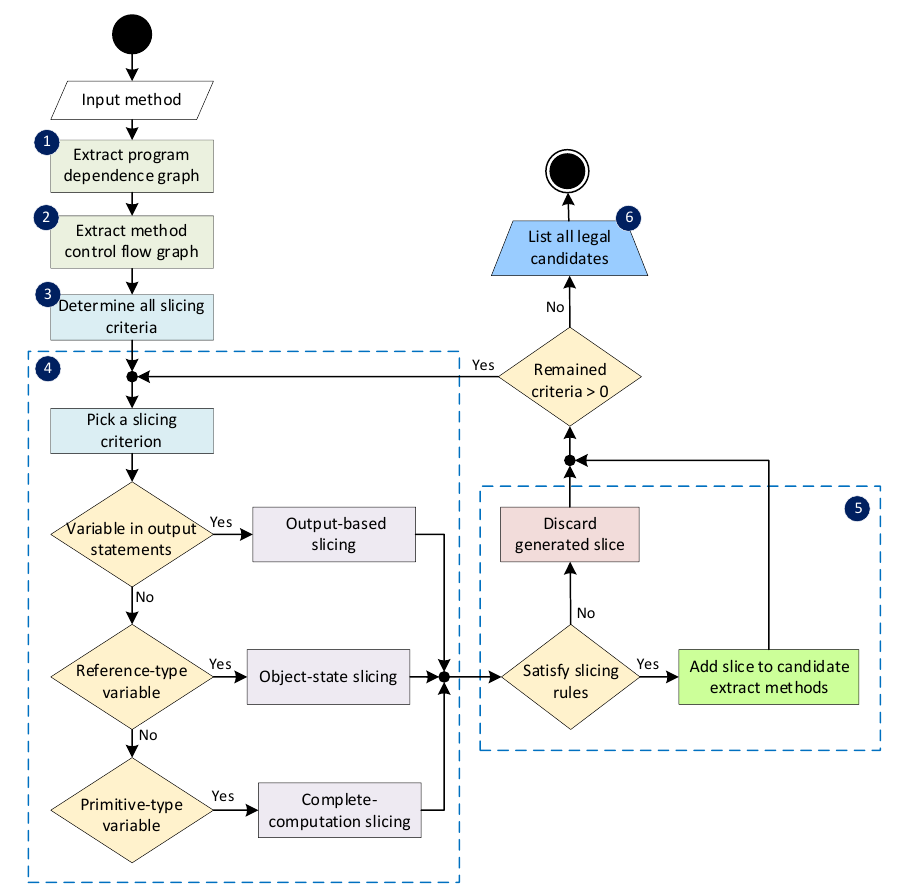}
    \caption{Flowchart depicting the methodology of SBSRE tool.}
    \label{fig:2}
\end{figure*}

    Figure \ref{fig:2} shows an overview of our approach. In steps 1 and 2, the required artifacts for identifying refactoring opportunities and applying them are extracted from the source code. These artifacts are the program dependence graph (PDG) and control flow graph (CFG), respectively, created for the program and the given method. In step 3, all slicing criteria for applying the slicing algorithms are extracted and saved in a list. In step 4, a slicing criterion is selected from the list, and according to the target statement and variable type, the appropriate slicing algorithm is applied to it. Each slice is considered an extract method candidate. In step 5, the extract method candidates are checked with a set of nine inhibitory rules to ensure whether the program behavior is preserved after applying the refactoring or not. If any rule is not satisfied, the candidate refactoring is discarded. Otherwise, it is added to the legal refactorings list. Steps 4 and 5 are repeated for all slicing criteria extracted in step 3. Finally, in step 6, applicable extract method refactorings to achieve single responsibility are suggested to the user. The details of each step are described in the remaining parts of Section \ref{sbsre}.

        \subsection{Extracting PDG and CFG}
        
        Our approach to identifying single responsible block-based slices relies on two essential artifacts, program dependence graph (PDG) \citep{ottenstein1984program} and control flow graph (CFG) \citep{allen1970control}.  The former is used to compute the slices of the method variables, while the latter is used to determine the block-based regions for backward slicing computation. The definition and use of a variable in different statements are determined with the help of PDG \citep{hotta2012identifying}, which is necessary for computing the backward slice of a given variable. Each node in the PDG indicates a statement that defines or uses a variable. There are two types of edges in the PDG, control dependence and data dependence edges. The data dependence edges demonstrate the flow of data between statements, and the control dependency edges represent the control conditions on which the execution of a node depends \citep{ferrante1987program}. In CFG, each node represents a basic block, and the edges show the path between the basic blocks. Indeed, the CFG represents all the paths within a method that may be traversed when the program is executed \citep{orailoglu1986flow}. To construct the PDG, we traverse the program abstract syntax tree (AST), considering each statement as a node of PDG. The statements with a shared variable are connected to each other such that the direction of the connection shows the dependency of the destination node on the source node. The CFG is constructed for a given method using the block-partitioning algorithm \citep{aho2007compilers}.

        \subsection{Determining slicing criteria}\label{sec:criteria}

        The slice is defined for a slicing criterion $C(x,v)$, where $x$ is a statement in method $m$, and $v$ is a variable in $x$. A static slice for $C(x,v)$ includes all the statements that can affect the value of variable $v$ at statement $x$ for any possible input \citep{weiser1984program}. Concerning the method flow direction, in backward slicing, a slice contains all statements and control predicates that may affect a variable at a given point \citep{de2001program}. To perform backward slicing, we first determine all required slicing criteria. Our approach uses two types of slicing criteria subject to different slicing algorithms. The first type is the criteria whose statements are output instructions. The second type is criteria that are obtained from other instructions. We specify output instructions to clarify the differences between different slicing criteria used for slicing a method.

\subsubsection{Slicing output instructions} \label{sec:outputCriteria}
A clear understanding of "output instruction" is essential within the context of our Output-based Slicing algorithm, where the goal is to start backward slicing from an output instruction. Green et al. \citep{green2009introduction} have summarized the output variables of the C programming language to calculate slice-base cohesion metrics. We look for output variables in which computational results of a given method are exposed outside its body. Therefore, statements that send information outside the method body are considered output instructions.
        
We identify five categories forming the output instructions in Java language, including return statement, printing to the output stream and any similar instruction, modifying the state of an object, writing in a file or inserting in the database, and calling a method without getting the result back or changing its input parameters. Table \ref{tab:instructions} describes the definitions of different categories of output instructions with an example for each instruction. Output instructions have been highlighted in each code snippet, and the output variable is denoted by the variable named \texttt{result}.
        
\begin{figure}[]
    \centering
    \includegraphics[width=0.35\linewidth]{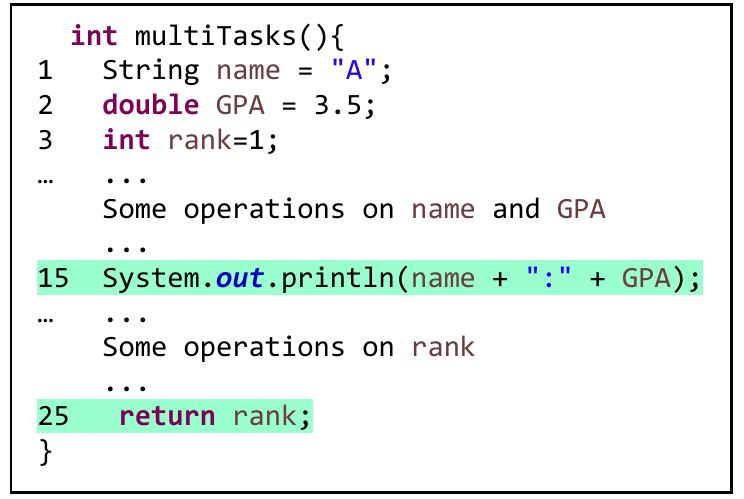}
    \caption{Illustration of a method with multiple \emph{output instructions and variables} used as slicing criteria in our approach.}
    \label{fig:3}
\end{figure}

After identifying the output instructions of a method, a criterion is created for each variable used in the output instructions. Then a backward slicing is performed based on each of these criteria. If more than one variable is used in an output instruction, a criterion is considered for each of them. For example, in Figure \ref{fig:3}, the method \texttt{multiTasks} has two output instructions on lines 15 and 25. In the first output instruction (line 15), two variables, \texttt{name} and \texttt{GPA}, are used. Therefore, the criteria for this method are \texttt{(15, name)}, \texttt{(15, GPA)}, and \texttt{(25, rank)}. As evident, the last slicing criterion belongs to the second output statement on line 25. We used the bindings of the AST nodes to make a very precise detection of output statements. To this aim, the input software codes should be fully compilable.

\begin{table*}[h!]
\footnotesize
	\centering
	\caption{Output instructions with examples.}
		\begin{tabular}{m{2cm}  m{5cm} m{8cm} }
			\hline	
			Output instruction type & Description & Example  \\
			\hline	\hline
			Calling method without getting the results back & A method is called, but the result is not returned to the original method. & 
\begin{lstlisting} 

void calculate(int input){
...
}
void multiply(int[]input){
  int n = input.lenght;
  int result= 1;
  for(int i=0 ; i ++ ; i<n){
    result = result *input[i];}
\end{lstlisting}
\vspace{-\baselineskip}
\begin{lstlisting}[backgroundcolor=\color{green}]
calculate(result);
\end{lstlisting}
\vspace{-\baselineskip}
\begin{lstlisting}
}
\end{lstlisting}\\	\hline
	Modifying class attributes or global variables & Changes that are directly occurring on a class field or global variable. & 
\begin{lstlisting}	
int resultClass;
void multiply(int[]input){
  int n = input.lenght;
  int result= 1;
  for(int i=0 ; i ++ ; i<n){
    result = result *input[i];}
\end{lstlisting}
\vspace{-\baselineskip}
\begin{lstlisting}[backgroundcolor=\color{green}]
  this.resultClass = result;
\end{lstlisting}
\vspace{-\baselineskip}
\begin{lstlisting}
}
\end{lstlisting}
    \\					
	\hline	
	Printing to the output stream &	Any instruction that prints or displays a value. &
\begin{lstlisting}	
void multiply(int[] input){
  int n = input.lenght;
  int result= 1;
  for(int i=0 ; i ++ ; i<n){
    result = result *input[i];}
\end{lstlisting}
\vspace{-\baselineskip}
\begin{lstlisting}[backgroundcolor=\color{green}]
  System.out.print(result);
\end{lstlisting}
\vspace{-\baselineskip}
\begin{lstlisting}
}
\end{lstlisting}
    \\
    \hline
    Writing an output in a file or inserting it into a database &	Any insertion of data in a text file or database tables. &
\begin{lstlisting}	
void multiply(int[]input){
  int n = input.lenght;
  int result= 1;
  for(int i=0 ; i ++ ; i<n){
    result = result*input[i];}
  FileWriter fw = new FileWriter ("address.txt");
  BufferedWriter bw = new BufferedWriter(fw);
\end{lstlisting}
\vspace{-\baselineskip}
\begin{lstlisting}[backgroundcolor=\color{green}]
  bw.write(result);
\end{lstlisting}
\vspace{-\baselineskip}
\begin{lstlisting}
}
\end{lstlisting}
    \\
	\hline
	Typical return statement & Return the value calculated by the method.&
\begin{lstlisting}	
int multiply(int[] input){
  int n = input.lenght;
  int result= 1;
  for(int i=0 ; i ++ ; i<n){
    result = result * input[i];}
\end{lstlisting}
\vspace{-\baselineskip}
\begin{lstlisting}[backgroundcolor=\color{green}]
  return result;
\end{lstlisting}
\vspace{-\baselineskip}
\begin{lstlisting}
}
\end{lstlisting} \\\hline
\end{tabular}
	\label{tab:instructions}
\end{table*}


\subsubsection{Slicing criteria for other statements}

In order to achieve a method with a single responsibility, it is essential to consider the variables in other instructions in addition  with output instructions. The main reason is the ability to identify extract method opportunities in long methods with single output instruction and single output variable. To this aim, we add two types of slicing criteria used by JDeodorant \citep{tsantalis2011identification}. First, the primitive or reference type variables defined within a method are considered the input of the complete computation slicing algorithm \citep{tsantalis2011identification}. For each variable, $v$, in the method body, the slicing criterion contains the variable, $v$, and the last statement, which defines $v$. Second, the reference type variables defined or used in the method body that refer to an object whose state is affected by at least one statement. This statement contains an appropriate method call or a direct field modification. Such references form the input of the object state slicing algorithm \citep{tsantalis2011identification}.

\subsection{Modified Block-based slicing}\label{sec:block-based}

This section describes a modified version of the block-based slicing algorithm proposed by Maruyama \citep{maruyama2001automated} to work on methods with multiple output instructions. First, this section explains the simple backward slicing technique, then the block-based slicing approach, and finally the modification we applied to it.

\subsubsection{Backward slicing}\label{sec:Backward}

The simple backward-slicing technique starts with a slicing criterion consisting of a statement and a variable for extracting the backward slice. The Backward slicing algorithm starts from the statement where the variable exists and then traces back all the statements that directly or indirectly affect the value of that variable back to the beginning of the method.
Figure \ref{fig:4} illustrates the \texttt{deleteParent} method of the Weka tool \citep{Weka2011} that returns the variable \texttt{iparent2}.  The backward slice for the slicing criterion (13, \texttt{iparent2}) is as follows. Starting from statement 13, we look for statements that affect the value of \texttt{iparent2}. The first statement that affects the value of \texttt{iparent2} is $\{6\}$, where the value of \texttt{iparent} is assigned to \texttt{iparent2}. So, whatever value \texttt{iparent} has, \texttt{iparent2} will also have it. Therefore, we have to look for statements that affect the value of \texttt{iparent} as well. The if-statement $\{5\}$ has a control dependency on the value of \texttt{iparent2}, as it determines whether statement $\{6\}$ will be executed or not. Statement $\{4\}$ also specifies the initial value of \texttt{iparent2}. Similarly, if we trace back, statements $\{1, 2, 3\}$ affect the value of \texttt{iparent}, and, consequently, the value of \texttt{iparent2}. As a result, the backward slice for the slicing criterion (13, \texttt{iparent2}) is the set of statements $\{1, 2, 3, 4, 5, 6\}$. A simple backward slicing technique can be used for extract method refactoring.  However, we may not consider statements that determine the value of a variable as a potential method to be extracted when they include the entire body of a method. In such cases, the extracted method would include all statements in the long method, which is not the desired refactoring.

\subsubsection{Block-based slicing}
    
Block-based slicing is a slicing technique that gives several valid method extraction candidates for each slicing criterion \citep{maruyama2001automated}. Block-based slicing limits the expansion\footnote{ Expansion in backward slicing refers to tracing backward from a slicing criterion (which includes a statement and a variable within it) to discover all statements that affect the variable's value.} of a slice within some valid boundaries by creating block-based regions within the body of a method during slicing.  The block-based slicing approach divides a method into smaller units based on the program's control and data flow as regions. Backward slicing starts from a statement in a method and traces back to the beginning of it to identify a set of statements as a slice. The union set of the basic blocks corresponding to the slice's statements is called the slice basic block set. A slice basic block set is a subset of the basic blocks of the method. In backward slicing, the tracing back process may stop at a statement other than the starting point of the method, resulting in a region. In another term, a region is a consecutive set of basic blocks that is a subset of a slice basic block set.

As mentioned in section \ref{sec:Backward} the backward slice of slicing criterion (13, \texttt{iparent2}) is the set $\{1, 2, 3, 4, 5, 6\}$. The mentioned set cannot be extracted as a separate method because the sets of statements $\{1, 2, 3\}$ are being used in other statements such as $\{7, 8, 9, 10\}$.
    
One solution is to keep the set of statements $\{1,2,3\}$ in the original method and duplicate them in the extracted method. This solution rarely becomes valuable when a method is long, and it contains many non-extractable statements since duplicate statements cause other bad smells, e.g., code clone. A better solution is to use block-based slicing in which a slice can expand in the different possible regions. Using block-based slicing, if there is an opportunity to extract a method that does not require duplicating the statements, it is identified and suggested to the user. In addition, the extract method opportunities that lead to code duplication are detected, and the user can select the best extract method candidate from them. Indeed, all possible extract method refactoring opportunities generated by expanding slices to the different regions are recommended. 
    
\begin{figure*}[]
    \centering
    \includegraphics[width=0.75\linewidth]{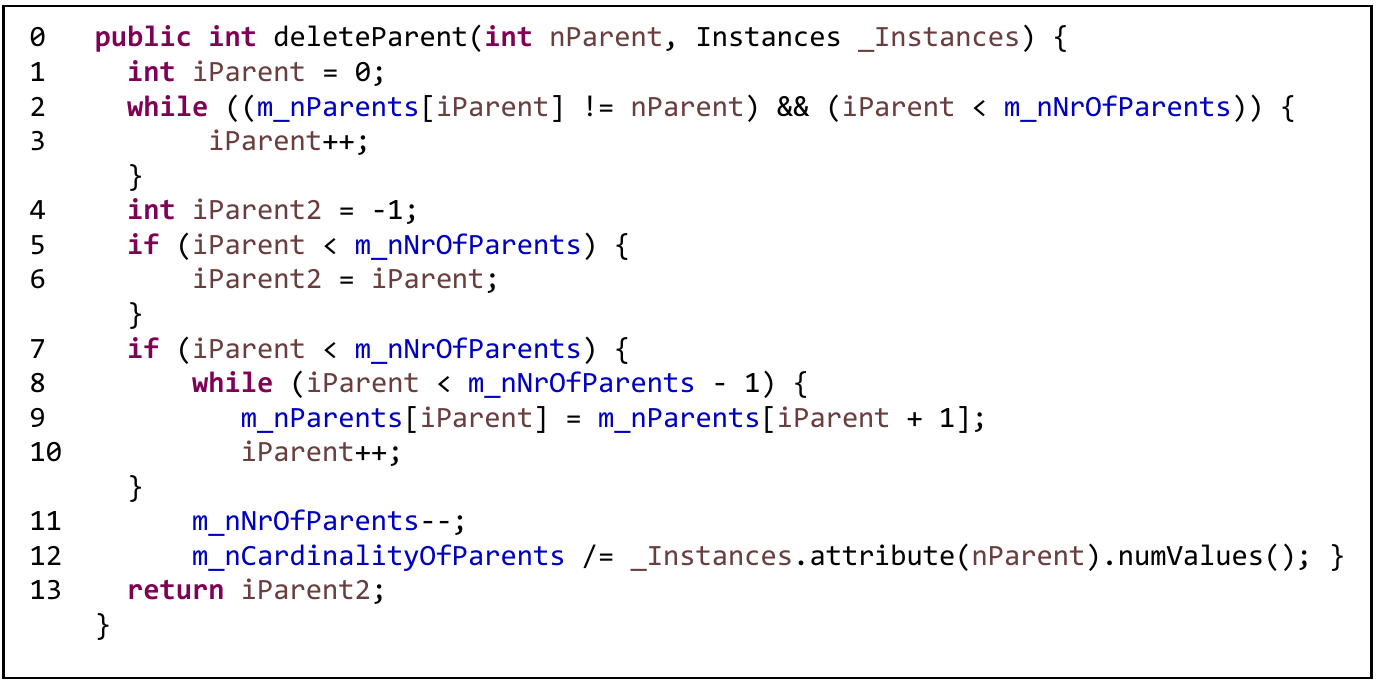}
    \caption{The \texttt{deleteParent} method with PDG node numbers to demonstrate the application of block-based slicing.}
    \label{fig:4}
\end{figure*}

Block-based slicing requires two sets of basic blocks to determine the slicing region, the reachable blocks, and the dominated blocks, computed using CFG and PDG \citep{maruyama2001automated}. We describe the computation of the reachable blocks, the dominated blocks, and block-based slicing for the \texttt{deleteParent} method illustrated in  Figure \ref{fig:4} in the next section.

\subsubsection{Identifying block-based regions}

\begin{figure}[!t]
    \centering
    \includegraphics[width=0.70\linewidth]{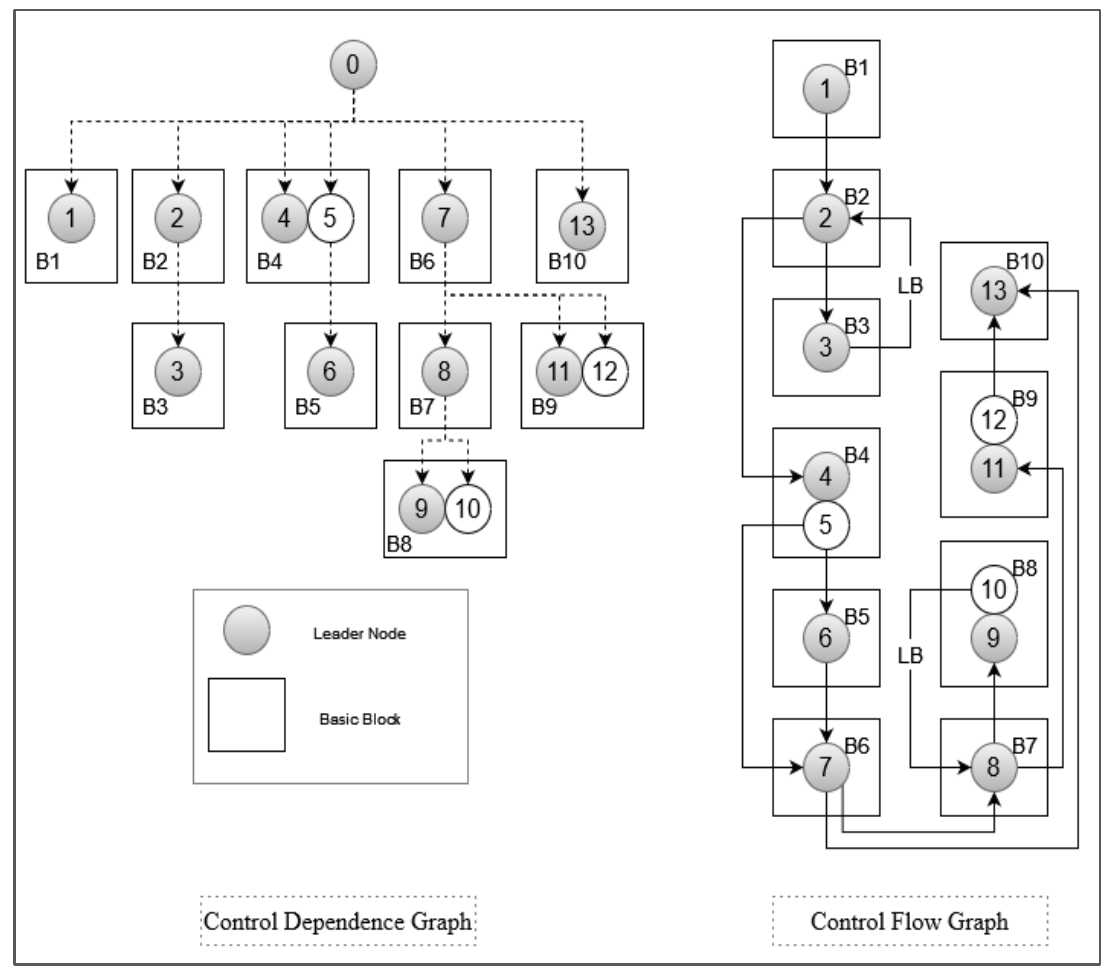}
    \caption{CDG and CFG for method \texttt{deleteParent}, shown in Figure \ref{fig:4} }
    \label{fig:5}
\end{figure}

Figure \ref{fig:5} shows the decomposition of the control flow graph (CFG) into basic blocks and the control dependence graph (CDG) of the \texttt{deleteParent}method. The CDG is constructed from PDG by removing the data dependency edge and keeping the control dependency edges \citep{ferrante1987program}. The CDG represents the nesting of statements inside a method, assuming that the code does not include unstructured control flow or exception flow \citep{tsantalis2011identification}. The following steps are used to determine the block-based regions for basic block 5 containing statement 6:

\textbf{Step1. Computing reachable blocks} \citep{maruyama2001automated}: Reachable blocks for the basic block, $B_{n}$, denoted by $Reach(B_{n})$, are the basic blocks that can be reached from $B_{n}$ without traversing loopback edges in the CFG. For example, the reachable blocks for the basic block, $B_{5}$, in the control flow graph shown in Figure \ref{fig:5} is $\{B_{5}, B_{6}, B_{7}, B_{8}, B_{9}, B_{10}\}$.
        
\textbf{Step 2. Computing dominated blocks} \citep{maruyama2001automated}:

 Dominated blocks for the basic block, $B_{n}$, denoted by $Dom(B_{n})$. In a control dependence graph (CDG), if there is a direct or indirect (transitive) edge from basic block $B_{r}$ to basic block $B_{n}$, it means that basic block $B_{n}$ is control dependent on block $B_{r}$. When basic block $B_{n}$ is control dependent on block $B_{r}$, the set of blocks that are control dependent on block $B_{r}$, is called $Dom(B_{n})$. For example, the dominated blocks for the basic block, $B_{5}$, in the CDG of Figure \ref{fig:5} is basic block $B_{5}$, because $B_{5}$ is the only  basic block that is control dependent on $B_{4}$. As another example, $Dom(B_{7})$ is set $\{B_{7}, B_{8}, B_{9}\}$, because $B_{7}$ is dominated by $B_{6}$, and $B_{8}$ and $B_{9}$ are dominated by $B_{6}$ too. Using domination blocks to determine the slicing region helps to preserve the method's behavior. As mentioned in the previous section, regions are subsets of a slice, and according to them, instead of a whole, some part of a backward slice can be determined. Therefore, domination blocks guarantee that all statements in a region are dominated by same basic block. In other words, all of them are in the same Code Block (e.g., loop statements, conditional statements).
        
\textbf{Step 3. Computing boundary blocks} \citep{maruyama2001automated}:
To calculate boundary blocks for statements in a basic block,$B_{x}$ , first, the sets of reachable blocks, $Reach(B_{n})$, and dominated blocks, $Dom(B_{n})$, for each basic block, $B_{n}$, in method, $m$, is computed. Afterward, the set of boundary blocks for the statement, $x$, which belongs to the basic block, $B_{x}$, are computed as follows. If the basic block, $B_{x}$, is included in the set, $Reach(B_{n}) \cap Dom(B_{n})$, then the basic block, $B_{n}$, is added to the set of boundary blocks for the statement, $x$. For example, the set of boundary blocks of statement 6 in Figure \ref{fig:5}, which belongs to $B_{5}$, are $\{B_{1}, B_{2}, B_{4}, B_{5}\}$. It means that the intersection of  Reach and Dom corresponding to  $B_{1}$, $B_{2}$, $B_{4}$, and $B_{5}$ includes basic block $B_{5}$.
        
\textbf{step 4. Determining block-based regions}
\citep{maruyama2001automated}:
Finally, the block-based region, $R(B_{n})$, for boundary block, $B_{n}$, is the set of nodes that belong to $Reach(B_{n})$. A region is part of a method in which the backward slice can be extended without affecting other parts of the method. The largest region to expand any slice is the entire body of a method, but with the help of the block-based technique, more limited regions may be created. In our example, the basic block $B_{5}$ has four boundary blocks. The region for these four boundary blocks are as follows:

\begin{itemize}
    \item $R(B_{1}) = \{1,2,3,4,5,6,7,8,9,10,11,12,13\}$ statements.
    \item $R(B_{2}) = \{2,3,4,5,6,7,8,9,10,11,12,13\}$ statements.
    \item $R(B_{4}) = \{4,5,6,7,8,9,10,11,12,13\}$ statements.
    \item $R(B_{5}) = \{6,7,8,9,10,11,12,13\}$ statements.
\end{itemize}

The backward slice of variable \texttt{iparent2} defined in  statement 6 is then computed by starting from the statement and expanding at each region, $R(B_{1})$, $R(B_{2})$, $R(B_{4})$, and $R(B_{5})$ which results in four different slices  as follows:
\begin{itemize}
    \item The backward slice expanded at $R(B_{1})$ contain $\{1,2,3,4,5,6\}$ statements.
    \item The backward slice expanded at $R(B_{2})$ contain $\{2,3,4,5,6\}$ statements.
    \item The backward slice expanded at $R(B_{4})$ contain $\{4,5,6\}$ statements.
    \item The backward slice expanded at $R(B_{5})$ contain $\{6\}$ statements.
\end{itemize}

Multiple extract method suggestions can be obtained for a variable in different basic blocks by using block-based slicing. Nevertheless, such regions are insufficient to expand the backward slicing for method extraction in cases where there is more than one refactoring opportunity inside a basic block. Indeed, the smallest region in block-based slicing is a basic block. For example, if a variable is used in two output instructions inside a basic block, its backward slice expands the entire body of the basic block, which may not be desirable. To address this challenge, we propose a new version of the block-based slicing algorithm in which if a method has more than one output instruction, the code fragment extraction is limited to the region between two output instructions.

\subsection{Output-based slicing}

\begin{figure*}[h!]
    \centering
    \includegraphics[width=0.80\linewidth]{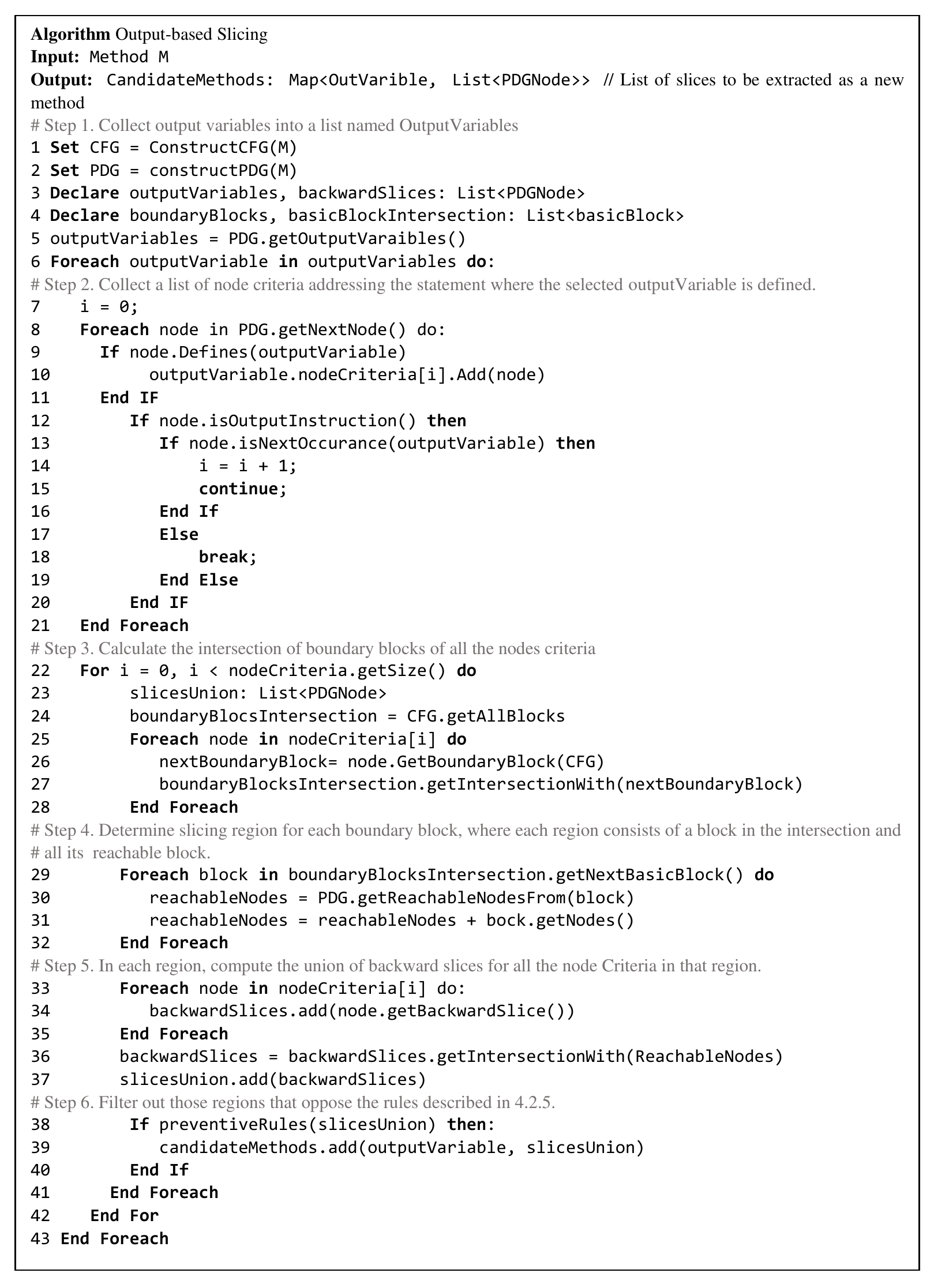}
    \centering\caption{Pseudocode of output-based slicing.}
    \label{fig:6}
\end{figure*}

As discussed earlier in this paper, if a method outputs the value of more than one primitive or reference type (object) variable, we consider it as having more than one responsibility. We describe our Output-based Slicing algorithm to achieve single responsibility while identifying extract method candidates. The Output-based Slicing algorithm is shown in Figure \ref{fig:6} and contains six steps described in the following:
    
In Step 1, a list of output variables/ objects is collected. The remaining steps are repeated in a loop for each output variable, \texttt{OutputVariable}. In Step 2, a list, \texttt{NodeCriteria}, addressing all the statements defining or assigning a value to the next output variable, is collected. If there is more than one output instruction for a specific output variable, then a different criteria list is created for each instruction. In Step 3, the boundary blocks are calculated for each node in the \texttt{NodeCriteria} list. Afterward, the intersection of the boundary blocks, \texttt{boundaryBlocksIntersection}, is calculated. Step 4 determines the reachable region, $R(B_{n})$, from each basic block, $B_{n}$, in \texttt{boundaryBlocksIntersection}. Step 5 computes the backward slice for each node criteria in $R(B_{n})$. The union of these backward slides is considered a candidate for extracting a method. Finally, in Step 6, if a candidate does not violate the rules for controlling the method's behavior and usefulness (discussed in Section \ref{sec:tool}), it is considered a final method extraction proposal.

The statements in an extracted method candidate are moved to a new method during the extract method application. The extracted method is called in the line of the first statement of the selected candidate. It returns the variable on which slicing is performed to the caller method. Not all statements in a generated candidate can be moved to the extracted method during the application phase in all situations. These statements should be repeated on both the original and extracted method to preserve the semantics and behavior of the program after refactoring. A statement can be extracted if and only if it does not appear in the slice of any of the remaining method statements. Otherwise, it is a duplicated or indispensable statement \citep{tsantalis2011identification, maruyama2001automated}.

\subsection{A running example}

\begin{figure*}[h!]
    \centering
    \includegraphics[width=0.8\linewidth]{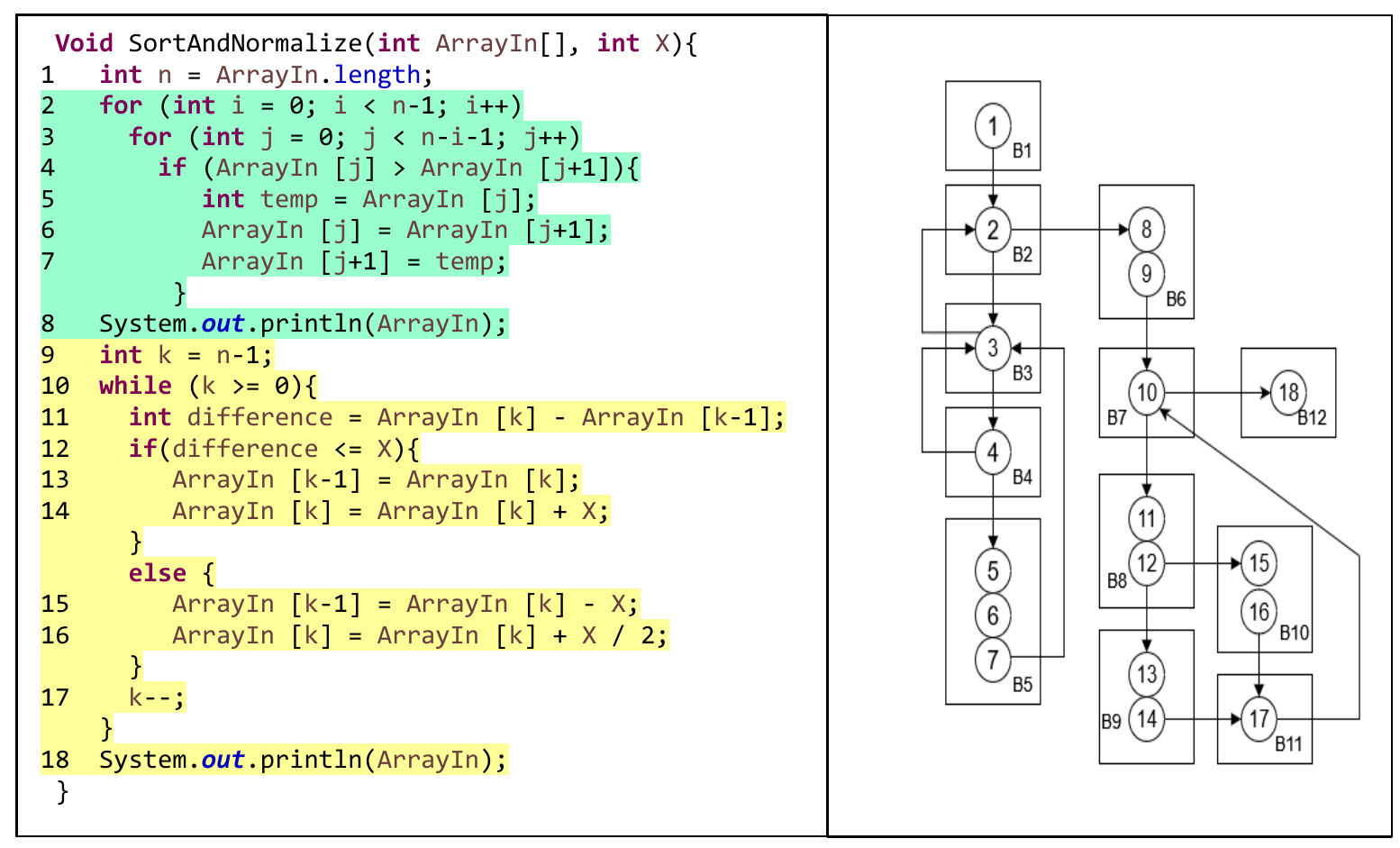}
    \centering\caption{An example to show the capability of the output-based slicing algorithm in extracting single responsible methods.}
    \label{fig:7}
\end{figure*}

Figure \ref{fig:7} illustrates \texttt{SortAndNormalize} method along with the corresponding control flow graph decomposed into basic blocks. It is observed that the method performs two distinct responsibilities of array sorting and kind of array normalization. The result of each responsibility is sent to the output stream separately using the \texttt{System.out.println} in statements 8 and 18 highlighted in Figure \ref{fig:7}. Therefore, our output-based slicing algorithm is used to extract each of these responsibilities. The output variable \texttt{ArrayIn} is used in two different output instructions (statements 8 and  18). Each output instruction can be extracted separately as an extract method refactoring candidate. The output-base slicing algorithm is applied as follows:
    
\textbf{Step 1}: The \texttt{ArrayIn} variable is identified as the output variable.

\textbf{Step 2}: Statements that define the \texttt{ArrayIn} variable are specified as \textit{nodeCriteria}=$\{\{6, 7\}, \{13, 14, 15, 16\}\}$.

\textbf{Step 3}: The sets of boundary blocks for each statement are defined, and the intersection of the sets of boundary blocks is $\{\{B1, B2, B3, B4, B5\}, \{B1, B2, B6, B7, B8\}\}$.
    
\textbf{Step 4}: The region $R(B_{n})$ for each boundary block is:
\begin{itemize}
    \item $B_{1}:\{1, 2, 3, 4, 5, 6, 7, 8, 9, 10, 11, 12, 13, 14, 15, 16, 17, 18\}$
    \item $B_{2}:\{2,3,4,5,6,7,8,9,10,11,12,13,14,15,16,17,18\}$
    \item $B_{3}:\{3,4,5,6,7\}$
    \item $B_{4}:\{4,5,6,7\}$
    \item $B_{5}:\{5,6,7\}$
    \item $B_{6}:\{8,9,10,11,12,13,14,15,16,17,18\}$
    \item $B_{7}:\{10,11,12,13,14,15,16,17,18\}$
    \item $B_{8}:\{11,12,13,14,15,16,17\}$
\end{itemize}
    
\textbf{Step 5.}: For the first output instruction (node 8) the union of backward slices in each region, which is considered as an extract method candidate, is as follows:
    
\begin{itemize}
    \item Extract method candidate based on region $B_{1}:\{1,2,3,4,5,6,7\}$
    \item $B_{2}:\{2,3,4,5,6,7\}$
    \item $B_{3}:\{3,4,5,6,7\}$
    \item $B_{4}:\{4,5,6,7\}$
    \item $B_{5}:\{5,6,7\}$
    \item $B_{6}:\{9,10,11,12,13,14,15,16,17\}$
    \item $B_{7}:\{10,11,12,13,14,15,16,17\}$
    \item $B_{8}:\{11,12,13,14,15,16\}$
\end{itemize}

The final suggestions of the SBSRE tool are presented in step 5, where the first five suggestions correspond to the first output instruction and the rest of them correspond to the second one. According to our results on the above running example, the proposed output-based slicing algorithm can recommend extract method candidates for the same variable by distinguishing between output instructions. It is impossible for existing slicing algorithms, specifically the complete-computation slice and object-state slice \citep{tsantalis2011identification}, to find such suggestions since they do not consider output instructions.

In Figure \ref{fig:7}, the \texttt{SortAndNormalize} method only contains local variables. Therefore, JDeodorant uses the complete-computation slicing algorithm to identify extract method opportunities. This algorithm attempts to provide a computation slice of a variable in block-based regions belonging to dominated blocks as an opportunity to extract the method. The statements that defined the variable \texttt{ArrayIn} are in basic blocks $\{B_{5}, B_{9}, B_{10}\}$. The boundary blocks for basic blocks $\{B_{5}, B_{9}, B_{10}\}$ are $\{B_{1}, B_{2}\}$. The complete-computation algorithm suggested two extract method opportunities for the \texttt{SortAndNormalize} method as follows:
    
\begin{itemize}
    \item Suggestion 1: Lines 1-7 and 9-17 belonging to the $R(B_{1})$
    \item Suggestion 2: Lines 2-7 and 9-17 belonging to the $R(B_{2})$
\end{itemize}

Applying the above suggestion results in appearing two output instructions (lines 8 and 18) consecutively in the remained method, which changes the program behavior. Therefore, JDeodorant \citep{tsantalis2011identification} does not offer any extract method.

\subsection{Rules on behavior preservation and usefulness of suggested refactorings} \label{sec:Rules}

According to Fowler and Beck \citep{Fowler1999}, the behavior preservation of the program is an essential requirement of every refactoring. This section describes a set of rules as the extract method refactoring preconditions concerning the behavior preservation and usefulness of the suggested refactoring. The first four rules are mainly concerned with behavior preservation of the program, and the remaining two rules are used to prevent recommending unuseful extract method candidates. In addition, our tool considers three preconditions rules applied by existing automated refactoring approaches \citep{tsantalis2011identification, komondoor2000semantics}. According to the results of our experiments (see Section \ref{evaluation}),  the use of six proposed preconditions, along with three other ones, highly improves the accuracy of our tool, SBSRE, in extracting single responsible methods from the body of long methods.

\subsubsection{Extracting output instructions regarding their types}

Backward slices of variables used in output instructions are computed to determine the responsibilities of a method. Statements that include output instructions, such as printing and inserting data in a text file or database, must be moved to the extracted method. Through this transfer, output instructions of the original method are reduced, and the responsibilities are assigned to the distinct methods. However, the return statement is an exception due to its specific behavior. A method gets back to its caller method whenever a return statement is executed, and the remaining statements are not executed. The semantics and run-time behavior of a method highly depends on return statements. Hence, a return statement cannot be moved to the extracted method.
        
\vspace{0.5 cm}
\hspace{0 cm}
\fbox{\begin{minipage}{16.25cm}
\textbf{Rule 1}: All output instructions except the return instructions are extracted along with backward slices of the variables used in them. For return statements the backward slice of the output variables are extracted to a new method while the return statement remains in the original method.
\end{minipage}}

\subsubsection{Extracted method parameters without initial values}

When the backward slice is computed, it is considered the extracted method and is called from the original method. If an extracted code fragment contains a variable that is not declared inside the extracted method, the variable must be passed as a parameter to the extracted method. All input parameters must be initialized before calling the extracted method. Consider the \texttt{getMaximimOrMinimum} method shown in Figure \ref{fig:8}. Suppose the extract method suggestion is applied to the nested regions \texttt{$(R_{2}, R_{3})$}. In this case, the extracted method lacks the variable declaration. Indeed, the variable \texttt{result} must be passed as a parameter to the extracted method, which is impossible and rejected by the second rule.
        
\vspace{0.5 cm}
\hspace{0 cm}
\fbox{\begin{minipage}{16.25cm}
\textbf{Rule 2}: A slice proposed for method extraction must not contain a variable that is not declared inside the slice and not been initialized. Otherwise, it is rejected from the candidate refactoring list. 
\end{minipage}}

\subsubsection{Slicing criteria with final variables}

\begin{figure}[h!]
   \centering
    \includegraphics[width=0.55\linewidth]{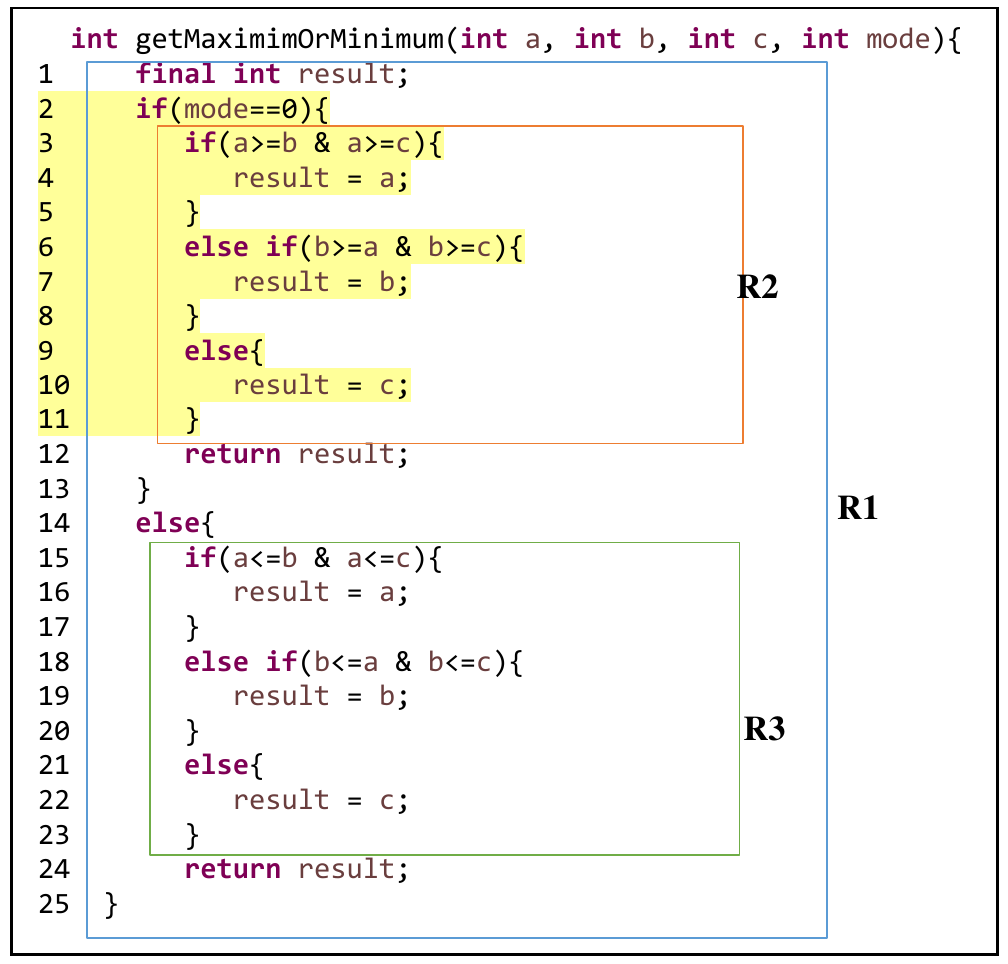}
    \centering\caption{A method with a final variable in its output instructions.}
    \label{fig:8}
\end{figure}

The final variables are constant in the Java language, meaning that they can only be defined once, and their value cannot be changed again. Consider the \texttt{getMaximimOrMinimum} method shown in Figure \ref{fig:8}. This method has two return statements in which the final variable result is used. At least two extract method recommendations can be provided regarding each of these statements. If we apply the extract method candidate in the region, $R_{1}$, considering any return statements, the remaining part of the method will produce an error due to defining the final variable more than one time. For instance, the extracted method candidate based on the result variable of line 12, highlighted in Figure \ref{fig:8}, leads to defining the final variable, \texttt{result}, in region $R_{1}$. After applying this candidate refactoring, the \texttt{result} variable is defined once for getting the output of the extracted method, which is called in line 1, and three times on lines 16, 19, and 22 in the remaining statements. We propose a rule to prevent the recommendation of such candidates.

\vspace{0.5 cm}
\hspace{0 cm}
\fbox{\begin{minipage}{16.25 cm}
\textbf{Rule 3}: The final variable should not be used in the node criterion of backward slices to extract the method, except in the cases that in the original method, the final variable is not defined under any conditions.
\end{minipage}}


\subsubsection{Local variable declaration with the same identifiers in different blocks}

\begin{figure}[h!]
    \centering
    \includegraphics[width=0.60\linewidth]{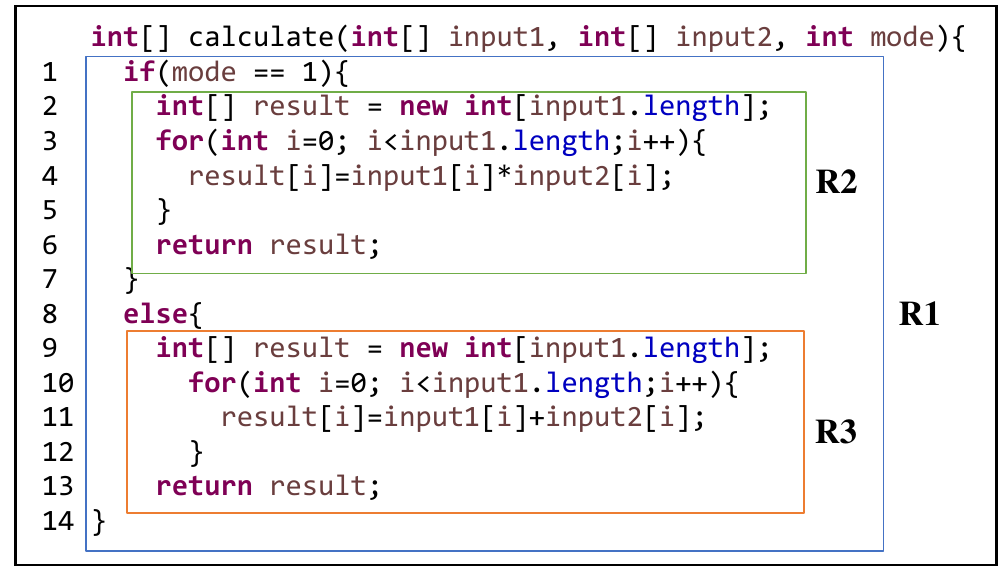}
    \centering\caption{An example of a method contains local variables with identical identifiers at different blocks.}
    \label{fig:9}
\end{figure}

Applying a candidate extract method may lead to declaring variables with the same identifier's name when the extracted statements contain a variable deceleration of the same name in an inner block. This situation causes a semantic error and must be avoided. For example, in Figure \ref{fig:9}, the \texttt{calculate} method has local variables with the same identifier in different blocks along with two return statements. For each of these two return statements, two extract methods candidates are suggested based on block-based slicing on the \texttt{result} variable. For Statement 6: (First candidate: \{$1, 2, 3, 4, 5$\}, Second candidate: {2, 3, 4, 5}) and for Statement 13: (First candidate: \{$1, 8, 9, 10, 11, 12$\}, Second candidate: \{$9, 10, 11, 12$\}). Applying the first extract method suggestion for both return statements produces a semantic error in the remained method. The extracted method is called in region $R_{1}$, leading to the \texttt{result} variable being defined twice, one in Line 1 and another in Line 9. The fourth rule prevents suggesting such extract method candidates. 


        \vspace{0.5 cm}
        \hspace{0 cm}
        \fbox{\begin{minipage}{16.25cm}
        \textbf{Rule 4}: The nesting level of the extracted statements must be greater than the nesting level of the most nested common block of variables with the same identifier’s name.
        \end{minipage}}

\subsubsection{Original and extracted methods expose a single responsibility}

\begin{figure}[h!]
    \centering
    \includegraphics[width=0.55\linewidth]{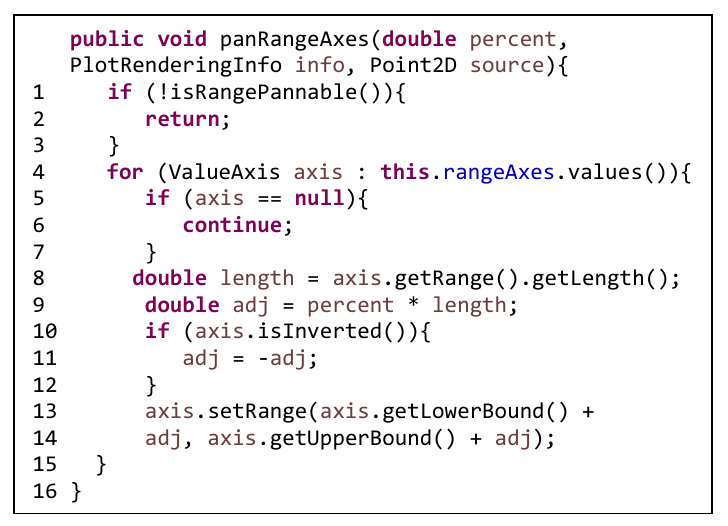}
    \centering\caption{The \texttt{panRangeAxes} method of the JFreeChart project without any extract method refactoring opportunity.}
    \label{fig:10}
\end{figure}

\begin{figure}[h!]
    \centering
    \includegraphics[width=0.85\linewidth]{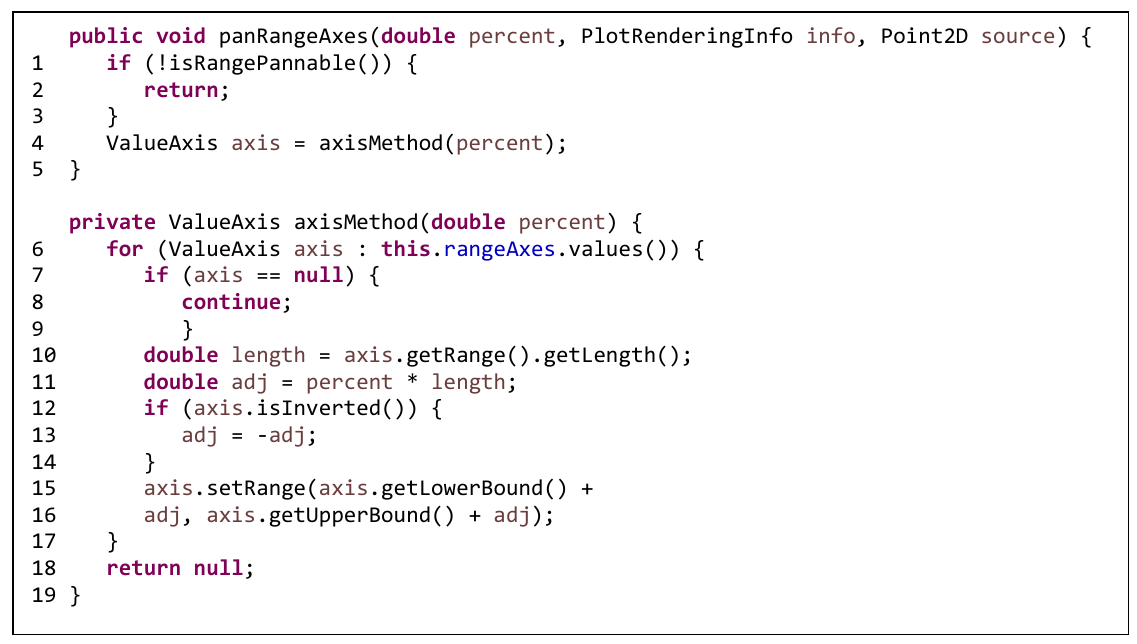}
    \centering\caption{The refactored version of the \texttt{panRangeAxes} method is shown in Figure \ref{fig:10}}
    \label{fig:11}
\end{figure}

The method extraction proposal must separate a part of the method that performs a specific task to extract a method with meaningful responsibility from the original method. In addition, the remaining part of the original method must also have a particular purpose or responsibility. Figure \ref{fig:10} shows the \texttt{plot.CategoryPlot.panRangeAxes} method of JFreeChart \citep{JFreeChart2017} project. All three tools, JDeodorant \citep{tsantalis2011identification}, SEMI \citep{charalampidou2016identifying}, and JExtract \citep{silva2014recommending}, suggest lines (4 to 15) as the extracted method, which is not acceptable as a method with separate responsibility. Figure \ref{fig:11} shows the method \texttt{panRangeAxes} after applying extract method refactoring. The extracted method is \texttt{axisMethod} which performs the corresponding calculation of the remaining method. In this example, the remaining method does nothing except call the extracted method. The fifth rule prevents suggesting such useless extracted method candidates by limiting the minimum counts of the definitions of variables in the remained method. We observed that this rule effectively prevents useless extract method suggestions (Section \ref{sec:rq2}).

\vspace{0.5cm}
        \hspace{0cm}
        \fbox{\begin{minipage}{16.25cm}
        \textbf{Rule 5}: There must be at least one statement in a long method's body that defines a variable, except the extracted statements. 
\end{minipage}}


\subsubsection{High overlapped output instructions slices}

\begin{figure}[h!]
    \centering
    \includegraphics[width=0.6\linewidth]{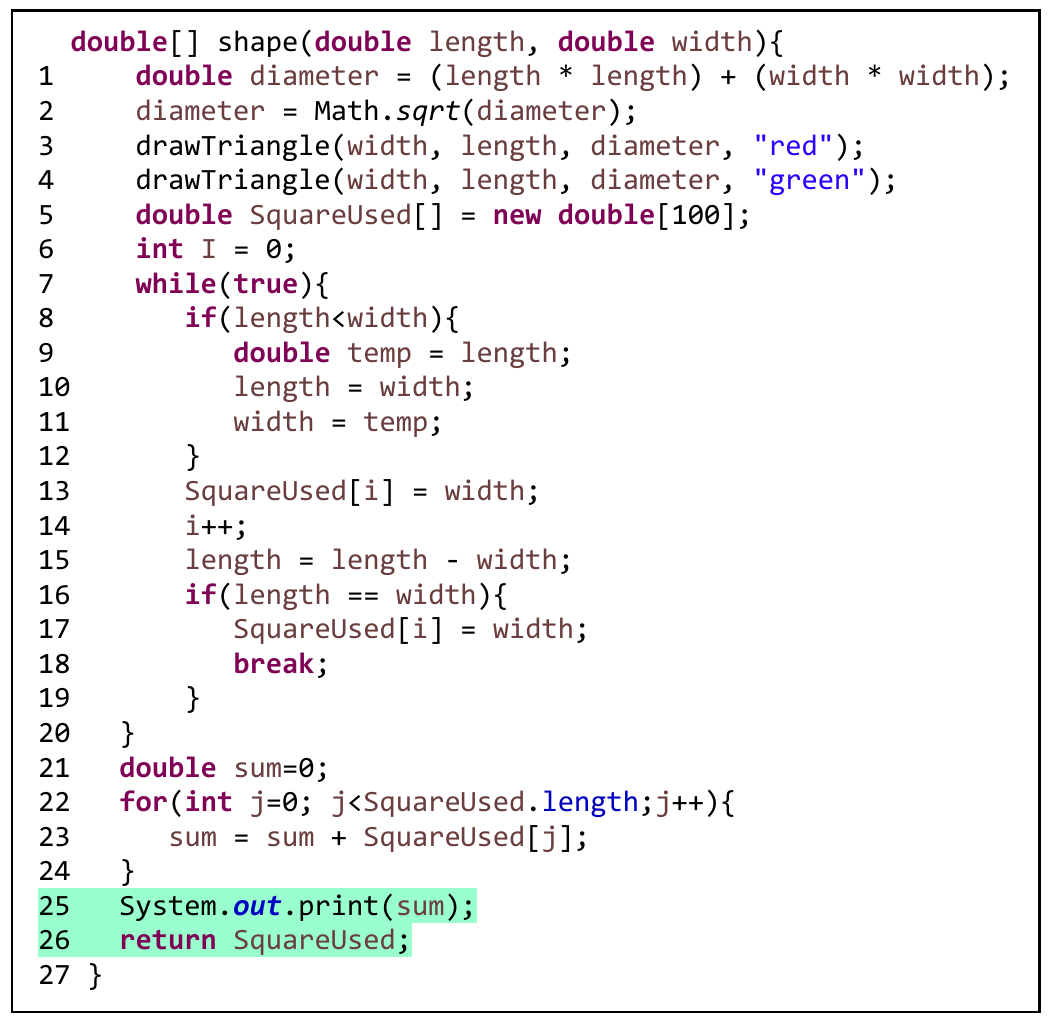}
    \centering\caption{The \texttt{shape} method contains two output instructions with highly overlapped slices.}
    \label{fig:12}
\end{figure}

As mentioned earlier, the method's statements cooperate to fulfill the responsibility of that method. Methods with multiple output instructions are suspected of having more than one responsibility. Nevertheless, not all methods with more than one output instruction violate the single responsibility principle (SRP). Figure \ref{fig:12} illustrates the \texttt{shape} method with two output instructions highlighted in lines 25 and 26.  The backward slice of the output instruction of line 25 includes lines 5 to 25, and the backward slice of the output instruction of line 26 includes lines 5 to 20. Therefore, Assigning separate responsibilities to these output instructions is impossible due to a high overlap of computed slices based on these instructions. Although the full slice of computation for the output instruction 25 (lines 10-25) overlaps a lot with another output instruction, a part of it (lines 21-25) is its non-overlapping part, which is suggested by the SBSRE tool due to the use of the block-based slicing technique.

Inspired by slice-based cohesion metrics \citep{green2009introduction}, we define \textit{slice overlap} as the ratio of the common statements to all statements in output instructions slices to identify methods that do not require refactoring based on a common computation of output instructions. The sixth rule avoids recommending highly overlapped extract method candidates.

\vspace{0.5 cm}
        \hspace{0 cm}
        \fbox{\begin{minipage}{16.25cm}
        \textbf{Rule 6}: The slice corresponding to the output instruction which has the largest common statements with other output instructions is considered as a extract method candidate when the \textit{slice overlap} is high (i.e., $>$ 0.75).
\end{minipage}}
    
\subsubsection{Other rules used in SBSRE}
  In addition to the proposed rules, we consider the three rules mentioned in related works to ensure that the method's behavior is preserved after applying the extract method refactoring. Table \ref{tab:rules} summarizes these rules.

\begin{table}[h!]
\footnotesize
	\centering
	\caption{Rules for maintaining method's behavior after refactoring.}
		\begin{tabular}{m {4cm}  m {12cm}}
\hline	
        Rule	& Description  \\ 	\hline\hline	
        \textbf{Rule 7}: Statements affecting the state of an object are duplicated. & The statements in both the extraction method and the remained method (duplicated statements) must not contain a set of defined composite field variables. Composite variables are variables whose first part is an object reference existing in the original method and the second part is a field \citep{tsantalis2011identification}.\\
        \hline
        \textbf{Rule 8}: Statements that contain a class instance creation are duplicated. & Statements creating the object must not be duplicated. By repeating creating an object in both the remaining and extracted methods, each reference points to a different object in memory, threatening the preservation of the program behavior \citep{tsantalis2011identification}.\\
        \hline
        \textbf{Rule 9}: Preservation of existing anti-dependences & The existence of anti-dependence between the statements extracted from the method and those remaining in the method will change the method's behavior. Consider that statement A is supposed to remain in the method, and statement B is to be extracted. There is an anti-dependency between statement A and statement B if variable x is used in statement A and defined in statement B, and there is a control flow path from A to B \citep{tsantalis2011identification, komondoor2000semantics}.If statement A, which defines the value of variable x, is extracted, the value of variable x will be different from what it was supposed to be when used in statement B. As a result, the program's behavior will be affected after the refactoring. \\
        \hline
        \end{tabular}%
	\label{tab:rules}
\end{table}

\subsection{Tool support SBSRE}
\label{sec:tool}

\begin{figure*}[!t]
    \centering
    \includegraphics[width=0.75\linewidth]{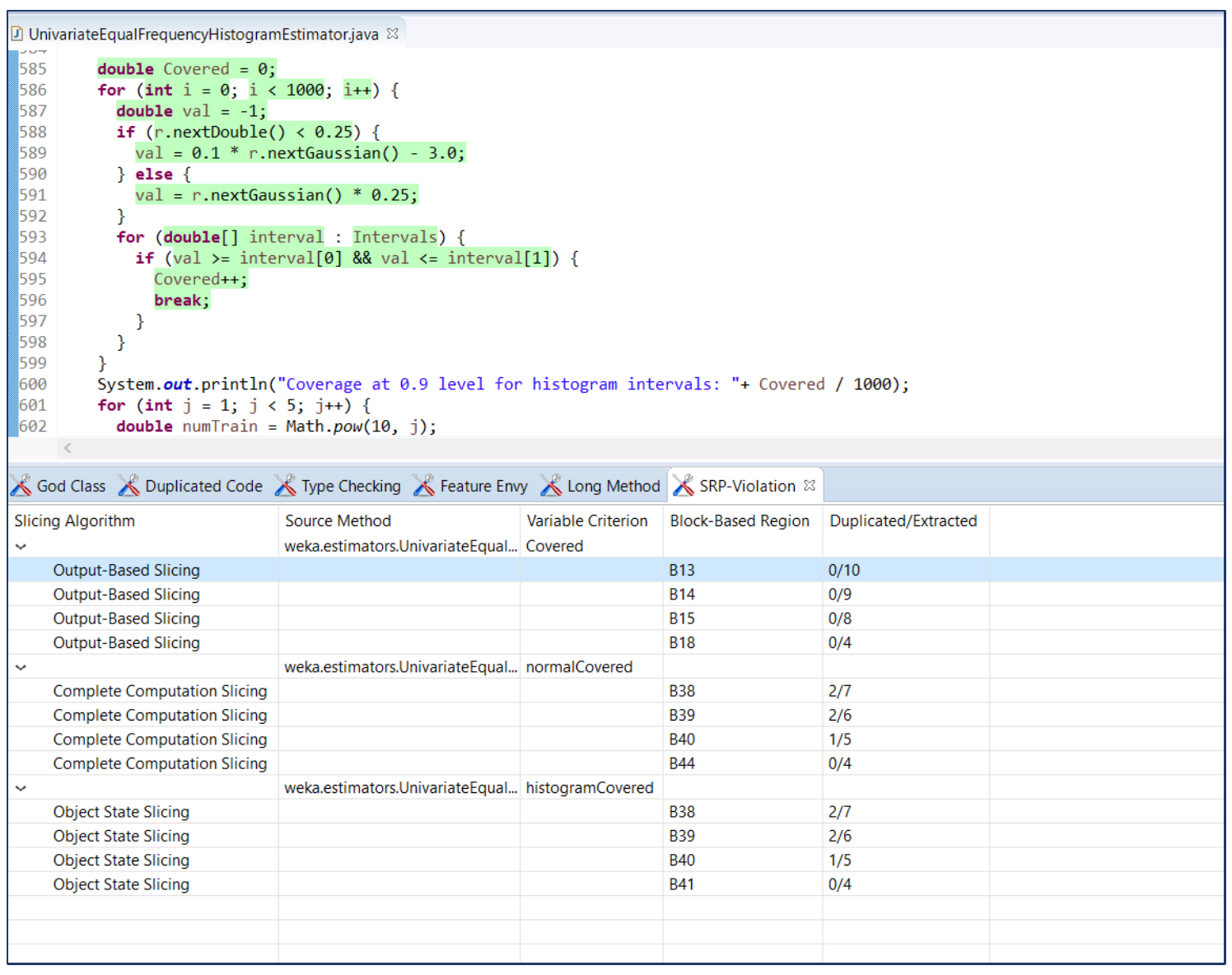}
    \centering\caption{SBSRE tool supports detecting SRP violations in long methods.}
    \label{fig:13}
\end{figure*}

  Our approach is supported by an open-source tool, SBSRE, integrated as an Eclipse plugin to be effectively used by the developers. The tool source code and documentation are publicly available on our GitHub repository \textit{\href{https://github.com/Alireza-Ardalani/SBSRE}{github.com/Alireza-Ardalani/SBSRE.}} The SBSRE program extends the JDeodorant Eclipse plugin \citep{tsantalis2018ten} by adding the SRP-Violation panel that uses output-base slicing and enhances complete-compute computation slice and object state slice algorithms. As the JDeodorant tool can apply extract method refactoring, any extract method opportunity provided by the SBSRE tool can be applied automatically too.  Figure \ref{fig:13} shows the GUI of the proposed SBSRE tool in the Eclipse environment. The extract method candidates identified by our tool are demonstrated to the programmer in the SRP-Violation panel. The first column of this panel shows the slicing algorithm by which the extract method opportunities are identified. The user can select the most appropriate refactoring recommendation based on the information illustrated in the panel. Moreover, users can adjust the ratio of duplicated statements in extract method suggestions. This way, they can obtain only non-duplication extract method suggestions.
    
\section{Evaluation}
\label{evaluation}

This section validates whether our proposed extract method refactoring tool, SBSRE, outperforms the existing state of art techniques provided by JDeodorant \citep{tsantalis2011identification}, JExtract \citep{silva2014recommending}, SEMI \citep{charalampidou2016identifying}, GEMS \citep{xu2017gems}, and Segmentation \citep{tiwari2022identifying}. The experimental results have been evaluated automatically, based on various software metrics, and manually by human experts. These evaluations have been done on eight open-source projects, which is the most comprehensive evaluation in the field of method extraction. Appropriate statistical tests were performed to indicate the significance level of the improvements obtained by our tool. On average, our experimental results show at least a 29.6\% improvement in precision and a 12.1\% improvement in recall of identifying refactoring opportunities compared to the state-of-the-art automated extract method approaches. Furthermore, our tool improves cohesion metrics by an average of 20\% after refactoring. The section is organized as follows. Section \ref{sec:rq} five research questions that drive our experiments, whose settings have been detailed in Section \ref{sec:exp-set}. Afterward, the results of the experiments regarding each research question are discussed in Section \ref{sec:exp-res}.

\subsection{Research questions}
\label{sec:rq}

We developed six research questions (RQs) based on the goals mentioned in our proposal to conduct our experiments and discuss the results:

    \begin{itemize}
        \item \textbf{RQ1:} \textit{How common are methods with multiple output instructions?}
        \item \textbf{RQ2:} \textit{What is the precision and recall of our proposed approach compared to the existing extract method refactoring recommendation approaches?}
        \item \textbf{RQ3:} \textit{How effective is the proposed output-based slicing algorithm on methods with more than one output instruction compared to the existing extract method refactoring tools?}
        \item \textbf{RQ4:} \textit{Compared to the remaining parts of the original method, does the extracted code fragment, recognized as a distinct method, have a single separate responsibility?}
        \item \textbf{RQ5:} \textit{What are the impacts of the proposed extract method refactoring on slice-based cohesion metrics?}
         \item \textbf{RQ6:} \textit{What are the impacts of the proposed extract method refactorings by our tool on the program complexity metrics?}
    \end{itemize}

    RQ1 aims to check the frequency of methods with multiple output instructions in the used dataset during evaluation. RQ2 aims to study the effectiveness of our proposed extract method refactoring approach in finding the long method code smells and fixing them in the different codebases compared to the state-of-the-art tools. RQ3 and RQ4 aim to evaluate the effectiveness of SBSRE on the identification of single responsible methods according to our proposal. Finally, RQ5 and RQ6 aim to assess the impact of applying the recommended refactoring by our tool on systems quality metrics, namely cohesion and complexity metrics.

\subsection{Experimental setting}
\label{sec:exp-set}

Specific case studies were designed to answer each research question outlined in Section \ref{sec:rq}. To answer RQ2, we evaluated the precision and recall of our approach on the GEMS \citep{xu2017gems} dataset. GEMS authors \citep{xu2017gems} have created a dataset that consists of extract method refactorings found in open-source projects, which were confirmed by the developers who applied them. This dataset has been used to evaluate and compare GEMS \citep{xu2017gems} with JDeodorant  \citep{tsantalis2011identification}, JExtract \citep{silva2014recommending}, and SEMI \citep{charalampidou2016identifying} extract method recommendation tools and is publicly available at \textit{\href{https://www.comp.nus.edu.sg/~specmine/gems/gems$\_$datasets.tar.gz.}{https://www.comp.nus.edu.sg/~specmine/gems/gems$\_$datasets.tar.gz.}}
    
    The GEMS dataset includes methods from five Java open-source projects, Junit, JHotDraw, MyWebMarket, SelfPlanner, and WikiDev, constituting the most comprehensive and validated dataset to evaluate the extract method refactoring tools to the best of our knowledge. Table \ref{tab:gems}  provides details of the benchmark. We compared our approach with JDeodorant \citep{tsantalis2011identification}, JExtract \citep{silva2014recommending}, SEMI \citep{charalampidou2016identifying}, GEMS \citep{xu2017gems}, and Segmentation \citep{tiwari2022identifying}  on the GEMS dataset as they are the current state-of-the-art tools in extract method recommendation.
    
\begin{table}[!t]
\footnotesize
	\centering
	\caption{Statistics of the subject systems in the GEMS dataset.}
		\begin{tabular}{ p{2cm}  p{1cm}  p{1cm} p{1cm}  p{1cm} p{2cm} }
			\hline \hfil 
			Project	& Version & ~Classes &	Methods &	Lines of code &	Refactoring opportunities \\
			\hline\hline
			Junit & 3.8.0  &	102 & 23& 3861 &31\\
			JHotDraw & 5.2.0	& 192	& 61 &	9844 &	81\\
			MyWebMarket & 1.0.0 &18 &22 &2007 &28\\
			SelfPlanner	& 1.5.2 & 295 &	30 & 18128 &47\\
			WikiDev	& 2.0.0	& 50 &	17 &6417 &	36\\
            \cmidrule(lr){3-6}
			Total&  &  657&  153 & 40257 & 223\\
			\hline
         \end{tabular}
	\label{tab:gems}
\end{table}

    To answer the remaining research questions (RQ3 to 6), experiments were designed and performed on three software systems, JFreeChart \citep{JFreeChart2017}, Weka \citep{Weka2011}, and Apache Xerces \citep{Xerces2011}. We prefer selecting these Java projects because they are open-source, updated regularly, and large enough to represent sufficient extract method opportunities \citep{mkaouer2016use}  and the scalability of our approach. These projects have been well documented and have a good level of understandability, making them appropriate to be evaluated by our human experts regarding the responsibility of the classes and methods. Furthermore, source code availability makes it possible to reproduce the experimental results. Table \ref{tab:gems2} shows the details of each project.

     As discussed in Section 5.3, the selected projects contain 8118 methods with multiple output statements. The relative frequency of methods with multiple output statements in these projects is 45\% which is higher than multiple outputs methods in the GEMS dataset, the largest extract method refactoring benchmark. Indeed, about 38.6\% of methods in the GEMS dataset are methods with multiple output statements. Using these projects for evaluating the proposed approach, we can ensure the reliability and generalizability of reported results.

\begin{table*}[!t]
\footnotesize
	\centering
	\caption{Statistics of the subject systems studied in addition to the GEMS dataset.}
		\begin{tabular}{ l  p{1cm}  p{6cm} p{1cm}  p{1cm} p{2cm}}
			\hline
	    	\hfil Project&	Version&	Application domain&	Classes&	Methods&	Lines of code\\
			\hline\hline
		    \hfil JFreeChart \citep{JFreeChart2017}	&  1.5.0 &  A comprehensive free chart library for the Java &  1032&  11015 &  132038\\ 
			\hfil Weka \citep{Weka2011} &	3.8.0&	A data mining software toolkit&	3634&	26858&	356294\\
		    \hfil Apache Xerces-J \citep{Xerces2011} &	2.12.0&	A collection of software libraries for parsing, validating, serializing, and manipulating XML.&	1136&	10839&	163525\\
			\hline
         \end{tabular}
	\label{tab:gems2}
\end{table*}

RQ3 discusses the capability of our proposed approach regarding output statements. Therefore, we focused on methods with multiple output instructions to answer RQ3. To do this, first, we selected all methods with more than one output instruction from projects in \ref{tab:gems2}. Afterward, to create the ground-truth refactoring dataset for this experiment, we asked five experts including three experienced Java developers, a Ph.D. candidate, and an M.Sc. student in computer science who are familiar with refactoring techniques to determine the refactoring opportunities in the selected methods. Accordingly, 58 methods were selected by experts to answer  RQ3. The source code and details of these methods are available at \textit{\href{https://github.com/Alireza-Ardalani/SBSRE/tree/main/dataSet/MultipleOutput}{https://github.com/Alireza-Ardalani/SBSRE/tree/main/dataSet/MultipleOutput.}}

To answer RQ4, RQ5, and RQ6, two packages with more lines of code and higher cyclomatic complexity than other packages were selected from each project of Table \ref{tab:gems2}. All methods of these six packages were used for the evaluations. Our proposed tool and the JDeodorant tool \citep{tsantalis2018ten} were used to identify and refactor the methods that expose multiple responsibilities. It should be noted that we only used SBSRE and JDeodorant \citep{tsantalis2018ten} in experiments for RQ4 and RQ5. The reason is that the other extract method refactoring tools, i.e., SEMI \citep{charalampidou2016identifying}, JExtract \citep{silva2014recommending}, GEMS \citep{xu2017gems},  and Segmentation \citep{tiwari2022identifying} cannot automatically apply suggested refactoring operations to the source code. In most cases, suggested extraction lines must be manually pruned before applying to the source code, and these manual changes affect the trustability of evaluation. Our proposed tool could identify 210 methods, and JDeodorant identified 163 methods with at least one refactoring opportunity in the six packages belonging to the projects mentioned in Table \ref{tab:gems2}. Finally, to answer the impacts of the proposed extract method refactorings on program complexity in RQ6, 210 methods found by our tool in selected packages have been analyzed.

\def\niknik{\makecell{210 (SBSRE)\\ 163 (JDeodorant)}}
\def\nikk{\makecell{153}}
\def\nikkk{\makecell{58}}
\def\nikkkk{\makecell{210}}
\def\nikkkkk{\makecell{18040}}
\begin{table*}[h!]
\footnotesize
	\centering
	\caption{Summary of the data sources and experts’ participation for each research question.}
		\begin{tabular}{ m{1cm} m{4cm} m{5cm} m{2cm} m{1.5cm}}
			\hline
	    	RQ &	Project under evaluation & How selected & \# Method used & Using expert \\
			\hline\hline
		    
                RQ1 &	JFreeChart, Weka, Apache Xerces-J and GEMS dataset &	All the methods of JFreeChart, Weka, Apache Xerces-J, and the GEMS dataset & \nikkkkk	& No\\
                
                RQ2	& GEMS dataset	& Collected by the GEMS' authors &	\nikk & No\\ 
			RQ3 &	JFreeChart, Weka, Apache Xerces-J &	Methods with multiple        output instructions &	\nikkk &	Yes\\
		    RQ4 & JFreeChart, Weka, Apache Xerces-J & Methods from the six           packages with the most line of code and cyclomatic complexity	&        \niknik & Yes\\
                RQ5 &	JFreeChart, Weka, Apache Xerces-J &	Methods from the six packages with the most line of code and cyclomatic complexity & 	\niknik & No \\
                RQ6	& JFreeChart, Weka, Apache Xerces-J	& Methods from the six packages with the most line of code and cyclomatic complexity &	\nikkkk	& No \\
			\hline
         \end{tabular}
	\label{tab:Summarizes}
\end{table*}

Table \ref{tab:Summarizes} summarizes the experimental setting used to answer each research question. It also shows the usage of experts' opinions for each research question. Although we have tried to make evaluation results as independent of experts as possible, in two cases, experts' opinions have been used to ensure the applicability of our approach. Five experts have been invited to participate in the evaluations of RQ3 and RQ4. Three of them have more than ten years of Java programming experience with a background in software refactoring techniques. The other two members were students with an academic background in clean code principles whose theses were related to software refactoring.  We asked our academic colleagues to recommend students with refactoring experience. After the interview, two students were invited to participate in the evaluation of this study. Using refactoring tools in industry and on real projects is one of their valuable benefits. Hence, we asked some software project managers we knew in well-known companies to assign some of their Java programmers who were experts in refactoring techniques. During a workshop, evaluators were introduced to this study and briefed on the evaluations they were expected to perform. All five experts have installed the SBSRE tool and participated in this study's evaluation for one month. To participate in RQ3, independent evaluators were tasked with identifying opportunities for the extract method based on their judgment without using SBSRE or other tools. This was termed as a “True Occurrence” (TO) for methods with multiple output instructions. To participate in RQ4, 623 extract method suggestions from the SBSRE tool and 583 from the JDeodorant tool were examined to ascertain whether these suggestions each carried a distinct responsibility.

\subsection{Experiments and Results}
\label{sec:exp-res}

The details and results of our experiments in the settings mentioned above are discussed in the following sub-sections.

\subsubsection{Results for RQ1}

\begin{figure}[h!]
     \centering
         \begin{subfigure}{0.6\textwidth}
              \centering
         
              \includegraphics[width= 1 \linewidth]{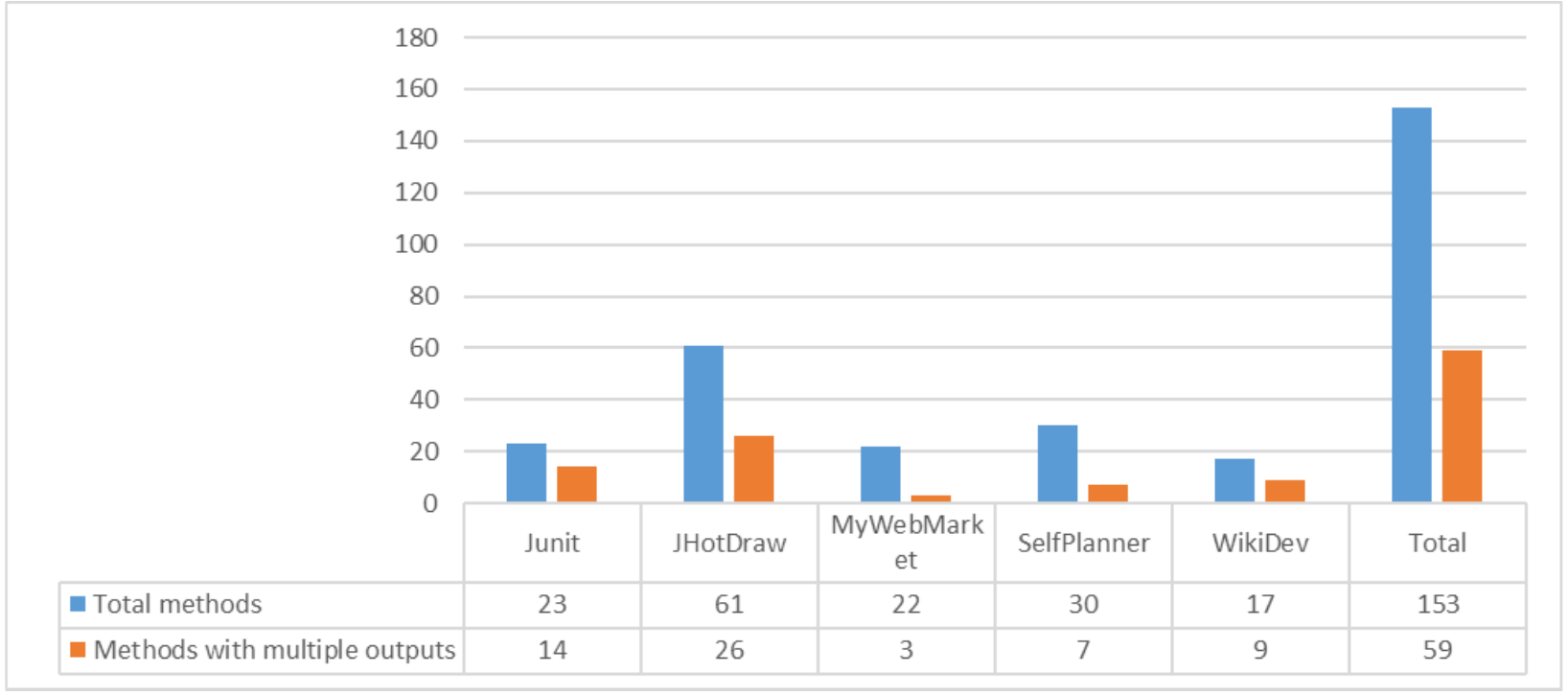}
               \caption{}
         \end{subfigure}
     
         \begin{subfigure}{0.6\textwidth}
            \centering
             \includegraphics[width= 1 \textwidth]{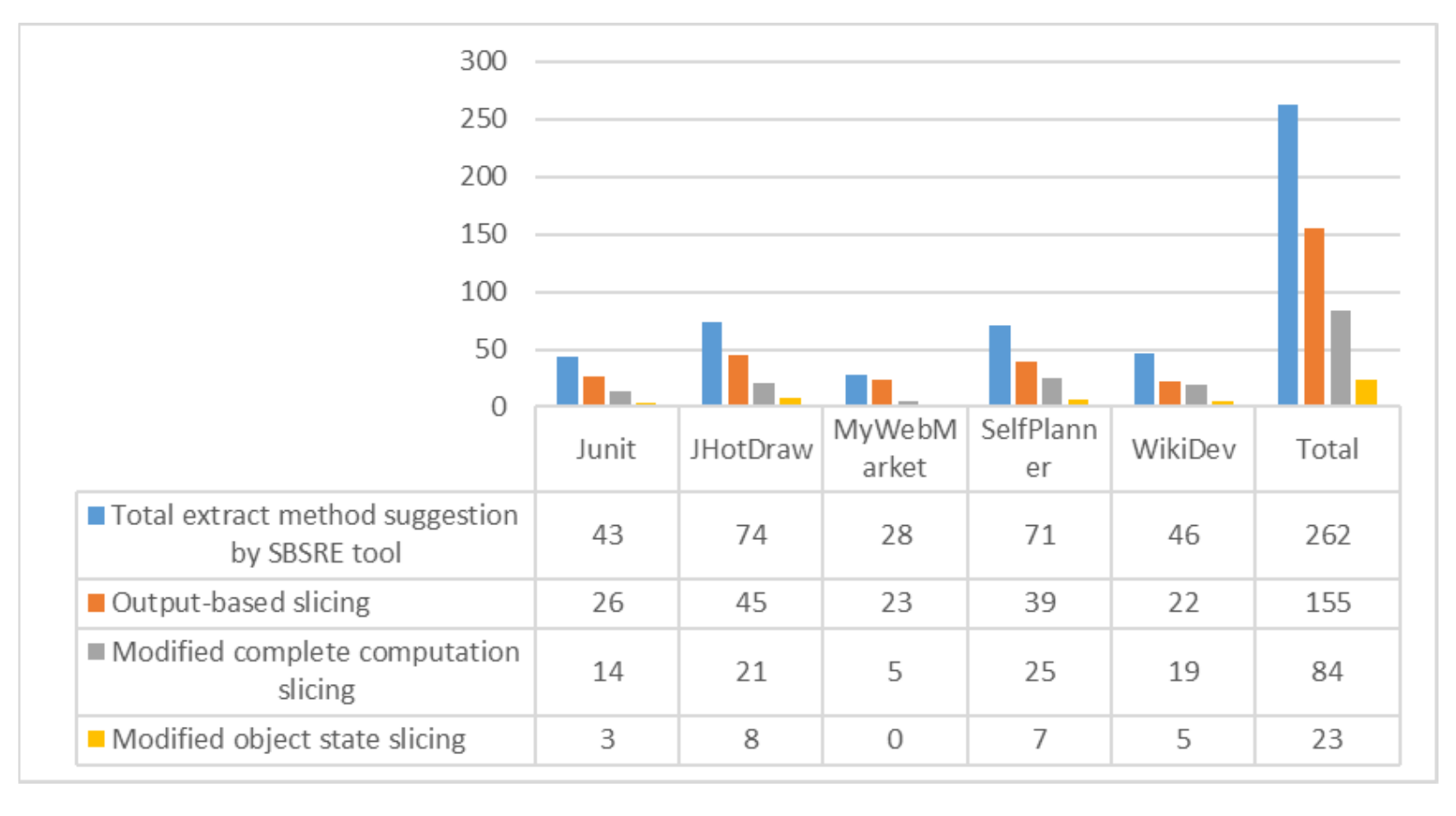}
             \caption{}
     \end{subfigure}
     
        \caption{(a) Frequency of multi-Output methods in the GEMS Data-set. (b) Frequency of slicing algorithms utilization in SBSRE on the GEMS data-set. }
        \label{fig:RQ1}
\end{figure}

Providing refactoring opportunities for methods with multiple output instructions is the main purpose of the SBSRE tool. Hence, this research question investigated the frequency of methods with multiple output instructions. Figure \ref{fig:RQ1},  shows the frequency of methods with more than one output instruction related to the GEMS dataset. According to Figure \ref{fig:RQ1}, nearly 38.6\% of GEMS dataset methods have more than one output instruction. Furthermore, the output-based slicing algorithm generates 59.1\% of the extract method suggestions provided by the SBSRE tool.

\begin{figure}[h]
     \centering
         \begin{subfigure}{0.8\textwidth}
              \centering
         
              \includegraphics[width=0.8\linewidth]{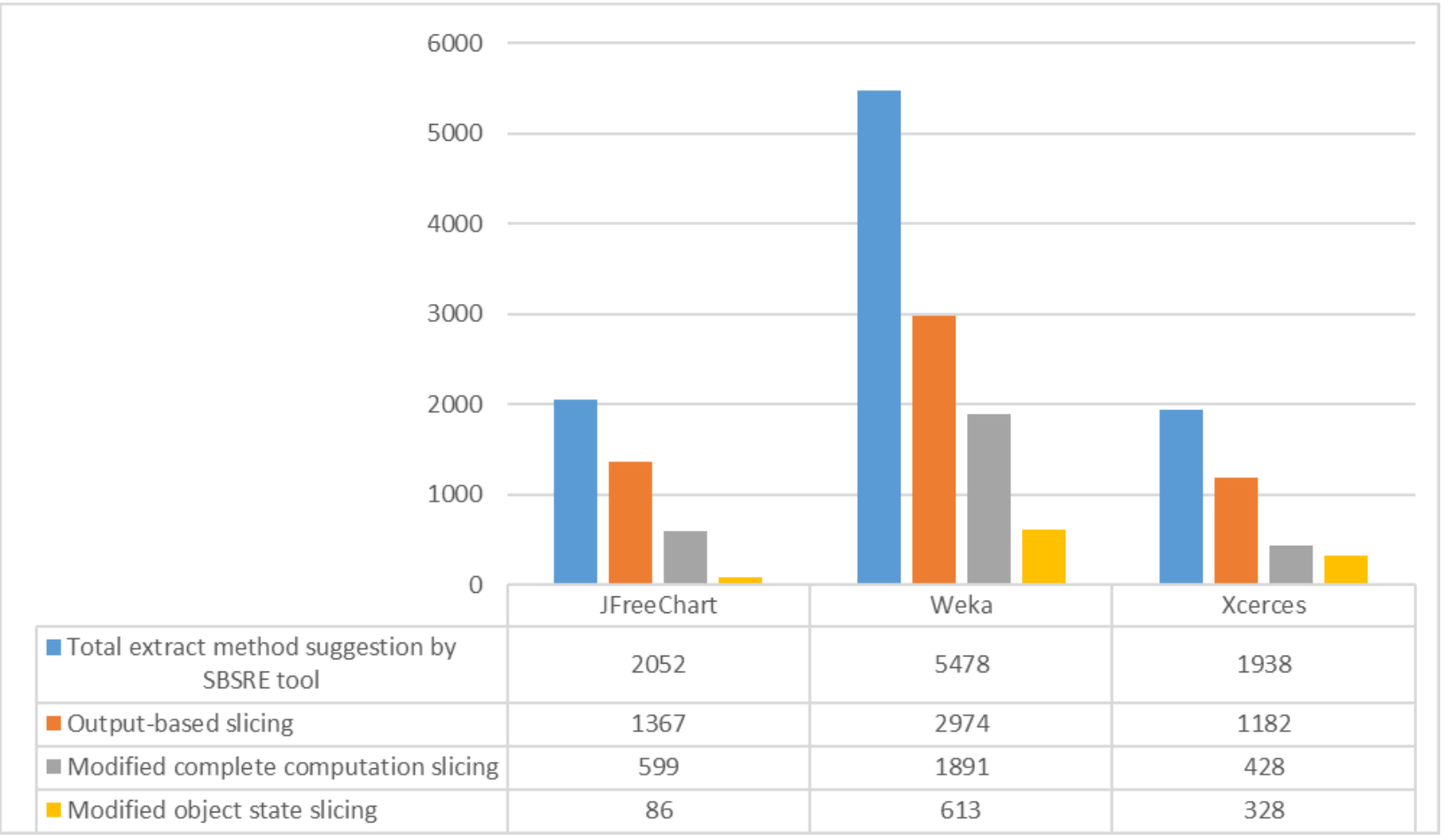}
               \caption{}
         \end{subfigure}
     
         \begin{subfigure}{0.8\textwidth}
            \centering
             \includegraphics[width=0.7\textwidth]{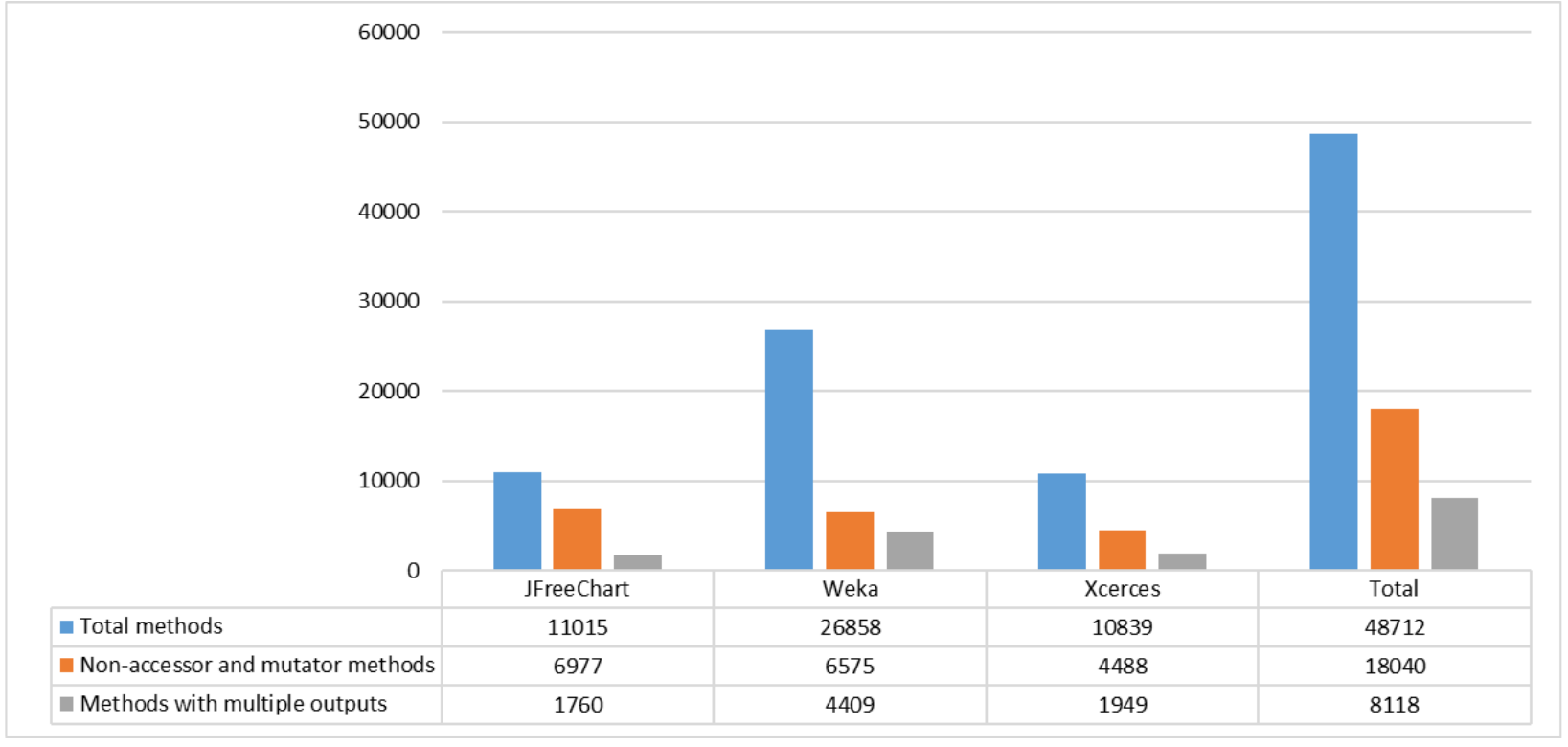}
             \caption{}
     \end{subfigure}
     
        \caption{(a) Frequency of multi-Output methods in JFreeChart, Weka, and Apache Xerces projects. (b) Frequency of slicing algorithms utilization in SBSRE on JFreeChart, Weka, and Apache Xerces projects.}
        \label{fig:RQ11}
\end{figure}

 Figure \ref{fig:RQ11} shows the frequency of methods with more than one output instruction in JFreeChart, Weka, and Apache Xerces projects. This table shows that the number of 8118 methods out of the total of 18040 (45\%) methods belonging to these projects have multiple output instructions. Furthermore, the output-based slicing algorithm generates 58.3\% of the extract method suggestions provided by the SBSRE tool.

 From Figure \ref{fig:RQ1}, and \ref{fig:RQ11}, it can be concluded that methods with multiple output instructions are not rare and exist numerously in software projects. More than half of the extract method suggestions provided by SBSRE tools are related to output-based slicing algorithms. Such a result shows that SBSRE focuses primarily on methods with multiple output instructions and simultaneously tries to discover all extract method opportunities with specific responsibilities in a comprehensive manner by using modified complete computation and object state slicing algorithms.

\subsubsection{Results for RQ2}

The results of evaluating different extract methods tools on the GEMS dataset \citep{xu2017gems} are presented and compared using precision, recall, and f-measure metrics to answer RQ2. We applied our tools and the other five extract method refactoring tools to the methods in the GEMS dataset. Table \ref{tab:confusion} shows the total number of the true occurrences (TO), true-positive (TP), false-positive (FP), and false-negative (FN) cases obtained by the extract method suggestion tools. In Table \ref{tab:confusion}, the second column, "TO," indicates the number of actual refactoring opportunities in the GEMS dataset. Indeed, "TO" represents a set of statements from a method identified as a potential candidate for method extraction. We assess the extraction method suggested by each tool for every method in the GEMS dataset against the corresponding "TO" to determine whether they match or not. The third column indicates the total number of refactoring suggested by each tool for each project. The fourth column, "TP," refers to the number of correctly identified refactoring by each tool. If the suggested method extraction by a tool matches with "TO," it is considered as a "TP." The fifth column, "FP," denotes the number of refactoring opportunities claimed by the software tool but not approved in the dataset. If the proposed method extraction does not match with "TO," it is considered as "FP." Finally, the sixth column, "FN," is the number of refactoring opportunities available,"TO", in the dataset but has not been recognized by the software tool. The precision (Pr), recall (Re), and F-measure (F1) of each refactoring tool on the GEMS dataset have been computed using data in Table \ref{tab:confusion}. The results of evaluation metrics are shown in Table \ref{tab:evaluation}.
        
        It should be noted that "TO," "TP," "FP," and "FN" metrics can also be calculated for each method separately since a method may have more than one refactoring opportunity. Therefore, we compute two types of the \textit{overall} and \textit{average} scores for each evaluation metric, i.e., precision, recall, and f-measure metrics. The \textit{overall} score for each evaluation metric is computed using the values of "TP,” "FP," and "FN” obtained by investigation of all methods. The \textit{average} score is calculated by averaging method-level precision, recall, and f-measure. The computation of overall and average scores for the precision and recall metrics is given by the following equations:
        
\begin{equation}
    \text {Pr}_{o v e r a l l }=\frac{\sum_{m \in  M e t h o d s } T P_{m}}{\sum_{m \in M e t h o d s} T P_{m}+\sum_{m \in  M e t h o d s } FP_{m}}
    \label{eq1}
\end{equation}

\begin{equation}
\text{Pr}_{a v e r a g e }=\frac{1}{\mid M e t h o d s\mid } \sum_{m \in  M e t h o d s } {Pr}_{m}
\label{eq:2}
\end{equation}

\begin{equation}
\text{Re}_ {o v e r a l l }=\frac{\sum_{m \in M e t h o d s} T P_{m}}{\sum_{m \in M e t h o d s} T P_{m}+\sum_{m \in M e t h o d s} FN_{m}}
\label{eq:3}
\end{equation}

\begin{equation}
\text{Re}_ {a v e r a g e}=\frac{1}{\mid \text { Methods } \mid} \sum_{m \in M e t h o d s} Re_{m}
\label{eq:4}
\end{equation}

In Equations \eqref{eq1}, \eqref{eq:2}, \eqref{eq:3} and \eqref{eq:4}, Methods denotes a set of all evaluated methods in the dataset, $Pr_{m}$ is the precision computed for the method, $m$, and $Re_{m}$ is the recall computed for the method, $m$. F-measure is the harmonic mean of precision and recall.

The best values of each metric in Table \ref{tab:evaluation} are bolded. It is observed from Table \ref{tab:evaluation} that our proposed tool, SBSRE, achieved the highest performance in terms of the computed evaluation metrics compared to JDeodorant \citep{tsantalis2011identification}, SEMI \citep{charalampidou2016identifying}, JExtract \citep{silva2014recommending}, GEMS \citep{xu2017gems}, and Segmentation \citep{tiwari2022identifying}. The second-best values of each metric are shown in the italic style. JExtract has achieved the highest recall rate compared to JDeodorant, SEMI, and GEMS since it has suggested the most method extraction opportunities. However, its precision is lower than the JDeodorant due to a relatively high false-positive rate. We observed that the GEMS tool did not suggest any refactoring candidate in many cases, which led to the weak performance of the tool compared to the other tools. 

Output-based slicing algorithm and proposed behavior preservation rules enabled SBSRE to discover more correct extract method opportunities (TP) while providing fewer incorrect extract method opportunities (FP). According to Table \ref{tab:evaluation}, SBSRE improves the average precision and recall of extract method refactoring by a minimum of 29.6 and 12.1\%, respectively. The overall precision and recall are also improved by 27\% and 28\%. Finally, our approach improves the F-measure of extract method refactoring on the GEMS dataset by a minimum of 24.6\% and 30.7\%, respectively, for the average and overall scores.

We used the independent t-test with a 99\% confidence level to determine whether the improvements by our approach compared to the other approach were statistically significant or not. The \textit{null} hypothesis, H0, was defined as no difference between the mean of performance metrics for each pair consisting of our proposed approach and one of the JDeodorant \citep{tsantalis2011identification}, SEMI \citep{charalampidou2016identifying}, JExtract \citep{silva2014recommending},  GEMS \citep{xu2017gems} approaches, and Segmentation \citep{tiwari2022identifying}. The \textit{alternative} hypothesis, H1, was defined as the mean of the performance metrics obtained by our approaches is greater than each of the other approaches. A p-value that is less than or equal to $\alpha (\le 0.01)$ means that H0 is rejected. However, a p-value that is strictly greater than $\alpha (> 0.01)$ means the opposite, indicating results are entirely obtained by chance. Since SBSRE is compared five times in the tests to other tools, the Bonferroni correction have been applied to alpha (it is divided by 5). Table \ref{tab:statistical} shows the statistical test results between our tool and the other tools, considering the different measures reported in Table \ref{tab:evaluation}. It is observed that all p-values are less than 0.01. Therefore, the precision, recall, and F-measure of our tool are significantly higher than JDeodorant \citep{tsantalis2011identification}, SEMI \citep{charalampidou2016identifying}, JExtract \citep{silva2014recommending}, GEMS \citep{xu2017gems}, and and Segmentation \citep{tiwari2022identifying}.
 Moreover, the effect size value of each statistical tests according to the Cohen's d formula is shown in the parenthesis with label "ES"  in Table \ref{tab:statistical}. The effect size value for each statistical test is higher than 0.5 except one, indicating a large or medium difference between precision, recall, F1 achieved by our tools, SBSRE and other tools \citep{cohen2013statistical}.

\begin{table}[!h]
\footnotesize
	\centering
	\caption{The overall confusion matrix elements obtained by executing different refactorin tool on the GEMS dataset.}
		\begin{tabular}{ m{1.5cm}  m{1cm}  m{3cm} m{0.8cm}  m{0.8cm} m{0.8cm}}
			\hline
	    	Tool&	\# TO& \# Suggestion by tool&	\# TP&	\# FP&	\#FN\\
			\hline\hline
		    SBSRE& 223& 	262&	152&	110&	71\\ 
			JDeodorant&	223&	102&	52&	50&	171\\
		    JExtract& 223 &		432&	101&	331&	122\\
		    SEMI& 223&	330&	81&	249&	142\\
		    GEMS& 223 & 186&	44&	142&	179\\
                Segmentation& 223 & 151 & 60 & 91 & 163 \\
			\hline
         \end{tabular}
	\label{tab:confusion}
\end{table}

\begin{table*}
\small
	\centering
	\caption{The average and overall values of the evaluation metrics for each refactoring tool on the GEMS dataset \citep{xu2017gems}.}
		\begin{tabular}{l c c| c c|c c}
			\hline
			\multicolumn{1}{l}{Tool} &
	    	\multicolumn{2}{c}{Precision (\%)} &
            \multicolumn{2}{c}{Recall (\%)} &
            \multicolumn{2}{c}{F1 (\%)} \\
        \cmidrule(lr){2-7}
			 & Average& Overall&	Average&	Overall&	Average&	Overall\\
			 \hline \hline
			 SBSRE&	\textbf{55.6}&	\textbf{58.0}&	\textbf{65.2}&	\textbf{68.1}&	\textbf{57.5}&	\textbf{62.6}\\
			 JDeodorant& \textit{26.0}&	\textit{50.9}& 22.4& 23.3& 22.8 & \textit{31.9}\\
			 JExtract& 25.4&	23.3&	\textit{53.1}&	\textit{45.2}&	\textit{32.9}& 30.7 \\
			 SEMI&	19.4&	24.5&	40.7&	36.3&	25.2&	29.2\\
			 GEMS&	9.6&	23.6&	21.3&	19.7&	12.7&	21.4\\
              Segmentation & 22.2 & 41.0 & 27.1 & 27.8 & 24.4 & 33.1 \\
			 \hline
         \end{tabular}
	\label{tab:evaluation}
\end{table*}

\begin{table}
\scriptsize.
	\centering
	\caption{Result of the statistical test for the performance metrics reported in Table \ref{tab:evaluation}.}
		\begin{tabular}{l c c c}
			\hline
			\multicolumn{1}{l}{Tool} &
	    	\multicolumn{3}{c}{Independent t-test p-value $(\alpha = 0.01)$} 
            \\
 \cmidrule(lr){2-4}
			    &  Precision&	Recall&	 F1\\
			 \hline\hline
			 SBSRE v.s. JDeodorant &	$3.04\times10^{-10}$ (ES = 0.69) &	$3.80\times10^{-18}$ (ES = 1.01)&	$5.30\times10^{-18}$ (ES = 0.87) \\
			 SBSRE v.s. JExtract &	$1.70\times10^{-14}$ (ES = 0.87)& $2.72\times10^{-6}$ (ES = 0.26)&	$5.46\times10^{-14}$ (ES = 0.69)\\
			 SBSRE v.s. SEMI &	$5.42\times10^{-17}$ (ES = 1.03)&	$1.99\times10^{-5}$ (ES = 0.53)&	$1.20\times10^{-12}$ (ES = 0.89)\\
			 SBSRE v.s. GEMS&	$5.00\times10^{-27}$ (ES = 1.42)&	$1.60\times10^{-16}$ (ES = 1.05)&	$5.62\times10^{-24}$ (ES = 1.35)\\
              SBSRE v.s. Segmentation & $2.47\times10^{-13}$ (ES = 0.74) & $3.40\times10^{-11}$ (ES = 0.94)&
              $2.75\times10^{-15}$ (ES = 0.66) \\
			 \hline

         \end{tabular}
	\label{tab:statistical}
\end{table}

\begin{figure}[h]
    \centering
    \includegraphics[width=0.45\linewidth]{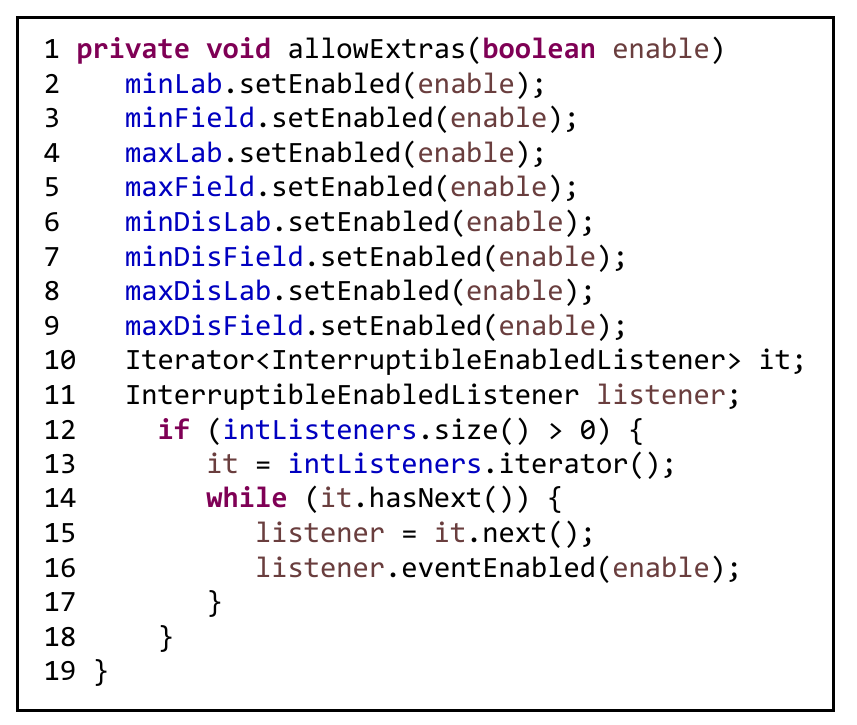}
    \centering\caption{An example to describe false-positive suggestion by \textbf{SBSRE}.}
    \label{fig:14}
\end{figure}

According to Table \ref{tab:confusion}, the SBSRE approach has been reasonably successful in reducing false positives due to leveraging various slicing techniques and prevention rules. We investigated some opportunities where SBSRE may yield false positives and tried to explain the reasons for which. SBSRE suggests extract method candidates using the output-based, complete-computation, and object-state slicing algorithms, all of them are based on enhanced block-based slicing discussed in Section \ref{sec:block-based} Consequently, a backward slice of a variable in a region may present as an extract method opportunity that is not accepted as a true occurrence. We observed that false-positive suggestions are not completely matched with the true occurrence. Indeed, they often cover the true occurrence refactoring opportunity partially, i.e., a subset of the true occurrence statements. For instance, Figure \ref{fig:14} shows \texttt{myplannerapp.TaskPanel.allowExtras} method from the \texttt{SelfPlanner} project. According to the GEMS dataset, statements 10 to 18  should be extracted from this method as a new method. SBSRE offers two extract method candidates as follows:

\begin{itemize}
    \item First suggestion: Lines 10 to 18,
    \item Second suggestion: Lines 10, 12, 13.
\end{itemize}

The first suggestion matches the true occurrence, while the second suggestion covers only a part of that which is not meaningful. It is reasonable that the user discards such sub-optimum refactoring recommendations. On the other hand, these candidates are yet legal due to passing all behavior preservation rules. We conclude that further inhibition rules are necessary to filter out systematically valid refactoring candidates, which are probably not meaningful from the developers' viewpoint.

\begin{table}[!h]
\footnotesize
	\centering
	\caption{The overlap percentage of the True Positive suggestions of each tool with others based on the GEMS dataset \citep{xu2017gems}.}
		\begin{tabular}{ m{1.5cm}  m{0.8cm}  m{1.5cm} m{0.8cm}  m{0.8cm} m{0.8cm} m{1.5cm}}
			\hline
	    	Tool &	SBSRE & JDeodorant &  SEMI &	JExtract &	GEMS & Segmentation\\
			\hline\hline
		    SBSRE & ----- &  34.2 &	38.8 &	46.7 &	23.6 & 28.2 \\ 
			JDeodorant &	100 & ----- &	38.4 &	50 & 25 & 36.5 \\
		    SEMI & 72.8 & 24.6 & ----- &	69.1 &	28.3 & 30.8 \\
		    JExtract & 70.2 &	25.7 &	55.4 &	----- &	25.7 & 31.8 \\
		    GEMS & 81.8  & 29.5 &	52.2 &	59.1 & ----- & 31.8\\
                Segmentation& 71.6 & 31.6 & 41.6 & 50 & 23.3 & ----- \\
			\hline
         \end{tabular}
	\label{tab:OverLapTP}
\end{table}

The last issue we explored on the GEMS dataset is the overlap of True Positive (TP) suggestions of each tool with others, shown in Table \ref{tab:OverLapTP}. To fill the values of Table \ref{tab:OverLapTP}, we calculated the percentage of common TP values of each tool with other tools out of the total number of TP values of that tool. For example, SBSRE and JDeodorant have 52 TP values in common. In the first row of Table \ref{tab:OverLapTP}, which is related to SBSRE, we divided 52 by the total number of TP values of SBSRE, which is 152. As a result, the value of the first row and the second column is equal to 34.2\%. For the second row and the first column, we divided 52 by the total number of TP values of JDeodorant, which is 52. This means that JDeodorant has a 100\% overlap with SBSRE. It can be said that all the correct suggestions (TP) that JDeodorant has discovered have also been discovered by our proposed tool, SBSRE, but JDeodorant has only been able to discover about a third of the TP suggestions of SBSRE. From the values in Table \ref{tab:OverLapTP}, it can be seen that all the tools overlap more than 70\% with SBSRE, which implies that if SBSRE is used instead of any tool, 70\% of their correct suggestions will be covered. On the other hand, no tool has been able to provide more than half of the correct suggestions provided by SBSRE. On average, JExtract and SEMI cover only 55\% and 45\% of the TP suggestions of other tools, respectively, which is lower than the average of SBSRE, which is 79\%.

\subsubsection{Results for RQ3}
\label{sec:rq2}

We invited experts to manually evaluate our tool on multiple-output methods and indicate the effectiveness of the proposed output-based slicing algorithm in identifying single responsible methods according to our proposal. To the best of our knowledge, there is no such data set, and we had to use experts' opinions. As shown in RQ1, the JFreeChart \citep{JFreeChart2017}, Weka \citep{Weka2011}, and Apache Xerces \citep{Xerces2011} projects contained 8,118 methods with more than one output instruction. Reviewing this volume of methods by the evaluators was very time-consuming, therefore they randomly selected 1\% of these methods (81 methods) for evaluation. Because these methods were chosen randomly, they may not have violated the single responsibility principle. Five independent evaluators have a consensus on 58 methods that violate the SRP. The main task of the independent evaluators was to determine the parts of 58 methods that should be extracted as true occurrences. Each of the evaluators independently determined the true occurrences of extract method opportunities. Then, they discussed their findings and reached a consensus on the final true occurrences for each method. 

\begin{figure*}[h!]
    \centering
    \includegraphics[width=0.75\linewidth]{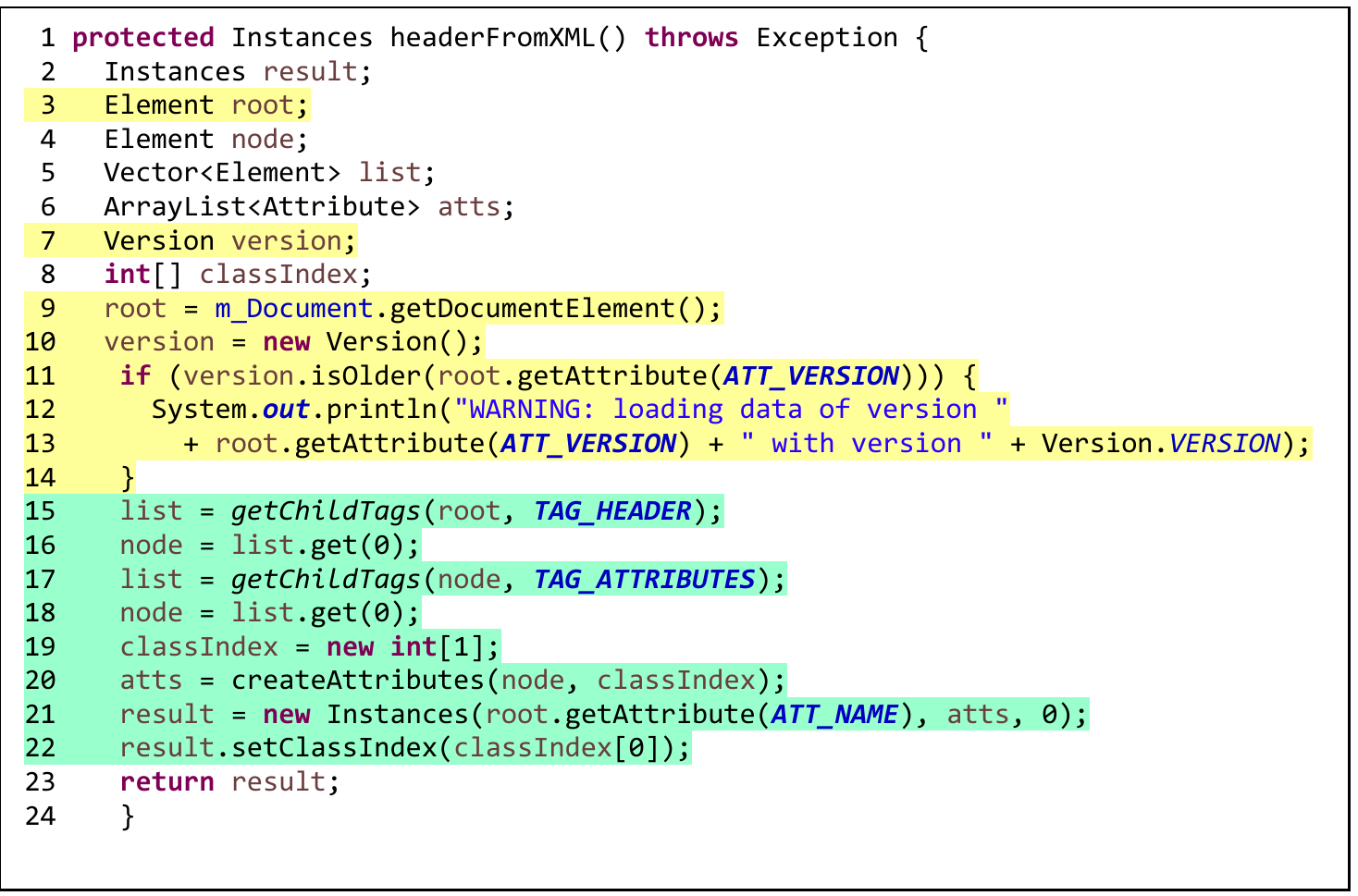}
    \centering\caption{The \texttt{headerFromXML} method in Weka with two output instructions.}
    \label{fig:15}
\end{figure*}

 Figure \ref{fig:15} shows the method \texttt{headerFromXM} from the Weka project \cite{Weka2011}. This method has two output instructions in line numbers 13 and 23. The final True Occurrence (TO) for this method are (9-14) and (16-22). These lines correspond to the computation of each output, and any extract method suggestion that matches them is considered as TP. Furthermore, if an extract method suggestion has more or fewer variable declarations than TO, it is also considered as TP. The reason is that the extracted method parameter can compensate for the missing variable declaration. For instance, SBSRE suggested line numbers (3 and 7 and 9-14) as an extract method. This suggestion has two extra variable declarations that are only used in the line numbers (9-14); therefore, it is labeled as TP. As another relaxation, our experts decided to accept one or two lines of code difference between TO and suggestion, as long as it does not change the program logic or behavior. For example, JDeodorant suggested line numbers (16-20). These lines are a valid subset of TO and do not affect the program logic, and their difference with TO is two lines of code.
 
 Table \ref{tab:refactoring} shows the detailed confusion matrix elements obtained by applying different extract method refactoring tools, including ours. All the refactoring tools have been applied to the same methods from the open-source projects JFreeChart \citep{JFreeChart2017}, Weka \citep{Weka2011}, and Apache-Xerces \citep{Xerces2011} to make a fair comparison. The first column of Table \ref{tab:refactoring} shows the method IDs whose complete names and codes are available at \textit{\href{https://github.com/Alireza-Ardalani/SBSRE/blob/main/dataSet/MultipleOutput/RQ2.md}{https://github.com/Alireza-Ardalani/SBSRE/blob/main/dataSet/MultipleOutput/RQ2.md}}. Other columns are defined similarly to the experiment in the previous section. "TO" (true-occurrences) indicates the number of proposed refactoring opportunities suggested by the independent expert. "TP" (true-positive) is a measure of how many refactoring opportunities have been claimed by both the expert and the refactoring tool in common. "FP" (false-positive) is a refactoring opportunity claimed by the software tool but not approved by the expert. Finally, "FN" (false-negative) is a refactoring opportunity claimed by experts but not the software tool. From the last row of Table \ref{tab:refactoring}, it can be concluded that, in total, our tool has discovered more correct refactoring opportunities than other tools in methods with multiple outputs. At the same time, it has offered relatively fewer unnecessary suggestions for refactoring such methods.
        
        Table \ref{tab:evaluation2} shows the average and overall accuracy of the identified refactoring opportunities by different refactoring tools. Equations \eqref{eq1} to \eqref{eq:4} have been used to compute the performance metrics similar to the previous experiment. To compute the average score for performance metrics, we first calculated metrics at the method level for each of the 58 methods based on their TP, FP, and FN, then averaged them. The \textit{overall} scores have been calculated using the values in the last row of Table \ref{tab:refactoring}, which is the sum of the TP, FP, and FN of all methods.
        
        The best values of each performance metric are bolded in Table \ref{tab:evaluation2}. The second best values of each metric are shown in the italic style. As shown in Table \ref{tab:evaluation2}, our approach is more accurate than JDeodorant \citep{tsantalis2011identification}, SEMI \citep{charalampidou2016identifying}, JExtract \citep{silva2014recommending}, GEMS \citep{xu2017gems}, and and Segmentation \citep{tiwari2022identifying} in extracting methods with a single responsibility according to experts' judgments. It is also observed that the SEMI recall is higher than JDeodorant, JExtract,  GEMS, and Segmentation in extracting single responsible methods from multiple output methods. However, the precision of JDeodornat is higher than other approaches. It concludes that slice-based extract method refactoring approaches recommends fewer candidates than other approaches, but most of them are meaningful. On the hand, approaches such as SEMI recommend more refactoring candidates. Among their candidates, more illegal refactoring appears compared to slice-based approaches. SBSRE improves the recall of the slice-based extract method refactoring approaches. 
        
        The proposed SBSRE tool pays specific attention to each output provided by a method. As a result, it achieves relatively higher recall, precision, and F-measure rates regarding both the overall and average scores among all other tools. The minimum improvement achieved by SBSRE is 40.8\% (72.1\% – 31.3\%) for the F-measure average and 28.8\% (70.8\% – 41.9\%) for the F-measure overall.
        Similar to the previous experiment, the independent t-test with a 99\% confidence level was used to determine whether the improvements were statistically significant or not.

\begin{table*}[h!]
\small
	\centering
	\caption{The average and overall accuracy of the proposed refactoring opportunities based on the expert's opinions.}
		\begin{tabular}{lcc|cc|cc}
			\hline
			\multicolumn{1}{l}{Tool} &
	    	\multicolumn{2}{c}{Precision (\%)} &
            \multicolumn{2}{c}{Recall (\%)} &
            \multicolumn{2}{c}{F1 (\%)} \\
            \cmidrule(lr){2-7}
			 & Average& Overall&	Average&	Overall&	Average&	Overall\\
			 \hline \hline
			 SBSRE&	\textbf{72.0}&	\textbf{66.6}&	\textbf{81.3}&	\textbf{75.5}&	\textbf{72.1}&	\textbf{70.7}\\
			 JDeodorant& \textit{36.3}&	\textit{62.2}& 29.7& 31.6& 230.9 & \textit{41.9}\\
			 JExtract& 26.7&	26.7&	43.5&	38.7&	\textit{31.3}& 31.5 \\
			 SEMI&	24.4&	27.2&	\textit{46.4}&	\textit{40.8}&	30.9&	32.6\\
			 GEMS&	08.6&	21.7&	17.2&	15.3&	11.2&	17.9\\
              Segmentation & 25.5 & 36.8 & 20.4 & 29.4 & 28.5 & 32.6 \\
			 \hline
         \end{tabular}
	\label{tab:evaluation2}
\end{table*}

Table \ref{tab:statistical2} shows the statistical test results for different measures reported in Table \ref{tab:evaluation2}. It is observed that all p-values are less than 0.01. Therefore, the precision, recall, and F-measure of our tool are significantly higher than JDeodorant \citep{tsantalis2011identification}, SEMI \citep{charalampidou2016identifying}, JExtract \citep{silva2014recommending}, GEMS \citep{xu2017gems}, and  Segmentation \citep{tiwari2022identifying} in extracting single responsible methods from methods with multiple outputs. Moreover, The effect size values mentioned in Table \ref{tab:statistical2} indicate a large ( > 0.5) difference between SBSRE  results and other tools.

\begin{table}[h!]
\scriptsize
	\centering
	\caption {Result of the statistical test for the performance metrics reported in Table \ref{tab:evaluation2}.}
		\begin{tabular}{lccc}
			\hline
			\multicolumn{1}{l}{\begin{tabular}{l}
                   Tool
            \end{tabular}} &
	    	\multicolumn{3}{c}{Independent t-test p-value $(\alpha = 0.01)$} 
            \\
            \cmidrule(lr){2-4}
			    &  Precision&	Recall&	 F1\\
			 \hline \hline
			 SBSRE v.s. JDeodorant &	$4.34\times10^{-7}$ (ES = 0.90) &	 $5.70\times10^{-10}$ (ES = 1.44)&	$1.20\times10^{-9}$ (ES = 1.19) \\
			 SBSRE v.s. JExtract &	$2.60\times10^{-14}$ (ES = 1.54) &	$1.20\times10^{-5}$ (ES = 0.99)&	$1.48\times10^{-11}$ (ES = 1.39)\\
			 SBSRE v.s. SEMI &	$5.42\times10^{-17}$ (ES = 1.74)&	$1.99\times10^{-5}$ (ES = 0.92)&	$1.21\times10^{-12}$ (ES = 1.47)\\
			 SBSRE v.s. GEMS&	$5.00\times10^{-27}$ (ES = 1.72)&	$1.60\times10^{-16}$ (ES = 2.99)&	$5.62\times10^{-24}$ (ES = 1.51)\\
              SBSRE v.s. Segmentation & $3.14\times10^{-12}$ (ES = 1.43)& $2.68\times10^{-7}$ (ES = 1.12)& $4.20\times10^{-14}$ (ES = 1.35) \\
			 \hline
         \end{tabular}
	\label{tab:statistical2}
\end{table}

To better understand the results, we consider the  \texttt{headerFromXML} method of the Weka project \citep{Weka2011} demonstrated in Figure \ref{fig:15}. The method contains two output instructions, a \textit{print} statement in line 12 and a \textit{return} statement in line 23. The following line numbers, enclosed in parenthesis, are suggested by each of the tools for extract method refactoring:

\begin{itemize}
    \item \textbf{Tool 1: SBSRE (proposed in this work).} Suggestion 1: (3 and 7 and 9-14), suggestion 2: (16-22).
    \item \textbf{Tool 2: JDeodorant} \citep{tsantalis2011identification}. Suggestion 1: (3-18), suggestion 2: (16-20).
    \item \textbf{Tool 3: JExtract} \citep{silva2014recommending}. Suggestion 1: (6-8), suggestion 2: (7-14), suggestion 3: (4-6).
    \item \textbf{Tool 4: SEMI} \citep{charalampidou2016identifying}. Suggestion 1: (7-21), suggestion 2: (5-14), suggestion 3: (8-14).
    \item \textbf{Tool 5: GEMS} \citep{xu2017gems}. No suggestion.
    \item \textbf{Tool 6: Segmentation} \citep{tiwari2022identifying}. No suggestion.
\end{itemize}

The first suggestion of the JDeodorant tool \citep{tsantalis2018ten} covers a large part of the method, line numbers 3 to 18, which is no different in terms of responsibility from the original method. Its second suggestion covers line numbers 15 to 20, tightly cooperating to compute the variable \texttt{atts} value. However, it does not consider the print statement, which is a responsibility of the method and displays some different results.
The first and the third suggestions of the JExtract tool \citep{silva2014recommending} are to extract the line numbers 6 to 8 and 4 to 6 as two different methods, which is wrong. The reason is that JExtract considers structural dependencies as a basis for method extraction. In this example, the lines 6 to 8 and 4 to 6 are syntactically related because they are all declarations. Similarly, the second suggestion, line numbers 7 to 15, includes an "if" construct and its prerequisites. The only difficulty concerning this suggestion is line number 8, which does not do anything with the "if" construct.
The first refactoring recommendation of the SEMI tool \citep{charalampidou2016identifying} includes almost all the lines of the method under study. The second and third SEMI suggestions for the extract method opportunity were to separate the lines affecting the values displayed by the print instruction. The problem is to consider line 8 as having an impact on the print without any connection to it. With our SBSRE tool, two different responsibilities can be separated properly by providing two slices for each output instruction highlighted in Figure \ref{fig:15}.

\subsubsection{Results for RQ4}
\label{sec:resrq3}

 Regarding RQ4, we use our tool to evaluate the identification of extract method opportunities that fulfill a specific responsibility. These opportunities can separate the responsibility from the method and make it a single responsible method.  The main question we try to answer in this experiment is whether the extracted method has a functionality, purpose, or responsibility distinct from the original method. According to Hora et al. \citep{hora2020characteristics}, the extracted and the original methods are slightly different in most cases and almost do one thing. Therefore, this experiment is necessary to measure the efficiency of our approach in extracting different responsibilities from the original method.
 
 This question has been answered with a similar experiment that Tsantalis et al. \citep{tsantalis2011identification} conducted with their refactoring tool, JDeodorant \citep{tsantalis2018ten}. After identifying all extract method candidates with our proposed tool and JDeodorant \citep{tsantalis2018ten} in six selected Java packages from JFreeChart \citep{JFreeChart2017}, Weka \citep{Weka2011}, and Apache Xcerces \citep{Xerces2011} , we asked experts to review and evaluate them to ensure they followed the single responsibility principle. The five independent evaluators have reviewed the extract method opportunities suggested by SBSRE and JDeodorant tools and labeled them as approved or disapproved in this experiment. Then, the final label for each suggestion by the majority vote of the evaluators was determined. Since we used the same independent evaluators in all the RQ3 and RQ4, any extract method suggestion provided by the SBSRE tool and JDeodorant in RQ3 labeled as false positive is also considered a disapproved suggestion in this experiment. Therefore, there is no inconsistency between these two experiments conducted by independent experts. According to Table \ref{tab:resultstab11}, 75\% of JDeodorant's and 91\% of SBSRE's recommendations were accepted and had a particular responsibility.  We apply the chi-square statistical test on Table \ref{tab:resultstab11} data in two manners. The first test was performed to indicate that the obtained results do not depend on the projects used in our experiments. The second test was performed to show that extracting single responsible methods depends on the selected refactoring tool. For the first one, the null hypothesis is that having distinct responsibility or not for extracting method suggestions provided by the SBSRE tool is independent of investigated projects. The chi-square test value in this test is 2.782. The critical value with two degrees of freedom and 0.005 significance level is 10.597. Since 2.782 is less than 10.597, the null hypothesis cannot be rejected. By a similar hypothesis, the chi-square test was applied to extract method suggestions provided by JDeodorant. The chi-square value is 6.99, less than 10.597, and the second null hypothesis is also cannot be rejected. Therefore, we conclude that the results obtained by SBSRE and JDeodorant are independent of the investigated projects.

Moreover, we conducted another chi-square test, with the null hypothesis that whether having distinct responsibility or not for extract method suggestions is independent of extract method tools. The corresponding chi-square value is 55.846. The critical value with one degree of freedom and 0.005 significance level is 7.879. Since 57.002 is greater than 10.597, the null hypothesis is rejected, and it concluded that there is a relationship between the tool used for extracting single responsibility and the ground truth responsibility approved by the experts. Indeed, using the best tool for extracting the single responsibility method is vital.

\begin{figure*}[]
    \centering
    \includegraphics[width=0.70\linewidth]{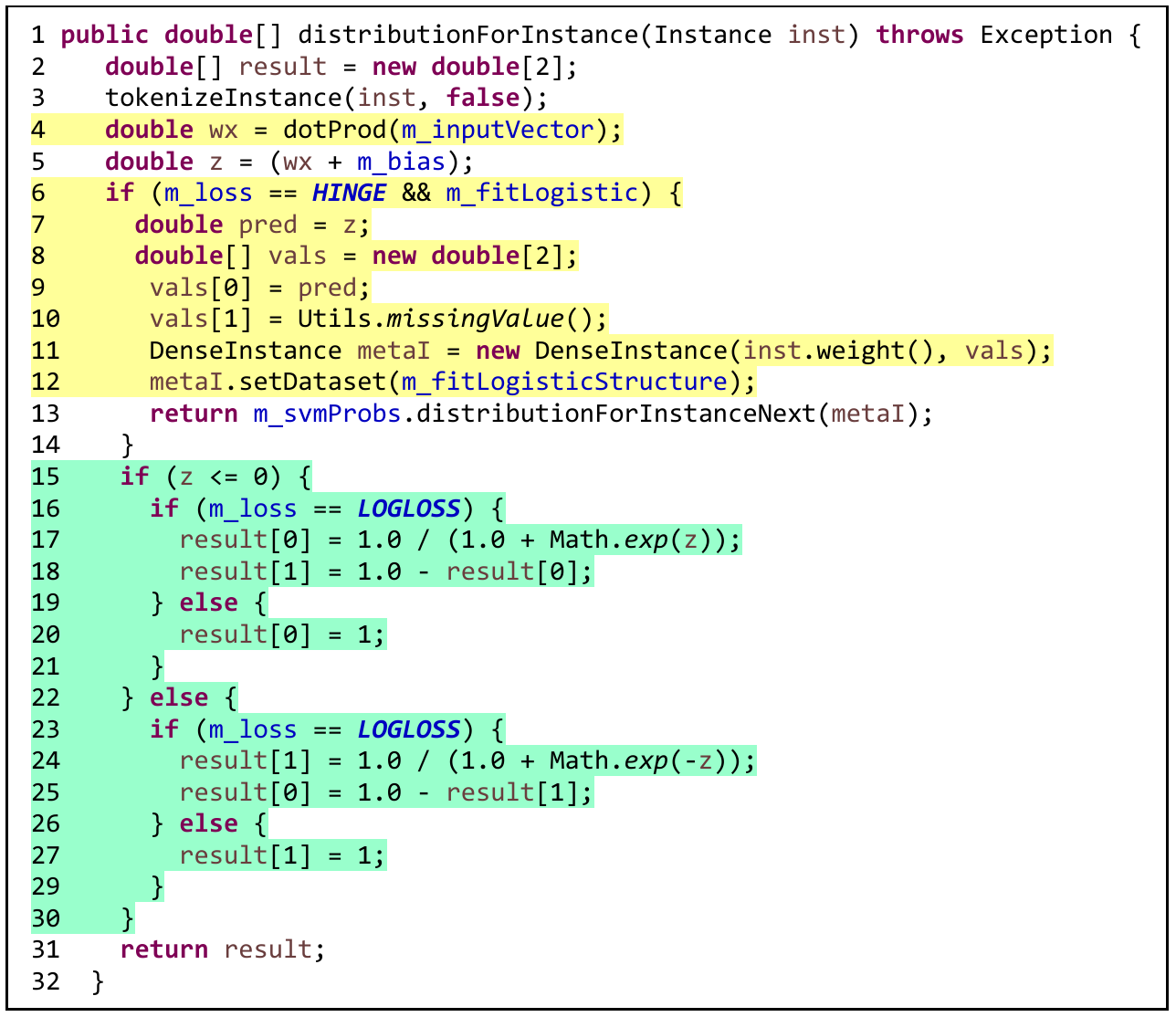}
    \centering\caption{\texttt{distributionForInstance} method in the Weka \citep{Weka2011} project with two distinct slicing criteria (21,\texttt{result}) and (13,\texttt{metaI}).}
    \label{fig:16}
\end{figure*}

\begin{table*}[] 
\scriptsize
	\centering
	\caption {Result of evaluation based on the experts' opinions.}
		\begin{tabular}{l c c c c c }
			\hline
			Project&	Tool&	Identified&	Having distinct responsibility&	Disapproved&	Acceptance rate\\
			\hline \hline
			\multirow{2}{*}{JfreeChart} & JDeodorant&	254&	186&	68&	73\% \\ & SBSRE&	251&	234&	17&	93\% \\
			\multirow{2}{*}{Weka} & JDeodorant&	110&	93&	17&	85\% \\ & SBSRE&	149&	135&	14&	91\% \\
			\multirow{2}{*}{Xserces} & JDeodorant&	219&	157&	62&	71\%\\ & SBSRE&	225&	200&	25&	89\% \\
			\cmidrule(lr){2-6}
			\multirow{2}{*}{Overall} & JDeodorant& 583&	436&	147&	75\%\\ & SBSRE&	625&	569&	56&	91\%\\
			\hline
         \end{tabular}
	\label{tab:resultstab11}
\end{table*}

The hypothesis behind this experiment, by which acceptable opportunities were selected, is as follows: The responsibility of software is to fulfill the goals at different levels of the goal hierarchy defined for the software. The fulfillment of requirements appearing as the leaves of the goal hierarchy is normally the responsibility of the methods. The responsibility of a method is revealed in its output. If there are several independent outputs, then the method may have several responsibilities. This point is not considered by other approaches such as JDeodorant \citep{tsantalis2018ten}. For example, in Figure \ref{fig:16}, the \texttt{distributionForInstance} method has two return instructions. The backward slices of these two return statements, apart from statements 4 and 5, are non-overlapping. Lines 3 to 12 are responsible for calculating the \texttt{"metaI"} variable, the value of which is eventually returned at line 13. Lines 15 to 30 compute the value of the result array, which is returned at line 31. Furthermore, the backward slice of the second return statement does not operate on the method's parameters, whereas the first return slice depends entirely on the input parameters. Moreover, when applying the extract method refactoring to a method with one or no output instruction, care should be taken to identify refactoring opportunities related to code fragments with a single distinct responsibility.

The independent experts reject the refactoring opportunities for the following reasons:

\textit{a.}	\textit{The refactoring opportunity covers a large chunk of the original method therefore, the remaining part gets limited responsibility after refactoring.}

\textit{b.} \textit{There is no specific purpose for the code fragment to be extracted as a method.}

\textit{c.}	\textit{The code fragment to be extracted as an independent method has a trivial functionality, and thus the extracted method would be useless.}

\textit{d.}	\textit{This code fragment will result in substantial code duplicates if it shares several statements with the original method.} 

Table \ref{tab:refactoring-rejection} shows the number of disapproved refactoring opportunities for each of the rejection reasons. It is observed that the most significant difference in the number of rejections resulting from our approach (SBSRE) compared to JDeodorant \citep{tsantalis2011identification} can be explained by the reason (a).

\begin{figure}[]
    \centering
    \includegraphics[width=0.75\linewidth]{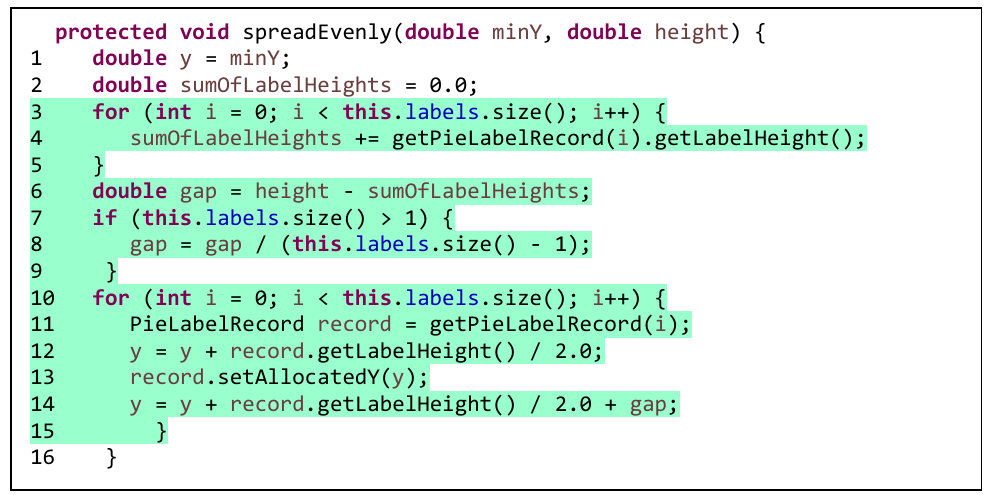}
    \centering\caption{The proposed extract method is highly similar to the original method.}
    \label{fig:17}
\end{figure}

Figure \ref{fig:17} illustrates the \texttt{plot.PieLabelDistributor.spreadEvenly} method of JfreeChart \citep{JFreeChart2017} project. JDeodorant \citep{tsantalis2018ten} recommends the extraction of the code fragment, highlighted in lines 3 to 15, as a distinct method, while our proposed approach, SBSRE, rejects it because the remaining part does not expose any meaningful task. The highlighted slice differs from the original method only in the declaration of variables in the beginning lines. Such candidates are rejected by Rule 5 in our tool. According to the evaluation results summarized in Table \ref{tab:resultstab11} and Table \ref{tab:refactoring-rejection}, it can be claimed that, in general, our proposed method has been relatively more successful in detecting single-responsibility code fragments compared to JDeodorant.

\begin{table}
    \footnotesize
    \centering
    \caption {Disapproved refactoring opportunities for each of the rejection reasons.}
    \begin{tabular}{l c c c c c }
        \hline
        Tool&	Disapproved refactorings &	(a)&	(b)&	(c)&	(d)\\
        \hline  \hline
        JDeodorant&	147&	101&	12&	16&	18 \\
        SBSRE&	56&	12&	13&	8&	23\\
        \hline
    \end{tabular}
    \label{tab:refactoring-rejection}
\end{table}

\subsubsection{Results for RQ5}

To investigate the impact of applied refactoring in Section \ref{sec:resrq3} on methods cohesion, we computed three slice-based cohesion metrics before and after applying the extract method refactoring recommended by our tool and JDeodorant \citep{tsantalis2018ten}. These metrics are calculated based on output values. The output values are typically categorized as follows \citep{green2009introduction}:

\begin{itemize}
    \item Return values, 
    \item Global variables values modified by the method,
    \item Parameters passed by reference to and modiﬁed by the method,
    \item Any variable printed, written to ﬁle, or otherwise output to the external environment.
\end{itemize}
        
Considering the above outputs provided by a method $M$, we use three slice-based cohesion metrics, namely Tightness, Overlap, and Coverage, to determine whether $M$ is cohesive \citep{ott1993slice}. The calculations formula for quality metrics of the method, $M$, are defined as follows:

\vspace{0.5cm}
\hspace{0cm}
\fbox{
    \begin{minipage}{16.2cm}
    Let $OutSet$ = The set of output values provided by the method, $M$, and $Slice_{\mathrm{i}}$ = slice for the variable  $\mathrm{i} \in OutSet$
            
    \begin{equation}
    SlicesIntersect =\bigcap_{\mathrm{i}}  Slice _{\mathrm{i}}, \forall \mathrm{i} \in OutSet 
    \label{eq5}
    \end{equation}
     
    \begin{equation}
    Tightness(M)=\mid  SlicesIntersect  \mid /  length(M)
    \label{eq6}
    \end{equation}   
        
    \begin{equation}
    Overlap(M)=\frac{\sum_{\mathrm{i}=1}^{\mid OutSet \mid} \frac{\mid  Slicesintersect \mid}{\mid  slice _{\mathrm{i}} \mid}}{\mid OutSet  \mid}
    \label{eq7}
    \end{equation}    
    
    \begin{equation}
    Coverage(M)=\frac{\sum_{\mathrm{i}=1}^{\mid OutSet  \mid} \frac{\mid Slices_{\mathrm{i}} \mid}{\mid Length(M) \mid}}{\mid  OutSet \mid}
    \label{eq8}
    \end{equation}       
    \end{minipage}
}
\vspace{0.25cm}

The $Tightness(M)$ metric \eqref{eq6} indicates the percentage of statements shared by different output slices of a method, $M$. Apparently, as the number of statements shared between the slices of a method increases, the method cohesion increases. The $Overlap(M)$ metric \eqref{eq7} is the average number of statements in common between the output variable slices. The more statements in common between the output slices of a method, the more cohesive the method is. A highly cohesive method is responsible for only a single task. Finally, the $Coverage(M)$ metric \eqref{eq8} indicates the percentage of the method, $M$, on average covered by its output slices. As the coverage of a method, $M$, approaches 1, the chance of that method having a single responsibility increase.
   
The long method smells detected by SBSRE and JDeodorant in the methods belonging to six packages of the JFreeChart \citep{JFreeChart2017}, Weka \citep{Weka2011}, and Apache-Xerces \citep{Xerces2011} are reported in Table \ref{tab:slice-based cohesion}. We applied extract method refactorings using both SBSRE and JDeodorant \citep{tsantalis2018ten} to remove the smells detected in these packages. Table \ref{tab:slice-based cohesion} presents the evaluation results of the methods based on those mentioned above, three slice-based cohesion metrics before and after the refactoring. The third column of Table \ref{tab:slice-based cohesion}, headed "Remain $-$ Original," shows the difference in the metric values between the original method before refactoring and the remaining part after applying the extract method refactoring. Column four, headed "extracted," represents metric values computed for the method separated from the original method after the extract method refactoring. Finally, the fifth column shows the average difference of metric values before and after applying refactoring. To calculate this column, the average metrics of the extracted and the remaining methods are computed and then subtracted from the original method. The significance of this column is that it indicates whether the proposed method extraction can reduce the metric values in both the remaining and the extracted methods. Table \ref{tab:newStaic1} shows the t-test of the values shown in Table \ref{tab:slice-based cohesion}. In this table, all the p-values are less than 0.01; therefore, the results are significantly different.

The evaluation results shown in the last column of Table \ref{tab:slice-based cohesion} show that SBSRE outperforms JDeodorant \citep{tsantalis2011identification}. Let us discuss why our approach outperforms the other ones. Apparently, when slicing out part of a method that is not relevant to the other slices of the method, the remaining parts will be relatively more cohesive. That is why we observe that after refactoring to the extract method, the cohesion metrics for all of the mentioned tools improve. The SBSRE tool outperforms JDeodorant because our extract method refactoring is based on the output slices, and the three evaluation metrics already used by the authors of JDeodorant \citep{tsantalis2011identification} consider the outputs provided by the method. The JDeodorant authors believe that using output variables can create high artificial values for coherence metrics, therefore instead of output variables, they preferred to use variables defined in the method's body to compute the slice-based cohesion metrics. Therefore, we also computed the metrics for all the variables declared in the body of the methods to make sure our comparison was fair. Table \ref{tab:slice-based cohesion2} shows the slice-based cohesion metrics considering all variables declared in the body of the methods. The evaluation results in Table \ref{tab:slice-based cohesion2} still reveal that SBSRE surpasses JDeodorant \citep{tsantalis2011identification}. Moreover, the t-test  and effect size shown in Table \ref{tab:newStaic2} illustrates there is  a significance difference between the results.
     
\begin{table*}
\footnotesize
	\centering
	\caption {Slice-based cohesion metrics considering output variables declared in the body method for JDeodorant and our proposed tool.}
		\begin{tabular}{l l c c c  }
			\hline
			Tool&	Metric&	Remain$-$Original& Extracted&	$\frac{\text{Extracted+Remain}}{2}-\text{Original}$\\
			\hline  \hline
			\multirow{3}{*}{JDeodorant} & Overlap&	+0.112&	0.931&	+0.193 
			\\ &Tightness &	+0.057&	0.918&	+0.284 \\ 
			& Coverage&	-0.032&	0.964&	+0.221\\
			\hline
			
    		\multirow{3}{*}{SBSRE tool} & Overlap&	+0.206&	0.984&	+0.259 \\ &Tightness & +0.132& 0.975&	+0.353 \\
    		& Coverage& -0.039&	0.981&	+0.267\\
    		\hline
    		
    		\multirow{3}{*}{Changes} & Overlap&	+0.094&	+0.053&	+0.066 \\ &Tightness & +0.075&	+0.057&	+0.069 \\
    		& Coverage& -0.007&	+0.017&	+0.046\\
    		\hline
    		
         \end{tabular}
	\label{tab:slice-based cohesion}
\end{table*}

\begin{table}[h!]
\scriptsize
	\centering
	\caption {The results of the statistical test for the performance metrics are reported in Table \ref{tab:slice-based cohesion}.}
		\begin{tabular}{lccc}
			\hline
			\multicolumn{1}{l}{\begin{tabular}{l}
                   Metric
            \end{tabular}} &
	    	\multicolumn{3}{c}{Independent t-test p-value $(\alpha = 0.01)$} 
            \\
            \cmidrule(lr){2-4}
			    &  Remain$-$Original &	Extracted&	 $\frac{\text{Extracted+Remain}}{2}-\text{Original}$\\
			 \hline \hline
			 Overlap &	$2.25\times10^{-4}$ (ES = 0.42)&	 $1.22\times10^{-3}$ (ES = 0.47 )&	$1.58\times10^{-6}$ (ES = 0.57) \\
			 Tightness &	$4.92\times10^{-6}$ (ES = 0.39)&	$3.56\times10^{-5}$ (ES = 0.44) &	$9.85\times10^{-6}$ (ES = 0.58)\\
			 Coverage &	$3.30\times10^{-4}$ (ES = 0.51)&	$7.36\times10^{-5}$ (ES = 0.63) &	$1.10\times10^{-5}$ (ES = 0.49)\\

			 \hline
         \end{tabular}
	\label{tab:newStaic1}
\end{table}

\begin{table*}
\footnotesize
	\centering
	\caption {Slice-based cohesion metrics considering all variables declared in the method body for JDeodorant and our proposed tool.}
		\begin{tabular}{l l c c c  }
			\hline
			Tool&	Metric&	Remain$-$Original& Extracted&	$\frac{\text{Extracted+Remain}}{2}-\text{Original}$\\
			\hline  \hline
			\multirow{3}{*}{JDeodorant}& Overlap& +0.181 & +0.828	&+0.274	
			\\ &Tightness &	+0.196& +0.638	&+0.231 \\ 
			& Coverage& +0.271	&+0.787	&+0.259 \\
			\hline
			
    		\multirow{3}{*}{SBSRE tool} & Overlap& +0.258	&+0.891	&+0.395	 \\ &Tightness &+0.235 &+0.696 &+0.316	 \\
    		& Coverage& +0.334 & +0.855	& +0.363	\\
    		\hline
    		
    		\multirow{3}{*}{Changes} & Overlap&	 +0.077 & +0.063	& +0.121\\
    		&Tightness & +0.039 & +0.058	&+0.085	\\
    		& Coverage&  +0.063 & +0.068	& +0.104	\\
    		\hline
    		
         \end{tabular}
	\label{tab:slice-based cohesion2}
\end{table*}

\begin{table}[h!]
\scriptsize
	\centering
	\caption {The results of the statistical test for the performance metrics are reported in Table \ref{tab:slice-based cohesion2}.}
		\begin{tabular}{lccc}
			\hline
			\multicolumn{1}{l}{\begin{tabular}{l}
                   Metric
            \end{tabular}} &
	    	\multicolumn{3}{c}{Independent t-test p-value $(\alpha = 0.01)$} 
            \\
            \cmidrule(lr){2-4}
			    &  Remain$-$Original &	Extracted&	 $\frac{\text{Extracted+Remain}}{2}-\text{Original}$\\
			 \hline \hline
			 Overlap &	$4.17\times10^{-6}$ (ES = 0.46) &	 $2.69\times10^{-5}$ (ES = 0.39)&	$8.62\times10^{-8}$ (ES = 0.65)\\
			 Tightness &	$4.22\times10^{-6}$ (ES = 0.53)&	$6.11\times10^{-4}$ (ES = 0.47)&	$7.29\times10^{-6}$ (ES = 0.69)\\
			 Coverage &	$5.25\times10^{-7}$ (ES = 0.64)&	$2.86\times10^{-5}$ (ES = 0.48)&	$3.24\times10^{-9}$ (ES = 0.70)\\

			 \hline
         \end{tabular}
	\label{tab:newStaic2}
\end{table}

\subsubsection{Results for RQ6}

The brain method is often considered the equivalent of a long method \citep{fontana2015towards}. Three software metrics have been mainly used to detect the brain method smell, lines of code (LOC), cyclomatic complexity (CC), and maximum nesting \citep{fontana2015towards}, \citep{lanza2007object}. An appropriate extract method refactoring is expected to decrease the value of these metrics as much as possible for both the extracted and the remaining parts of the refactored method \citep{lopes2022and}. In the final experiments, we study the impact of the refactoring suggested by our tool on lines of code, cyclomatic complexity, and maximum nesting metrics.

Table \ref{tab:average changes} shows the average changes to these metrics before and after refactoring. The second column shows the average difference between the metrics of the original methods and the remaining part of the methods after the refactoring. It is observed that the values of the three metrics reduce, indicating a reduction in complexity after extract method refactoring. In the third column, the average difference between the metrics of the original method and the average metric value for the remaining parts is shown. This column reveals that the complexities of the remaining and extracted parts of the methods are reduced on average. As a result, our SBSRE tool has successfully improved the main source code metrics related to long method code smell without loss of functionality.

\begin{figure*}[!h]
    \centering
    \includegraphics[width=0.70\linewidth]{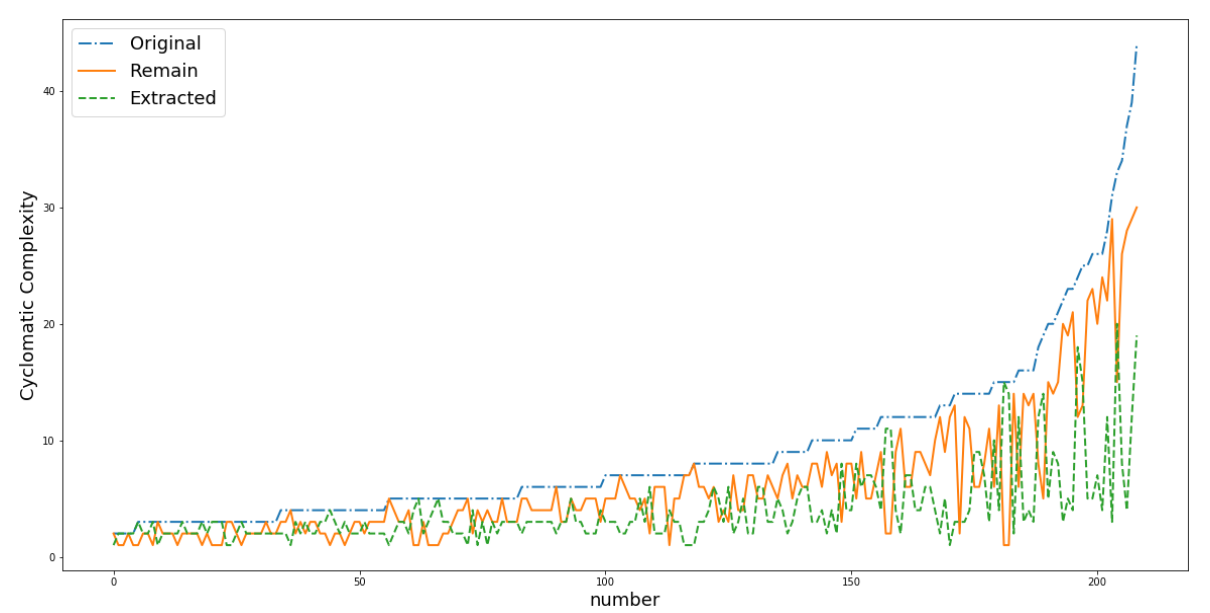}
    \centering\caption{Changes in cyclomatic complexity of the methods.}
    \label{fig:18}
\end{figure*}

\begin{figure*}[!h]
    \centering
    \includegraphics[width=0.70\linewidth]{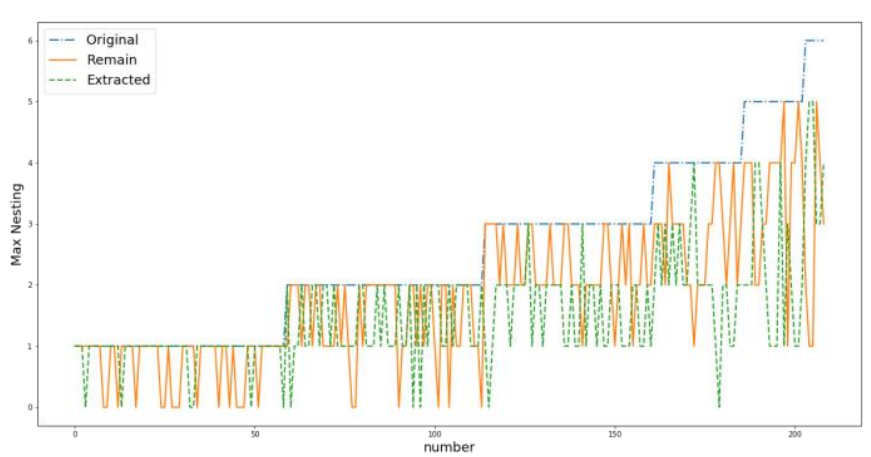}
    \centering\caption{Changes in maximum nesting level of the methods.}
    \label{fig:19}
\end{figure*}

Figure \ref{fig:18} and \ref{fig:19} illustrate the changes in the cyclomatic complexity and maximum nesting metrics of 210 methods belonging to the six packages of three Java projects, JFreeChart \citep{JFreeChart2017}, Weka \citep{Weka2011}, and Xerces \citep{Xerces2011}. In Figure \ref{fig:18}, the horizontal axis shows the long methods' IDs sorted by their cyclomatic complexity before refactoring, and the vertical axis shows the cyclomatic complexity values. In Figure \ref{fig:19}, the horizontal axis shows the long methods' IDs sorted by their maximum nesting level, and the vertical axis shows the corresponding maximum nesting values. As shown in Figure \ref{fig:19}, the cyclomatic complexity of the remained and extracted parts of all the methods has been reduced after applying extract method refactoring. Similarly, according to Figure \ref{fig:19}, the maximum nesting has decreased in a significant number of the refactored methods. 
        
\begin{figure}[h!]
    \centering
    \frame{\includegraphics[width=0.50\linewidth]{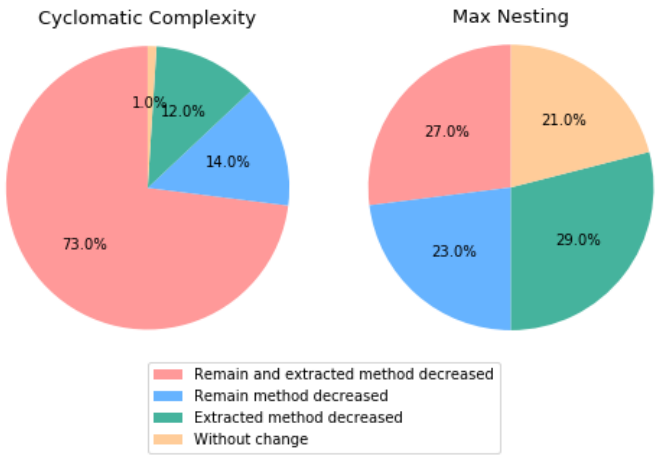}}
    \centering\caption{Overall changes in cyclomatic and maximum nesting metrics for extracted and remained methods.}
    \label{fig:20}
\end{figure}

Figure \ref{fig:20} shows an overall improvement of cyclomatic complexity and maximum nesting level metrics after refactoring separately. Four different possible changes occurred in the value of each source code metric after applying the extract method. First, the metric value is reduced in both the remained and extracted methods. Second, the metric value decreases in the remained method while it does not change in the extracted method. Third, the metric decreases in the extracted method and does not change in the remained method. Finally, the metric value remains fixed in both the remained and extracted methods. Figure \ref{fig:20} reveals that the values of the cyclomatic complexity metric of both the extracted and remaining method decrease in 73\% of refactoring applied by our tool. The same status is observed for the values of the maximum nesting metric in 27\% of the refactoring, which is more than the other three statuses where the metrics do not change or only change for one of the methods. We conclude that our tool suggests extract method refactoring candidates, which often leads to a decrease in the complexity metrics in both the extracted and remaining method. Therefore, our proposed tool is appropriate to decompose long method instances into single responsibility methods with minimum duplicated code.

\begin{table*}[h!]
\footnotesize
	\centering
	\caption {Average changes in cyclomatic and maximum nesting metrics after method extraction.}
		\begin{tabular}{lcc}
			\hline
			Metric&	Remain$-$Original &	$\frac{\text{Extracted+Remain}}{2}-\text{Original}$\\
			\hline  \hline
			Lines of code & -14.375 &  	-25.8875\\
			Cyclomatic complexity&	-3.575&	-5.5875\\
			Maximum nesting&	-0.725&	-1.1625\\
			\hline
         \end{tabular}
	\label{tab:average changes}
\end{table*}

\subsection{User Experience Survey}

\begin{figure}[h]
     \centering
         \begin{subfigure}{0.6\textwidth}
              \centering
         
              \includegraphics[width=0.8\linewidth]{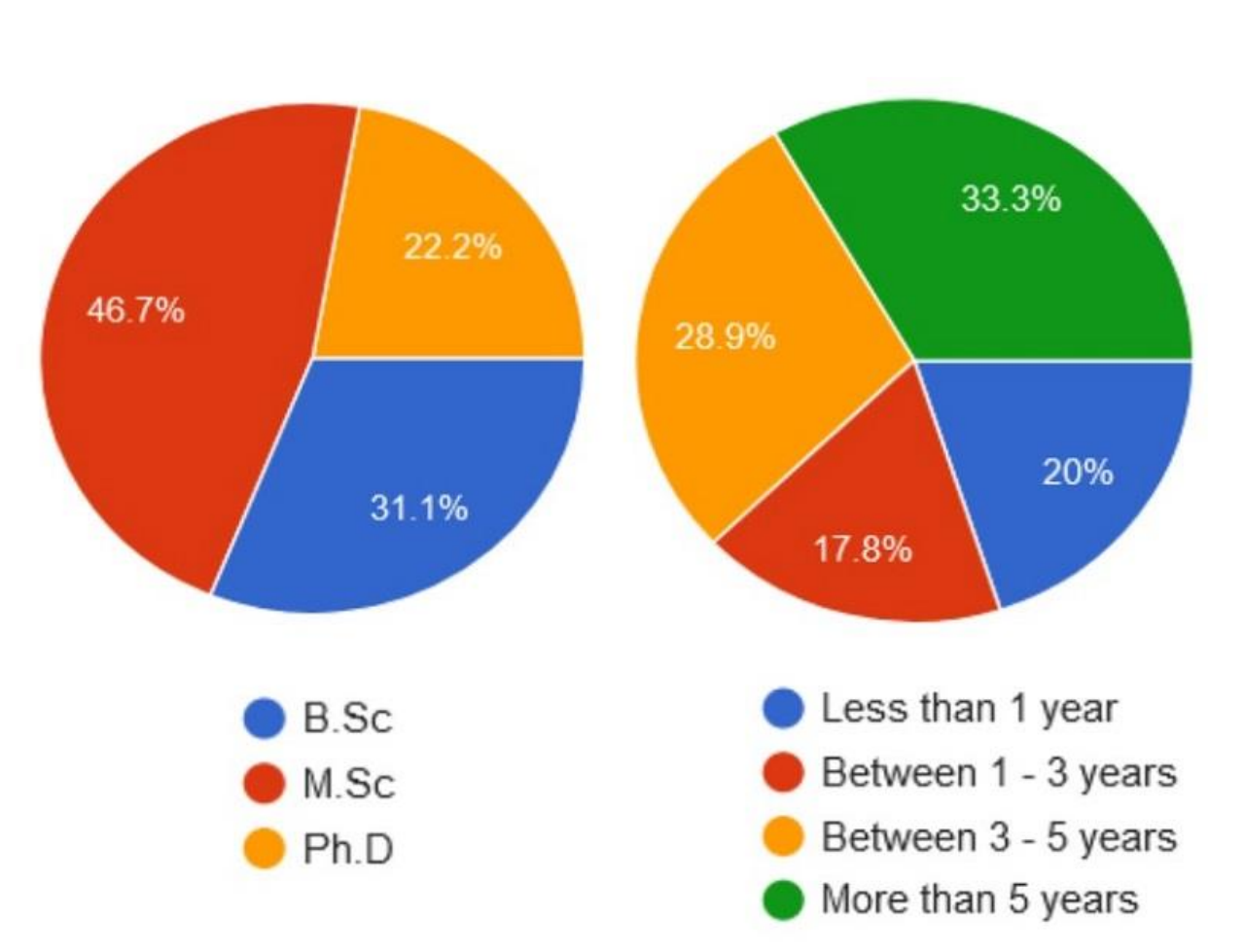}
               \caption{}
               \label{fig:surveya}
         \end{subfigure}
     
         \begin{subfigure}{0.6\textwidth}
            \centering
             \includegraphics[width=0.8\textwidth]{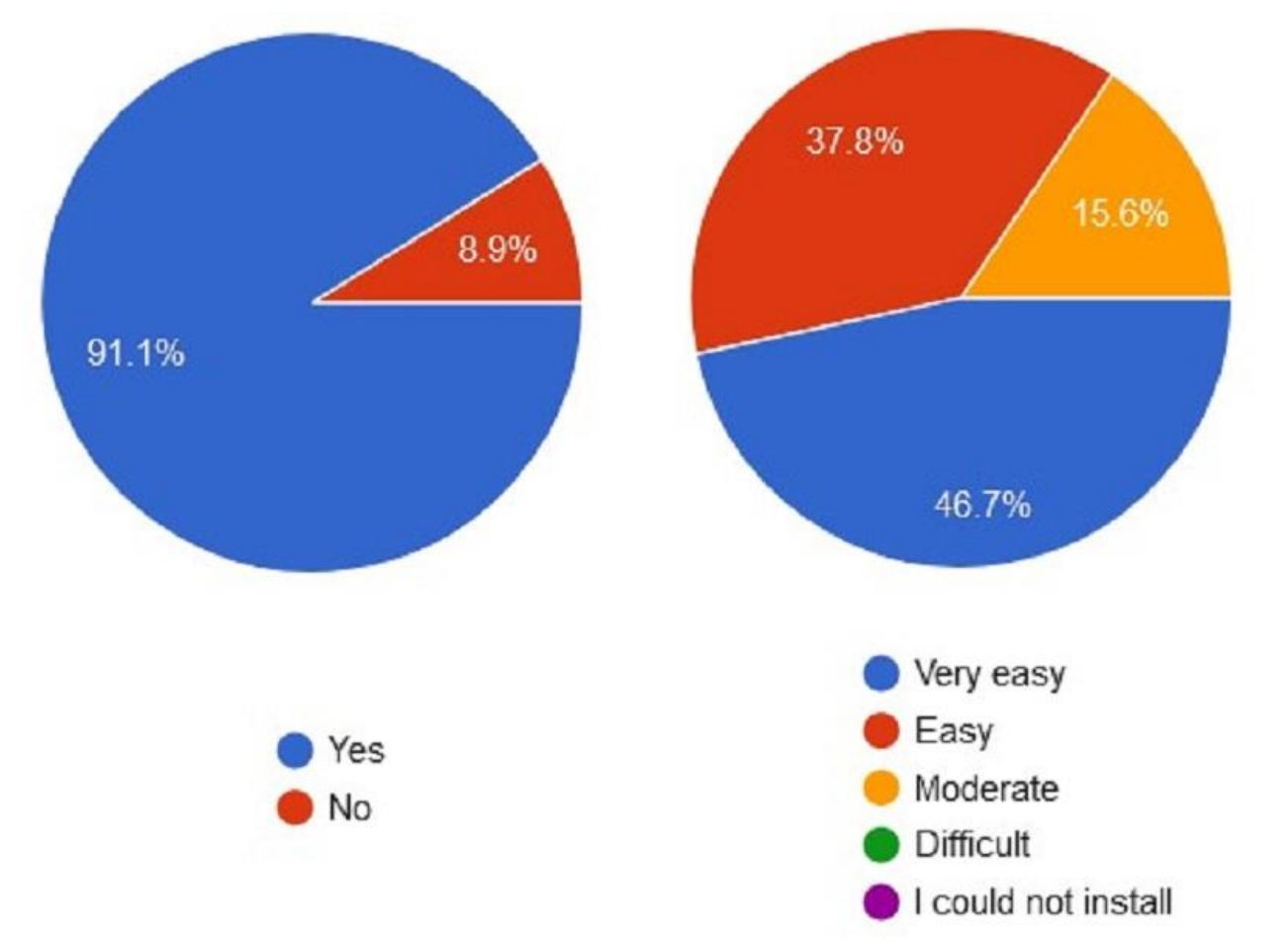}
             \caption{}
             \label{fig:surveyb}
         \end{subfigure}
     
        \caption{(a) Distribution of participants' academic degrees and Java experience. (b) The result of the participants' answers to Q1 and Q2.}
        \label{fig:survey1}
\end{figure}

This section outlines the online survey conducted to assess the usefulness and usability of the SBSRE tool. The survey was designed for experts within the author's universities. Additionally, we extended the invitation through social media platforms, such as Linkedin and Researchgate, to fill out the survey while experimenting with SBSRE. The questionnaire was created using Google Forms, and to ensure the suitability of the respondents, we included four qualifying questions: level of university degree, proficiency in Java programming, familiarity with SRP, and experience with extract method refactoring. In total, 45 experts completed the survey; all of them were acquainted with SRP, with only one exception regarding familiarity with the extract method. Figure \ref{fig:survey1} provides an overview of the participants' diverse range of academic degrees and years of Java expertise. The survey questions are as follows:

\begin{itemize}
        \item \textbf{Q1:} \textit{Consider a method that outputs (return, print, show, write in file/database, etc.) different values, each resulting from a different calculation in the method body. Do you agree that, in this case, a method violates the Single Responsibility Principle?}
        \item \textbf{Q2:} \textit{How do you evaluate installing the SBSRE plugin? Was it easy or difficult? }
        \item \textbf{Q3:} \textit{Do you find the extract method suggestions provided by the SBSRE plugin appropriate, useful, and meaningful? (Very appropriate: 10 - Not appropriate: 0)}
        \item \textbf{Q4:} \textit{Do you prefer using the SBSRE tool in future software development projects? (Highly prefer: 10 – Not prefer: 0).}
        \item \textbf{Q5:} \textit{Do you think automated refactoring tools like SBSRE are useful for programmers? (Very useful: 10 – Not useful: 0)}
    \end{itemize}

Figure \ref{fig:survey1} depicts that 91\% of the survey participants answered "yes" to Q1. This implies that having multiple output instructions in a method may indicate a violation of the Single Responsibility Principle, which is the main idea of the SBSRE tool. Moreover, Figure \ref{fig:survey1} reveals that, based on the responses to Q2, all survey participants were able to install the SBSRE tool, with none of them finding the installation process "Difficult." Approximately 15\% of participants rated the installation as "Moderate," while the majority described it as "Easy" or "Very Easy."

\begin{figure*}[!h]
    \centering
    \includegraphics[width=0.75\linewidth]{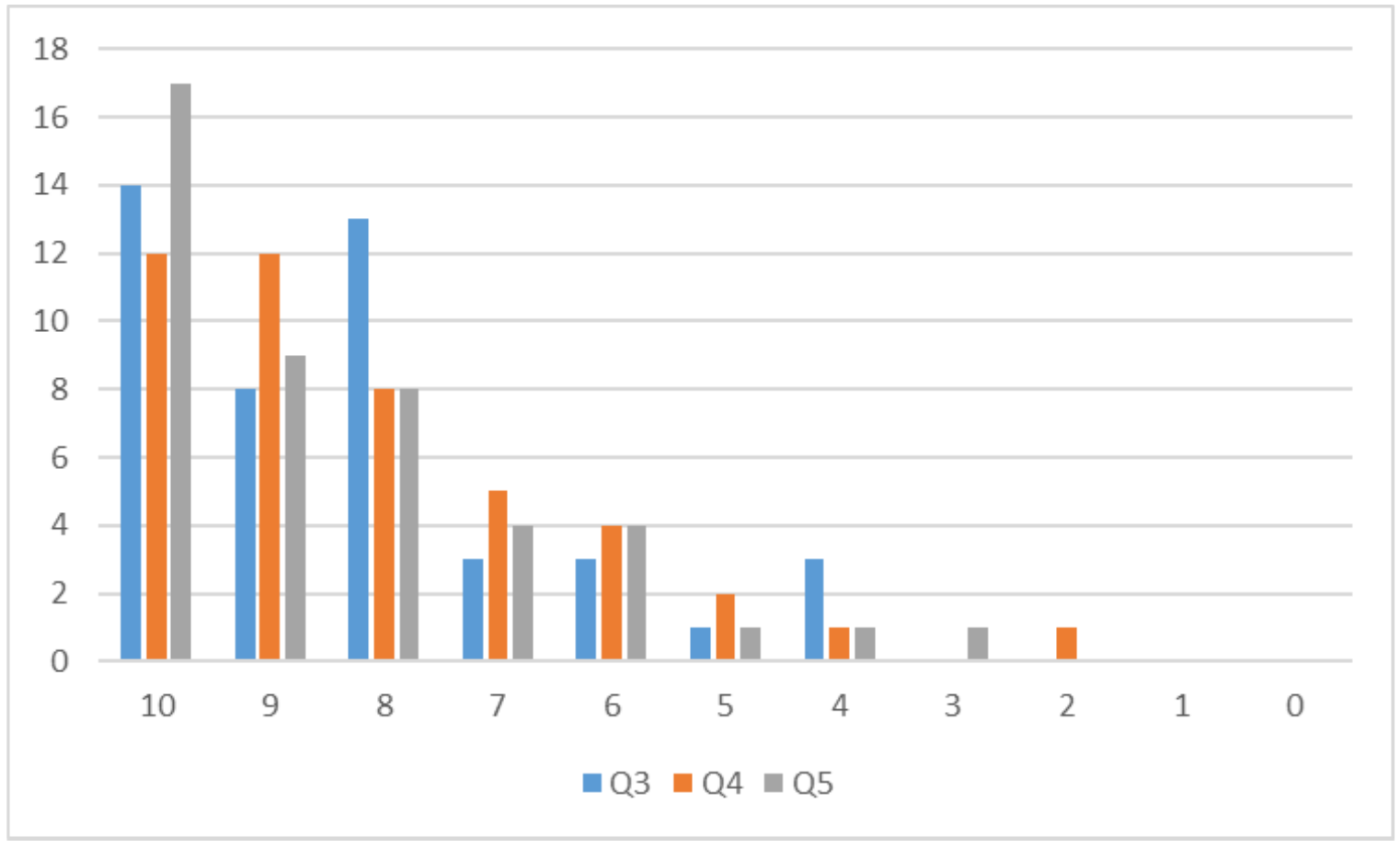}
    \centering\caption{The result of the participants' answers to Q3, Q4, and Q2.}
    \label{fig:survey2}
\end{figure*}  

Figure \ref{fig:survey2} shows the responses of participants to Q3, Q4, and Q5. On average, the extract method suggestions provided by the SBSRE tool received a score of 8.2 out of 10. This indicates that the extract method  suggestions from the SBSRE tool aligned well with the participants' apinion. Additionally, the high average score of 8.1 for Q4 suggests that participants expressed a strong inclination to incorporate this tool into their future work. Lastly, positive responses to Q5 demonstrate that method extraction tools hold practical value for programmers beyond academic or research-oriented context.

\section{Implications of Study}

The results of this study have several implications for different stakeholders, such as researchers, practitioners, educators, and tool builders, who are interested in the single responsibility principle (SRP) and its application at the method level. The SRP states that every module or class should have one and only one reason to change, and it is considered one of the fundamental principles of software design. However, applying the SRP and manually performing the extract method can be challenging, time-consuming, and error-prone, especially for large or legacy systems. Therefore, automated or semi-automated extract method tools such as SBSRE that can assist developers in performing extract method refactoring to achieve single-responsible methods are desirable.

This study shows that the SBSRE tool can effectively and efficiently apply the SRP at the method level and improve software quality attributes, such as cohesion, coupling, readability, maintainability, testability, and reusability. These results can encourage researchers to consider the SRP and its application at the method level as a relevant and important topic for software engineering research. The results can also provide researchers with a baseline and a benchmark for conducting further empirical studies on the SRP and its impact on software design and quality and other related topics, such as code smells, refactoring, and software metrics. Moreover, the results can inspire researchers to explore and evaluate other aspects and dimensions of the SRP, such as its applicability and benefits at different levels of abstraction (e.g., class, package, module), its relationship and interaction with other software design principles (e.g., open-closed, dependency inversion, interface segregation), and its trade-offs and challenges in different contexts and scenarios (e.g., legacy systems, distributed systems, agile development). Therefore, the SBSRE tool can be valuable for researchers exploring or evaluating the SRP and its implications for software engineering.

Similarly, the results of this study suggest that practitioners, developers, and tool builders can leverage the assistance of the SBSRE tool in applying the SRP consistently and effectively throughout the software development lifecycle. The results indicate that using the SRP at the method level can positively affect the design and structure of software systems, especially in terms of reducing complexity, increasing modularity, and enhancing readability. The results also demonstrate that the SBSRE tool can help practitioners, developers, and tool builders identify and eliminate SRP violations in their code and perform code reviews and inspections more efficiently and effectively. The results also show that the SBSRE tool can support them in resolving long method code smells, one of the most common and harmful code smells in software systems. Furthermore, the results reveal that the SBSRE tool has a fast response time, which makes it suitable for industrial projects where time consumption is essential. Therefore, the SBSRE tool can assist practitioners, developers, and tool builders in applying the SRP and improve their software systems' design and quality.

For educators, this study highlights the importance and benefits of teaching and learning the SRP and other related concepts in software engineering courses. The SBSRE tool can help students understand the rationale and benefits of applying the SRP and the challenges and trade-offs involved. The SBSRE tool can also enable students to practice and improve their skills in extract method refactoring through interactive exercises and feedback.
For tool builders, this study provides insights and guidelines on designing and implementing effective and efficient extract method tools to support the SRP at the method level. The SBSRE tool relies on static analysis of method graphs, such as AST, CFG, and PDG, to identify the optimal extraction candidates and perform the refactoring operations. The SBSRE tool also incorporates several heuristics and criteria to ensure the quality and consistency of the extracted methods, such as preserving the original behavior, and avoiding duplication. The SBSRE tool can serve as a reference or a benchmark for other extract method tools that aim to support the SRP at the method level.

We acknowledge that the size of the studied applications may not be representative of large industrial projects and that the SBSRE tool may face some challenges in handling the frequent changes and updates in the source code of large industrial projects. However, we believe that the SBSRE tool can still be used in large industrial projects where time consumption is essential, as it analyzes the source code at a method level and applies the extract method refactoring whenever there is more than one output-based slice conditioned to generate not many duplicates. This means that the SBSRE tool does not need to analyze the whole source code of a large industrial project but only the methods that are changed or updated in each incoming commit. Therefore, the SBSRE tool can (re)generate the required graphs for each changed or updated method without consuming too much time and resources.

\section{Threats to Validity and Limitations}
\label{validity}

The main threat to construct validity is the variety of output instructions that our approach should support. We investigated several Java programs and found five different types of output instructions typically used by Java methods. However, there is not a definite set of output instructions. Any new output instruction can be added easily to our tool without changing the other parts of the extract method algorithm. Another threat to construct validity is the limited set of precondition rules used to ensure the program's behavior preservation after refactoring. Unfortunately, the problem of behavior preservation through program transformation is undecidable \citep{mens2004survey}. Excessive precondition and inhibitor rules lead to overly strong conditions and the loss of appropriate refactoring opportunities \citep{mongiovi2017detecting}. Our tool only suggests refactoring candidates and asks the programmer to select the best one for applying such that minimizes the refactoring side effects.

The internal validity of our evaluations is mainly threatened by human experts used for evaluating the correctness and usefulness of the suggested refactoring in RQ3 and RQ4. We chose experts with at least three years of Java programming experience and knowledge in refactoring and clean code principles. The experts were asked to evaluate all refactoring candidates and report their judgment independently. We then used majority voting to make our final decision about each refactoring candidate suggested by the tools. We observed that expert opinions agreed with each other in most cases, indicating the reliability of the results. We have also selected open-source projects which are well-documented and have an acceptable level of understandability. Otherwise, developers cannot understand and contribute to them. Therefore, the expert's decisions were based on their understanding of methods, functionalities, and responsibilities. 

Regarding external validity, all evaluations reported in Section 5 are limited by examined methods and projects. We evaluated our approach on the GEMS dataset and three real-world Java projects with different application domains to demonstrate the usefulness and correctness of suggested refactorings under different conditions. However, further empirical evaluations are required to ensure the generalization of results. Specifically, for other programming languages, a similar implementation of our tool is needed to evaluate the applicability of our approach in achieving single responsibility.

Despite all the capabilities that the SBSRE tool has, it also has some limitations. It may not be able to handle complex or non-standard code structures or syntaxes, such as exceptions, lambdas, etc., which can affect the quality and correctness of the refactoring. Since the purpose of the SBSRE tool is to discover the responsibilities of the method, it may not be able to capture the semantic or functional aspects of the code fragments, such as the purpose, intention, or meaning of the code, which can affect its ability to find such responsibilities. Another limitation of the SBSRE tool is that the user has to choose which method suggestion to apply; therefore, SBSRE tool are categorized as semi-automated tools instead of fully automated.

\section{Conclusion}\label{conclusion}

Achieving single responsibility is an essential design principle when refactoring long methods to smaller ones by applying the extract method. Otherwise, the automatically extracted methods are not helpful from the viewpoint of developers. Output instructions, including the return, print, and write in a stream, distinctly reflect a method's responsibilities. Therefore, this paper uses slicing based on these statements to decompose a long method with multiple output instructions and variables into methods with a single responsibility. The Slice-based single responsibility extraction (SBSRE) tool improves cohesion metrics by an average of 20\% after refactoring. Compared to state-of-the-art tools, JDeodorant \citep{tsantalis2011identification}, JExtract \citep{silva2014recommending}, GEMS \citep{xu2017gems}, and SEMI \citep{charalampidou2016identifying}, the precision and recall of automated extract method refactoring are respectively improved by at least 29.6 and 12.1\%.

The Output-based Slicing algorithm often generates different slices based on different output instructions and variables. As a future work in this area, a search-based refactoring approach can be used to find the best refactoring solutions and compromise between high-level metrics affected by single responsibility, e.g., modularity \citep{fontana2017code} in large projects. To this aim, a sequence of refactoring candidates generated by our approach for different methods is considered a solution used in a genetic optimization process to maximize quality metrics.

The future work for the SBSRE tool can be considered from two perspectives. First, we plan to use machine learning-based approaches to identify long methods, which could improve our tool's precision and recall rates. Second, we intend to rank the suggested extract method refactorings, which could make our tool more automated and user-friendly.

\section*{Declarations}

\subsection*{Data Availability Statement}
\noindent The datasets generated and analyzed during the current study are available in  the SBSRE GitHub repository, \href{https://github.com/Alireza-Ardalani/SBSRE}{\emph{https://github.com/Alireza-Ardalani/SBSRE}}.

\subsection*{Funding}
\noindent
This study has received no funding from any organization. 

\subsection*{Conflict of Interest}
\noindent
All of the authors declare that they have no conflict of interest. 

\subsection*{Ethical Approval} 
\noindent
This article does not contain any studies with human participants or animals performed by any of the authors.

\begin{appendices}

\section{Extended Data}\label{secA1}

 Table \ref{tab:refactoring} shows the confusion matrix elements obtained by expert evaluations of different extract method refactoring tools, including the proposed SBSRE tool, JDeodorant, SEMI, JExtract, and GEMS.

\begin{table}
\scriptsize
	\centering
	\caption{The confusion matrix elements of the proposed refactoring opportunities based on the experts' opinions.}
		\begin{tabular}{l l| l l l l l |l l l l l |l l l l l}
			\hline
			\multicolumn{1}{c}{} &
			\multicolumn{1}{c}{} &
			\multicolumn{5}{c}{TP} &
	    	\multicolumn{5}{c}{FP} &
	    	\multicolumn{5}{c}{FN} \\
         \cmidrule(lr){3-17}
			  \begin{tabular}[m]{@{}c@{}}Method\end{tabular} & To &  \rotatebox[origin=c]{90}{SBSRE}& \rotatebox[origin=c]{90}{JDeodorant}& \rotatebox[origin=c]{90}{SEMI}& \rotatebox[origin=c]{90}{JExtract}&  \rotatebox[origin=c]{90}{GEMS}&  \rotatebox[origin=c]{90}{SBSRE}& \rotatebox[origin=c]{90}{JDeodorant}& \rotatebox[origin=c]{90}{SEMI}& \rotatebox[origin=c]{90}{JExtract}&  \rotatebox[origin=c]{90}{GEMS}&   \rotatebox[origin=c]{90}{SBSRE}& \rotatebox[origin=c]{90}{JDeodorant}& \rotatebox[origin=c]{90}{SEMI}& \rotatebox[origin=c]{90}{JExtract}&  \rotatebox[origin=c]{90}{GEMS} \\
			 \hline\hline
			 M1	& 1	& 1	& 0	& 1	& 0	& 0	& 2	& 0	& 2	& 2	& 0	& 0	& 1	& 0	& 1	& 1 \\
            M2	& 1	& 1	& 0	& 0	& 0	& 0	& 1	& 1	& 3	& 2	& 0	& 0	& 1	& 1	& 1	& 1 \\
            M3	& 2	& 2	& 2	& 2	& 2	& 0	& 0	& 0	& 1	& 1	& 0	& 0	& 0	& 0	& 0	& 2 \\
            M4	& 2	& 2	& 1	& 1	& 2	& 0	& 0	& 0	& 2	& 0	& 0	& 0	& 1	& 1	& 0	& 2 \\
            M5	& 2	& 0	& 0	& 0	& 0	& 0	& 0	& 0	& 2	& 3	& 0	& 2	& 2	& 2	& 2	& 2 \\
            M6	& 2	& 2	& 1	& 2	& 2	& 1	& 0	& 0	& 1	& 1	& 2	& 0	& 1	& 0	& 0	& 1 \\
            M7	& 1	& 1	& 0	& 1	& 0	& 0	& 0	& 0	& 1	& 3	& 0	& 0	& 1	& 0	& 1	& 1 \\
            M8	& 2	& 1	& 0	& 1	& 1	& 1	& 0	& 0	& 2	& 2	& 2	& 1	& 2	& 1	& 1	& 1 \\
            M9	& 2	& 2	& 0	& 0	& 0	& 0	& 0	& 0	& 2	& 2	& 0	& 0	& 2	& 2	& 2	& 2 \\
            M10	& 2	& 1	& 0	& 2	& 0	& 1	& 1	& 0	& 1	& 2	& 2	& 1	& 2	& 0	& 2	& 1 \\
            M11	& 1	& 1	& 0	& 0	& 0	& 1	& 0	& 0	& 3	& 2	& 2	& 0	& 1	& 1	& 1	& 0 \\
            M12	& 2	& 2	& 1	& 1	& 2	& 1	& 0	& 0	& 1	& 1	& 2	& 0	& 1	& 1	& 0	& 1 \\
            M13	& 1	& 1	& 1	& 1	& 0	& 0	& 1	& 0	& 2	& 2	& 0	& 0	& 0	& 0	& 1	& 1 \\
            M14	& 1	& 1	& 0	& 0	& 0	& 0	& 1	& 0	& 0	& 3	& 3	& 0	& 1	& 1	& 1	& 1 \\
            M15	& 1	& 1	& 0	& 0	& 1	& 0	& 0	& 0	& 0	& 2	& 0	& 0	& 1	& 1	& 0	& 1 \\
            M16	& 2	& 1	& 0	& 1	& 0	& 0	& 1	& 0	& 2	& 3	& 0	& 1	& 2	& 1	& 2	& 2 \\
            M17	& 1	& 1	& 0	& 0	& 1	& 1	& 0	& 0	& 3	& 2	& 2	& 0	& 1	& 1	& 0	& 0 \\
            M18	& 1	& 1	& 1	& 1	& 1	& 0	& 0	& 0	& 2	& 1	& 3	& 0	& 0	& 0	& 0	& 1 \\
            M19	& 1	& 1	& 0	& 1	& 1	& 0	& 0	& 0	& 1	& 2	& 0	& 0	& 1	& 0	& 0	& 1 \\
            M20	& 1	& 0	& 0	& 1	& 0	& 0	& 0	& 0	& 2	& 2	& 0	& 1	& 1	& 0	& 1	& 1 \\
            M21	& 1	& 1	& 0	& 0	& 1	& 0	& 0	& 0	& 3	& 2	& 0	& 0	& 1	& 1	& 0	& 1 \\
            M22	& 2	& 1	& 1	& 2	& 0	& 0	& 1	& 0	& 1	& 2	& 0	& 1	& 1	& 0	& 2	& 2 \\
            M23	& 4	& 1	& 1	& 1	& 1	& 0	& 0	& 0	& 2	& 1	& 3	& 3	& 3	& 3	& 3	& 4 \\
            M24	& 1	& 1	& 0	& 0	& 1	& 0	& 0	& 0	& 3	& 2	& 3	& 0	& 1	& 1	& 0	& 1 \\
            M25	& 1	& 1	& 0	& 0	& 1	& 0	& 0	& 0	& 3	& 2	& 0	& 0	& 1	& 1	& 0	& 1 \\
            M26	& 1	& 1	& 1	& 0	& 1	& 0	& 1	& 0	& 2	& 1	& 0	& 0	& 0	& 1	& 0	& 1 \\
            M27	& 1	& 1	& 0	& 0	& 1	& 0	& 2	& 2	& 3	& 2	& 0	& 0	& 1	& 1	& 0	& 1 \\
            M28	& 2	& 2	& 0	& 2	& 2	& 0	& 0	& 0	& 1	& 0	& 0	& 0	& 2	& 0	& 0	& 2 \\
            M29	& 1	& 0	& 0	& 1	& 0	& 0	& 1	& 1	& 2	& 3	& 0	& 1	& 1	& 0	& 1	& 1 \\
            M30	& 2	& 1	& 0	& 0	& 1	& 0	& 1	& 0	& 0	& 1	& 0	& 1	& 2	& 2	& 1	& 2 \\
            M31	& 2	& 2	& 1	& 1	& 1	& 1	& 1	& 1	& 2	& 1	& 2	& 0	& 1	& 1	& 1	& 1 \\
            M32	& 3	& 3	& 2	& 0	& 1	& 0	& 0	& 0	& 0	& 2	& 0	& 0	& 1	& 3	& 2	& 3 \\
            M33	& 3	& 1	& 0	& 1	& 1	& 0	& 2	& 2	& 2	& 2	& 0	& 2	& 3	& 2	& 2	& 3 \\
            M34	& 3	& 3	& 0	& 1	& 1	& 0	& 0	& 0	& 2	& 1	& 0	& 0	& 3	& 2	& 2	& 3 \\
            M35	& 2	& 2	& 1	& 1	& 0	& 0	& 3	& 1	& 2	& 2	& 0	& 0	& 1	& 1	& 2	& 2 \\
            M36	& 2	& 2	& 2	& 0	& 0	& 1	& 1	& 1	& 0	& 2	& 2	& 0	& 0	& 2	& 2	& 1 \\
            M37	& 2	& 1	& 0	& 1	& 0	& 0	& 0	& 0	& 2	& 2	& 0	& 1	& 2	& 1	& 2	& 2 \\
            M38	& 1	& 1	& 0	& 1	& 1	& 0	& 0	& 0	& 1	& 2	& 3	& 0	& 1	& 0	& 0	& 1 \\
            M39	& 3	& 3	& 3	& 0	& 0	& 0	& 1	& 1	& 3	& 2	& 0	& 0	& 0	& 3	& 3	& 3 \\
            M40	& 1	& 1	& 1	& 1	& 0	& 0	& 1	& 0	& 2	& 2	& 3	& 0	& 0	& 0	& 1	& 1 \\
            M41	& 2	& 2	& 0	& 0	& 0	& 1	& 2	& 2	& 3	& 2	& 2	& 0	& 2	& 2	& 2	& 1 \\
            M42	& 2	& 1	& 0	& 1	& 1	& 0	& 1	& 0	& 2	& 1	& 0	& 1	& 2	& 1	& 1	& 2 \\
            M43	& 1	& 1	& 0	& 0	& 1	& 0	& 2	& 2	& 2	& 1	& 0	& 0	& 1	& 1	& 0	& 1 \\
            M44	& 2	& 2	& 1	& 1	& 1	& 0	& 2	& 2	& 1	& 2	& 0	& 0	& 1	& 1	& 1	& 2 \\
            M45	& 1	& 1	& 0	& 1	& 0	& 0	& 0	& 0	& 2	& 2	& 3	& 0	& 1	& 0	& 1	& 1 \\
            M46	& 1	& 1	& 0	& 1	& 0	& 1	& 0	& 0	& 2	& 2	& 2	& 0	& 1	& 0	& 1	& 0 \\
            M47	& 1	& 1	& 0	& 1	& 1	& 0	& 0	& 0	& 2	& 2	& 3	& 0	& 1	& 0	& 0	& 1 \\
            M48	& 2	& 1	& 1	& 1	& 1	& 0	& 0	& 0	& 2	& 1	& 0	& 1	& 1	& 1	& 1	& 2 \\
            M49	& 1	& 1	& 1	& 0	& 1	& 1	& 2	& 2	& 3	& 2	& 2	& 0	& 0	& 1	& 0	& 0 \\
            M50	& 2	& 1	& 1	& 0	& 1	& 1	& 1	& 0	& 3	& 1	& 2	& 1	& 1	& 2	& 1	& 1 \\
            M51	& 2	& 1	& 1	& 1	& 1	& 1	& 1	& 0	& 2	& 1	& 2	& 1	& 1	& 1	& 1	& 1 \\
            M52	& 2	& 2	& 2	& 0	& 1	& 1	& 0	& 0	& 3	& 2	& 2	& 0	& 0	& 2	& 1	& 1 \\
            M53	& 1	& 1	& 1	& 1	& 0	& 1	& 0	& 0	& 2	& 3	& 2	& 0	& 0	& 0	& 1	& 0 \\
            M54	& 1	& 1	& 0	& 1	& 0	& 0	& 2	& 0	& 2	& 3	& 0	& 0	& 1	& 0	& 1	& 1 \\
            M55	& 3	& 1	& 1	& 0	& 0	& 0	& 0	& 0	& 0	& 3	& 0	& 2	& 2	& 3	& 3	& 3 \\
            M56	& 1	& 1	& 0	& 1	& 1	& 0	& 2	& 1	& 2	& 1	& 0	& 0	& 1	& 0	& 0	& 1 \\
            M57	& 4	& 1	& 0	& 0	& 0	& 0	& 0	& 0	& 3	& 2	& 0	& 3	& 4	& 4	& 4	& 4 \\
            M58	& 2	& 2	& 2	& 1	& 1	& 0	& 0	& 0	& 2	& 1	& 0	& 0	& 0	& 1	& 1	& 2 \\ 
            \cmidrule(lr){2-17}
            Overall	& 98	& 74	& 31	& 40	& 38	& 15	& 37	& 19	& 107	& 104	& 54	& 24	& 67	& 58	& 60	& 83 \\
			 \hline
         \end{tabular}
	\label{tab:refactoring}
\end{table}




\end{appendices}

\bibliography{references}

\begin{thebibliography}{}
\renewcommand{\doi}[1]{\url{https://doi.org/#1}}
\bibcommenthead

\bibitem [\protect \citeauthoryear {%
Aho%
, Sethi%
\BCBL {}\ \BBA {} Ullman%
}{%
Aho%
\ \protect \BOthers {.}}{%
{\protect \APACyear {2007}}%
}]{%
aho2007compilers}
\APACinsertmetastar {%
aho2007compilers}%
\begin{APACrefauthors}%
Aho, A.V.%
, Sethi, R.%
\BCBL {} Ullman, J.D.%
\end{APACrefauthors}%
\unskip\
\newblock
\APACrefYear{2007}.
\newblock
\APACrefbtitle {Compilers: principles, techniques, and tools} {Compilers:
  principles, techniques, and tools}\ (\BVOL~2).
\newblock
\APACaddressPublisher{}{Addison-wesley Reading}.
\PrintBackRefs{\CurrentBib}

\bibitem [\protect \citeauthoryear {%
Alcocer%
, Antezana%
, Santos%
\BCBL {}\ \BBA {} Bergel%
}{%
Alcocer%
\ \protect \BOthers {.}}{%
{\protect \APACyear {2020}}%
}]{%
alcocer2020improving}
\APACinsertmetastar {%
alcocer2020improving}%
\begin{APACrefauthors}%
Alcocer, J.P.S.%
, Antezana, A.S.%
, Santos, G.%
\BCBL {} Bergel, A.%
\end{APACrefauthors}%
\unskip\
\newblock
\APACrefYearMonthDay{2020}{}{}.
\newblock
{\BBOQ}\APACrefatitle {Improving the success rate of applying the extract
  method refactoring} {Improving the success rate of applying the extract
  method refactoring}.{\BBCQ}
\newblock
\APACjournalVolNumPages{Science of Computer Programming}{195}{}{102475,}
\newblock

\newblock

\PrintBackRefs{\CurrentBib}

\bibitem [\protect \citeauthoryear {%
Al~Dallal%
}{%
Al~Dallal%
}{%
{\protect \APACyear {2010}}%
}]{%
al2010measuring}
\APACinsertmetastar {%
al2010measuring}%
\begin{APACrefauthors}%
Al~Dallal, J.%
\end{APACrefauthors}%
\unskip\
\newblock
\APACrefYearMonthDay{2010}{}{}.
\newblock
{\BBOQ}\APACrefatitle {Measuring the discriminative power of object-oriented
  class cohesion metrics} {Measuring the discriminative power of
  object-oriented class cohesion metrics}.{\BBCQ}
\newblock
\APACjournalVolNumPages{IEEE Transactions on Software
  Engineering}{37}{6}{788--804,}
\newblock

\newblock

\PrintBackRefs{\CurrentBib}

\bibitem [\protect \citeauthoryear {%
Allen%
}{%
Allen%
}{%
{\protect \APACyear {1970}}%
}]{%
allen1970control}
\APACinsertmetastar {%
allen1970control}%
\begin{APACrefauthors}%
Allen, F.E.%
\end{APACrefauthors}%
\unskip\
\newblock
\APACrefYearMonthDay{1970}{}{}.
\newblock
{\BBOQ}\APACrefatitle {Control flow analysis} {Control flow analysis}.{\BBCQ}
\newblock
\APACjournalVolNumPages{ACM Sigplan Notices}{5}{7}{1--19,}
\newblock

\newblock

\PrintBackRefs{\CurrentBib}

\bibitem [\protect \citeauthoryear {%
Ampatzoglou%
\ \protect \BOthers {.}}{%
Ampatzoglou%
\ \protect \BOthers {.}}{%
{\protect \APACyear {2019}}%
}]{%
ampatzoglou2019applying}
\APACinsertmetastar {%
ampatzoglou2019applying}%
\begin{APACrefauthors}%
Ampatzoglou, A.%
, Tsintzira, A\BHBI A.%
, Arvanitou, E\BHBI M.%
, Chatzigeorgiou, A.%
, Stamelos, I.%
, Moga, A.%
\BDBL {}Kehagias, D.%
\end{APACrefauthors}%
\unskip\
\newblock
\APACrefYearMonthDay{2019}{}{}.
\newblock
{\BBOQ}\APACrefatitle {Applying the single responsibility principle in
  industry: modularity benefits and trade-offs} {Applying the single
  responsibility principle in industry: modularity benefits and
  trade-offs}.{\BBCQ}
\newblock
 \APACrefbtitle {Proceedings of the Evaluation and Assessment on Software
  Engineering} {Proceedings of the evaluation and assessment on software
  engineering}\ (\BPGS\ 347--352).
\PrintBackRefs{\CurrentBib}

\bibitem [\protect \citeauthoryear {%
Arcelli~Fontana%
, M{\"a}ntyl{\"a}%
, Zanoni%
\BCBL {}\ \BBA {} Marino%
}{%
Arcelli~Fontana%
\ \protect \BOthers {.}}{%
{\protect \APACyear {2016}}%
}]{%
arcelli2016comparing}
\APACinsertmetastar {%
arcelli2016comparing}%
\begin{APACrefauthors}%
Arcelli~Fontana, F.%
, M{\"a}ntyl{\"a}, M.V.%
, Zanoni, M.%
\BCBL {} Marino, A.%
\end{APACrefauthors}%
\unskip\
\newblock
\APACrefYearMonthDay{2016}{}{}.
\newblock
{\BBOQ}\APACrefatitle {Comparing and experimenting machine learning techniques
  for code smell detection} {Comparing and experimenting machine learning
  techniques for code smell detection}.{\BBCQ}
\newblock
\APACjournalVolNumPages{Empirical Software Engineering}{21}{}{1143--1191,}
\newblock

\newblock

\PrintBackRefs{\CurrentBib}

\bibitem [\protect \citeauthoryear {%
Brdar%
, Vlajkov%
, Slivka%
, Gruji{\'c}%
\BCBL {}\ \BBA {} Kova{\v{c}}evi{\'c}%
}{%
Brdar%
\ \protect \BOthers {.}}{%
{\protect \APACyear {2022}}%
}]{%
brdar2022semi}
\APACinsertmetastar {%
brdar2022semi}%
\begin{APACrefauthors}%
Brdar, I.%
, Vlajkov, J.%
, Slivka, J.%
, Gruji{\'c}, K\BHBI G.%
\BCBL {} Kova{\v{c}}evi{\'c}, A.%
\end{APACrefauthors}%
\unskip\
\newblock
\APACrefYearMonthDay{2022}{}{}.
\newblock
{\BBOQ}\APACrefatitle {Semi-supervised detection of Long Method and God Class
  code smells} {Semi-supervised detection of long method and god class code
  smells}.{\BBCQ}
\newblock
 \APACrefbtitle {2022 IEEE 20th Jubilee International Symposium on Intelligent
  Systems and Informatics (SISY)} {2022 ieee 20th jubilee international
  symposium on intelligent systems and informatics (sisy)}\ (\BPGS\ 403--408).
\PrintBackRefs{\CurrentBib}

\bibitem [\protect \citeauthoryear {%
Charalampidou%
, Ampatzoglou%
, Chatzigeorgiou%
, Gkortzis%
\BCBL {}\ \BBA {} Avgeriou%
}{%
Charalampidou%
\ \protect \BOthers {.}}{%
{\protect \APACyear {2016}}%
}]{%
charalampidou2016identifying}
\APACinsertmetastar {%
charalampidou2016identifying}%
\begin{APACrefauthors}%
Charalampidou, S.%
, Ampatzoglou, A.%
, Chatzigeorgiou, A.%
, Gkortzis, A.%
\BCBL {} Avgeriou, P.%
\end{APACrefauthors}%
\unskip\
\newblock
\APACrefYearMonthDay{2016}{}{}.
\newblock
{\BBOQ}\APACrefatitle {Identifying extract method refactoring opportunities
  based on functional relevance} {Identifying extract method refactoring
  opportunities based on functional relevance}.{\BBCQ}
\newblock
\APACjournalVolNumPages{IEEE Transactions on Software
  Engineering}{43}{10}{954--974,}
\newblock

\newblock

\PrintBackRefs{\CurrentBib}

\bibitem [\protect \citeauthoryear {%
Cheng%
, Sawyer%
, Bencomo%
\BCBL {}\ \BBA {} Whittle%
}{%
Cheng%
\ \protect \BOthers {.}}{%
{\protect \APACyear {2009}}%
}]{%
cheng2009goal}
\APACinsertmetastar {%
cheng2009goal}%
\begin{APACrefauthors}%
Cheng, B.H.%
, Sawyer, P.%
, Bencomo, N.%
\BCBL {} Whittle, J.%
\end{APACrefauthors}%
\unskip\
\newblock
\APACrefYearMonthDay{2009}{}{}.
\newblock
{\BBOQ}\APACrefatitle {A goal-based modeling approach to develop requirements
  of an adaptive system with environmental uncertainty} {A goal-based modeling
  approach to develop requirements of an adaptive system with environmental
  uncertainty}.{\BBCQ}
\newblock
 \APACrefbtitle {Model Driven Engineering Languages and Systems: 12th
  International Conference, MODELS 2009, Denver, CO, USA, October 4-9, 2009.
  Proceedings 12} {Model driven engineering languages and systems: 12th
  international conference, models 2009, denver, co, usa, october 4-9, 2009.
  proceedings 12}\ (\BPGS\ 468--483).
\PrintBackRefs{\CurrentBib}

\bibitem [\protect \citeauthoryear {%
Cohen%
}{%
Cohen%
}{%
{\protect \APACyear {2013}}%
}]{%
cohen2013statistical}
\APACinsertmetastar {%
cohen2013statistical}%
\begin{APACrefauthors}%
Cohen, J.%
\end{APACrefauthors}%
\unskip\
\newblock
\APACrefYear{2013}.
\newblock
\APACrefbtitle {Statistical power analysis for the behavioral sciences}
  {Statistical power analysis for the behavioral sciences}.
\newblock
\APACaddressPublisher{}{Academic press}.
\PrintBackRefs{\CurrentBib}

\bibitem [\protect \citeauthoryear {%
De~Lucia%
}{%
De~Lucia%
}{%
{\protect \APACyear {2001}}%
}]{%
de2001program}
\APACinsertmetastar {%
de2001program}%
\begin{APACrefauthors}%
De~Lucia, A.%
\end{APACrefauthors}%
\unskip\
\newblock
\APACrefYearMonthDay{2001}{}{}.
\newblock
{\BBOQ}\APACrefatitle {Program slicing: Methods and applications} {Program
  slicing: Methods and applications}.{\BBCQ}
\newblock
 \APACrefbtitle {Proceedings First IEEE International Workshop on Source Code
  Analysis and Manipulation} {Proceedings first ieee international workshop on
  source code analysis and manipulation}\ (\BPGS\ 142--149).
\PrintBackRefs{\CurrentBib}

\bibitem [\protect \citeauthoryear {%
Fayad%
, Hamza%
\BCBL {}\ \BBA {} S{\'a}nchez%
}{%
Fayad%
\ \protect \BOthers {.}}{%
{\protect \APACyear {2003}}%
}]{%
fayad2003pattern}
\APACinsertmetastar {%
fayad2003pattern}%
\begin{APACrefauthors}%
Fayad, M.E.%
, Hamza, H.%
\BCBL {} S{\'a}nchez, H.%
\end{APACrefauthors}%
\unskip\
\newblock
\APACrefYearMonthDay{2003}{}{}.
\newblock
{\BBOQ}\APACrefatitle {A pattern for an effective class responsibility
  collaborator (CRC) cards} {A pattern for an effective class responsibility
  collaborator (crc) cards}.{\BBCQ}
\newblock
 \APACrefbtitle {Proceedings Fifth IEEE Workshop on Mobile Computing Systems
  and Applications} {Proceedings fifth ieee workshop on mobile computing
  systems and applications}\ (\BPGS\ 584--587).
\PrintBackRefs{\CurrentBib}

\bibitem [\protect \citeauthoryear {%
Fernandes%
, Aguiar%
\BCBL {}\ \BBA {} Restivo%
}{%
Fernandes%
\ \protect \BOthers {.}}{%
{\protect \APACyear {2022}}%
}]{%
fernandes2022live}
\APACinsertmetastar {%
fernandes2022live}%
\begin{APACrefauthors}%
Fernandes, S.%
, Aguiar, A.%
\BCBL {} Restivo, A.%
\end{APACrefauthors}%
\unskip\
\newblock
\APACrefYearMonthDay{2022}{}{}.
\newblock
{\BBOQ}\APACrefatitle {A Live Environment to Improve the Refactoring
  Experience} {A live environment to improve the refactoring
  experience}.{\BBCQ}
\newblock
 \APACrefbtitle {Companion Proceedings of the 6th International Conference on
  the Art, Science, and Engineering of Programming} {Companion proceedings of
  the 6th international conference on the art, science, and engineering of
  programming}\ (\BPGS\ 30--37).
\PrintBackRefs{\CurrentBib}

\bibitem [\protect \citeauthoryear {%
Ferrante%
, Ottenstein%
\BCBL {}\ \BBA {} Warren%
}{%
Ferrante%
\ \protect \BOthers {.}}{%
{\protect \APACyear {1987}}%
}]{%
ferrante1987program}
\APACinsertmetastar {%
ferrante1987program}%
\begin{APACrefauthors}%
Ferrante, J.%
, Ottenstein, K.J.%
\BCBL {} Warren, J.D.%
\end{APACrefauthors}%
\unskip\
\newblock
\APACrefYearMonthDay{1987}{}{}.
\newblock
{\BBOQ}\APACrefatitle {The program dependence graph and its use in
  optimization} {The program dependence graph and its use in
  optimization}.{\BBCQ}
\newblock
\APACjournalVolNumPages{ACM Transactions on Programming Languages and Systems
  (TOPLAS)}{9}{3}{319--349,}
\newblock

\newblock

\PrintBackRefs{\CurrentBib}

\bibitem [\protect \citeauthoryear {%
Fontana%
, Ferme%
, Zanoni%
\BCBL {}\ \BBA {} Roveda%
}{%
Fontana%
\ \protect \BOthers {.}}{%
{\protect \APACyear {2015}}%
}]{%
fontana2015towards}
\APACinsertmetastar {%
fontana2015towards}%
\begin{APACrefauthors}%
Fontana, F.A.%
, Ferme, V.%
, Zanoni, M.%
\BCBL {} Roveda, R.%
\end{APACrefauthors}%
\unskip\
\newblock
\APACrefYearMonthDay{2015}{}{}.
\newblock
{\BBOQ}\APACrefatitle {Towards a prioritization of code debt: A code smell
  intensity index} {Towards a prioritization of code debt: A code smell
  intensity index}.{\BBCQ}
\newblock
 \APACrefbtitle {2015 IEEE 7th International Workshop on Managing Technical
  Debt (MTD)} {2015 ieee 7th international workshop on managing technical debt
  (mtd)}\ (\BPGS\ 16--24).
\PrintBackRefs{\CurrentBib}

\bibitem [\protect \citeauthoryear {%
Fontana%
\ \BBA {} Zanoni%
}{%
Fontana%
\ \BBA {} Zanoni%
}{%
{\protect \APACyear {2017}}%
}]{%
fontana2017code}
\APACinsertmetastar {%
fontana2017code}%
\begin{APACrefauthors}%
Fontana, F.A.%
\BCBT {}\ \BBA {} Zanoni, M.%
\end{APACrefauthors}%
\unskip\
\newblock
\APACrefYearMonthDay{2017}{}{}.
\newblock
{\BBOQ}\APACrefatitle {Code smell severity classification using machine
  learning techniques} {Code smell severity classification using machine
  learning techniques}.{\BBCQ}
\newblock
\APACjournalVolNumPages{Knowledge-Based Systems}{128}{}{43--58,}
\newblock

\newblock

\PrintBackRefs{\CurrentBib}

\bibitem [\protect \citeauthoryear {%
Fortunato%
}{%
Fortunato%
}{%
{\protect \APACyear {2010}}%
}]{%
fortunato2010community}
\APACinsertmetastar {%
fortunato2010community}%
\begin{APACrefauthors}%
Fortunato, S.%
\end{APACrefauthors}%
\unskip\
\newblock
\APACrefYearMonthDay{2010}{}{}.
\newblock
{\BBOQ}\APACrefatitle {Community detection in graphs} {Community detection in
  graphs}.{\BBCQ}
\newblock
\APACjournalVolNumPages{Physics reports}{486}{3-5}{75--174,}
\newblock

\newblock

\PrintBackRefs{\CurrentBib}

\bibitem [\protect \citeauthoryear {%
Fowler%
}{%
Fowler%
}{%
{\protect \APACyear {1999}}%
}]{%
Fowler1999}
\APACinsertmetastar {%
Fowler1999}%
\begin{APACrefauthors}%
Fowler, M.%
\end{APACrefauthors}%
\unskip\
\newblock
\APACrefYear{1999}.
\newblock
\APACrefbtitle {Refactoring: Improving the Design of Existing Code}
  {Refactoring: Improving the design of existing code}.
\newblock
\APACaddressPublisher{Boston, MA, USA}{Addison-Wesley}.
\PrintBackRefs{\CurrentBib}

\bibitem [\protect \citeauthoryear {%
Fundation%
}{%
Fundation%
}{%
{\protect \APACyear {2011}}%
}]{%
Xerces2011}
\APACinsertmetastar {%
Xerces2011}%
\begin{APACrefauthors}%
Fundation, A.%
\end{APACrefauthors}%
\unskip\
\newblock
\APACrefYearMonthDay{2011}{}{}.
\newblock
\APACrefbtitle {Apache Xerces-J 2.} {Apache xerces-j 2.}
\newblock
\begin{APACrefURL} {http://xerces.apache.org/index.html} \end{APACrefURL}
\PrintBackRefs{\CurrentBib}

\bibitem [\protect \citeauthoryear {%
Gilbert%
}{%
Gilbert%
}{%
{\protect \APACyear {2017}}%
}]{%
JFreeChart2017}
\APACinsertmetastar {%
JFreeChart2017}%
\begin{APACrefauthors}%
Gilbert, D.%
\end{APACrefauthors}%
\unskip\
\newblock
\APACrefYearMonthDay{2017}{}{}.
\newblock
\APACrefbtitle {JFree Chart.} {Jfree chart.}
\newblock
\begin{APACrefURL} {https://github.com/jfree/jfreechart} \end{APACrefURL}
\PrintBackRefs{\CurrentBib}

\bibitem [\protect \citeauthoryear {%
Green%
, Lane%
, Rainer%
\BCBL {}\ \BBA {} Scholz%
}{%
Green%
\ \protect \BOthers {.}}{%
{\protect \APACyear {2009}}%
}]{%
green2009introduction}
\APACinsertmetastar {%
green2009introduction}%
\begin{APACrefauthors}%
Green, P.%
, Lane, P.C.%
, Rainer, A.%
\BCBL {} Scholz, S.%
\end{APACrefauthors}%
\unskip\
\newblock
\APACrefYearMonthDay{2009}{}{}.
\newblock
{\BBOQ}\APACrefatitle {An introduction to slice-based cohesion and coupling
  metrics} {An introduction to slice-based cohesion and coupling
  metrics}.{\BBCQ}
\newblock

\newblock

\newblock

\PrintBackRefs{\CurrentBib}

\bibitem [\protect \citeauthoryear {%
Hadj-Kacem%
\ \BBA {} Bouassida%
}{%
Hadj-Kacem%
\ \BBA {} Bouassida%
}{%
{\protect \APACyear {2018}}%
}]{%
hadj2018hybrid}
\APACinsertmetastar {%
hadj2018hybrid}%
\begin{APACrefauthors}%
Hadj-Kacem, M.%
\BCBT {}\ \BBA {} Bouassida, N.%
\end{APACrefauthors}%
\unskip\
\newblock
\APACrefYearMonthDay{2018}{}{}.
\newblock
{\BBOQ}\APACrefatitle {A Hybrid Approach To Detect Code Smells using Deep
  Learning.} {A hybrid approach to detect code smells using deep
  learning.}{\BBCQ}
\newblock
 \APACrefbtitle {ENASE} {Enase}\ (\BPGS\ 137--146).
\PrintBackRefs{\CurrentBib}

\bibitem [\protect \citeauthoryear {%
Han%
, Kamber%
\BCBL {}\ \BBA {} Pei%
}{%
Han%
\ \protect \BOthers {.}}{%
{\protect \APACyear {2012}}%
}]{%
han2012data}
\APACinsertmetastar {%
han2012data}%
\begin{APACrefauthors}%
Han, J.%
, Kamber, M.%
\BCBL {} Pei, J.%
\end{APACrefauthors}%
\unskip\
\newblock
\APACrefYearMonthDay{2012}{}{}.
\newblock
{\BBOQ}\APACrefatitle {Data mining concepts and techniques third edition} {Data
  mining concepts and techniques third edition}.{\BBCQ}
\newblock
\APACjournalVolNumPages{University of Illinois at Urbana-Champaign Micheline
  Kamber Jian Pei Simon Fraser University}{}{}{,}
\newblock

\newblock

\PrintBackRefs{\CurrentBib}

\bibitem [\protect \citeauthoryear {%
Hora%
\ \BBA {} Robbes%
}{%
Hora%
\ \BBA {} Robbes%
}{%
{\protect \APACyear {2020}}%
}]{%
hora2020characteristics}
\APACinsertmetastar {%
hora2020characteristics}%
\begin{APACrefauthors}%
Hora, A.%
\BCBT {}\ \BBA {} Robbes, R.%
\end{APACrefauthors}%
\unskip\
\newblock
\APACrefYearMonthDay{2020}{}{}.
\newblock
{\BBOQ}\APACrefatitle {Characteristics of method extractions in Java: A large
  scale empirical study} {Characteristics of method extractions in java: A
  large scale empirical study}.{\BBCQ}
\newblock
\APACjournalVolNumPages{Empirical Software Engineering}{25}{}{1798--1833,}
\newblock

\newblock

\PrintBackRefs{\CurrentBib}

\bibitem [\protect \citeauthoryear {%
Hotta%
, Higo%
\BCBL {}\ \BBA {} Kusumoto%
}{%
Hotta%
\ \protect \BOthers {.}}{%
{\protect \APACyear {2012}}%
}]{%
hotta2012identifying}
\APACinsertmetastar {%
hotta2012identifying}%
\begin{APACrefauthors}%
Hotta, K.%
, Higo, Y.%
\BCBL {} Kusumoto, S.%
\end{APACrefauthors}%
\unskip\
\newblock
\APACrefYearMonthDay{2012}{}{}.
\newblock
{\BBOQ}\APACrefatitle {Identifying, tailoring, and suggesting form template
  method refactoring opportunities with program dependence graph} {Identifying,
  tailoring, and suggesting form template method refactoring opportunities with
  program dependence graph}.{\BBCQ}
\newblock
 \APACrefbtitle {2012 16th European Conference on Software Maintenance and
  Reengineering} {2012 16th european conference on software maintenance and
  reengineering}\ (\BPGS\ 53--62).
\PrintBackRefs{\CurrentBib}

\bibitem [\protect \citeauthoryear {%
Hubert%
}{%
Hubert%
}{%
{\protect \APACyear {2019}}%
}]{%
hubert2019implementation}
\APACinsertmetastar {%
hubert2019implementation}%
\begin{APACrefauthors}%
Hubert, J.%
\end{APACrefauthors}%
\unskip\
\newblock
\APACrefYear{2019}.
\unskip\
\newblock
\APACrefbtitle {Implementation of an automatic extract method refactoring}
  {Implementation of an automatic extract method refactoring}\
  \APACtypeAddressSchool {\BUMTh}{}{}.
\PrintBackRefs{\CurrentBib}

\bibitem [\protect \citeauthoryear {%
Komondoor%
\ \BBA {} Horwitz%
}{%
Komondoor%
\ \BBA {} Horwitz%
}{%
{\protect \APACyear {2000}}%
}]{%
komondoor2000semantics}
\APACinsertmetastar {%
komondoor2000semantics}%
\begin{APACrefauthors}%
Komondoor, R.%
\BCBT {}\ \BBA {} Horwitz, S.%
\end{APACrefauthors}%
\unskip\
\newblock
\APACrefYearMonthDay{2000}{}{}.
\newblock
{\BBOQ}\APACrefatitle {Semantics-preserving procedure extraction}
  {Semantics-preserving procedure extraction}.{\BBCQ}
\newblock
 \APACrefbtitle {Proceedings of the 27th ACM SIGPLAN-SIGACT symposium on
  Principles of programming languages} {Proceedings of the 27th acm
  sigplan-sigact symposium on principles of programming languages}\ (\BPGS\
  155--169).
\PrintBackRefs{\CurrentBib}

\bibitem [\protect \citeauthoryear {%
Lacerda%
, Petrillo%
, Pimenta%
\BCBL {}\ \BBA {} Gu{\'e}h{\'e}neuc%
}{%
Lacerda%
\ \protect \BOthers {.}}{%
{\protect \APACyear {2020}}%
}]{%
lacerda2020code}
\APACinsertmetastar {%
lacerda2020code}%
\begin{APACrefauthors}%
Lacerda, G.%
, Petrillo, F.%
, Pimenta, M.%
\BCBL {} Gu{\'e}h{\'e}neuc, Y.G.%
\end{APACrefauthors}%
\unskip\
\newblock
\APACrefYearMonthDay{2020}{}{}.
\newblock
{\BBOQ}\APACrefatitle {Code smells and refactoring: A tertiary systematic
  review of challenges and observations} {Code smells and refactoring: A
  tertiary systematic review of challenges and observations}.{\BBCQ}
\newblock
\APACjournalVolNumPages{Journal of Systems and Software}{167}{}{110610,}
\newblock

\newblock

\PrintBackRefs{\CurrentBib}

\bibitem [\protect \citeauthoryear {%
Lanza%
\ \BBA {} Marinescu%
}{%
Lanza%
\ \BBA {} Marinescu%
}{%
{\protect \APACyear {2007}}%
}]{%
lanza2007object}
\APACinsertmetastar {%
lanza2007object}%
\begin{APACrefauthors}%
Lanza, M.%
\BCBT {}\ \BBA {} Marinescu, R.%
\end{APACrefauthors}%
\unskip\
\newblock
\APACrefYear{2007}.
\newblock
\APACrefbtitle {Object-oriented metrics in practice: using software metrics to
  characterize, evaluate, and improve the design of object-oriented systems}
  {Object-oriented metrics in practice: using software metrics to characterize,
  evaluate, and improve the design of object-oriented systems}.
\newblock
\APACaddressPublisher{}{Springer Science \& Business Media}.
\PrintBackRefs{\CurrentBib}

\bibitem [\protect \citeauthoryear {%
Lopes%
\ \BBA {} Hora%
}{%
Lopes%
\ \BBA {} Hora%
}{%
{\protect \APACyear {2022}}%
}]{%
lopes2022and}
\APACinsertmetastar {%
lopes2022and}%
\begin{APACrefauthors}%
Lopes, M.%
\BCBT {}\ \BBA {} Hora, A.%
\end{APACrefauthors}%
\unskip\
\newblock
\APACrefYearMonthDay{2022}{}{}.
\newblock
{\BBOQ}\APACrefatitle {How and why we end up with complex methods: a
  multi-language study} {How and why we end up with complex methods: a
  multi-language study}.{\BBCQ}
\newblock
\APACjournalVolNumPages{Empirical Software Engineering}{27}{5}{115,}
\newblock

\newblock

\PrintBackRefs{\CurrentBib}

\bibitem [\protect \citeauthoryear {%
Martin%
}{%
Martin%
}{%
{\protect \APACyear {2003}}%
}]{%
martin2003agile}
\APACinsertmetastar {%
martin2003agile}%
\begin{APACrefauthors}%
Martin, R.C.%
\end{APACrefauthors}%
\unskip\
\newblock
\APACrefYear{2003}.
\newblock
\APACrefbtitle {Agile software development: principles, patterns, and
  practices} {Agile software development: principles, patterns, and practices}.
\newblock
\APACaddressPublisher{}{Prentice Hall PTR}.
\PrintBackRefs{\CurrentBib}

\bibitem [\protect \citeauthoryear {%
Martin%
}{%
Martin%
}{%
{\protect \APACyear {2009}}%
}]{%
martin2009clean}
\APACinsertmetastar {%
martin2009clean}%
\begin{APACrefauthors}%
Martin, R.C.%
\end{APACrefauthors}%
\unskip\
\newblock
\APACrefYear{2009}.
\newblock
\APACrefbtitle {Clean code: a handbook of agile software craftsmanship} {Clean
  code: a handbook of agile software craftsmanship}.
\newblock
\APACaddressPublisher{}{Pearson Education}.
\PrintBackRefs{\CurrentBib}

\bibitem [\protect \citeauthoryear {%
Maruyama%
}{%
Maruyama%
}{%
{\protect \APACyear {2001}}%
}]{%
maruyama2001automated}
\APACinsertmetastar {%
maruyama2001automated}%
\begin{APACrefauthors}%
Maruyama, K.%
\end{APACrefauthors}%
\unskip\
\newblock
\APACrefYearMonthDay{2001}{}{}.
\newblock
{\BBOQ}\APACrefatitle {Automated method-extraction refactoring by using
  block-based slicing} {Automated method-extraction refactoring by using
  block-based slicing}.{\BBCQ}
\newblock
 \APACrefbtitle {Proceedings of the 2001 symposium on Software reusability:
  putting software reuse in context} {Proceedings of the 2001 symposium on
  software reusability: putting software reuse in context}\ (\BPGS\ 31--40).
\PrintBackRefs{\CurrentBib}

\bibitem [\protect \citeauthoryear {%
Meananeatra%
, Rongviriyapanish%
\BCBL {}\ \BBA {} Apiwattanapong%
}{%
Meananeatra%
\ \protect \BOthers {.}}{%
{\protect \APACyear {2018}}%
}]{%
meananeatra2018refactoring}
\APACinsertmetastar {%
meananeatra2018refactoring}%
\begin{APACrefauthors}%
Meananeatra, P.%
, Rongviriyapanish, S.%
\BCBL {} Apiwattanapong, T.%
\end{APACrefauthors}%
\unskip\
\newblock
\APACrefYearMonthDay{2018}{}{}.
\newblock
{\BBOQ}\APACrefatitle {Refactoring opportunity identification methodology for
  removing long method smells and improving code analyzability} {Refactoring
  opportunity identification methodology for removing long method smells and
  improving code analyzability}.{\BBCQ}
\newblock
\APACjournalVolNumPages{IEICE TRANSACTIONS on Information and
  Systems}{101}{7}{1766--1779,}
\newblock

\newblock

\PrintBackRefs{\CurrentBib}

\bibitem [\protect \citeauthoryear {%
Mens%
\ \BBA {} Tourw{\'e}%
}{%
Mens%
\ \BBA {} Tourw{\'e}%
}{%
{\protect \APACyear {2004}}%
}]{%
mens2004survey}
\APACinsertmetastar {%
mens2004survey}%
\begin{APACrefauthors}%
Mens, T.%
\BCBT {}\ \BBA {} Tourw{\'e}, T.%
\end{APACrefauthors}%
\unskip\
\newblock
\APACrefYearMonthDay{2004}{}{}.
\newblock
{\BBOQ}\APACrefatitle {A survey of software refactoring} {A survey of software
  refactoring}.{\BBCQ}
\newblock
\APACjournalVolNumPages{IEEE Transactions on software
  engineering}{30}{2}{126--139,}
\newblock

\newblock

\PrintBackRefs{\CurrentBib}

\bibitem [\protect \citeauthoryear {%
Mkaouer%
, Kessentini%
, Bechikh%
, {\'O}~Cinn{\'e}ide%
\BCBL {}\ \BBA {} Deb%
}{%
Mkaouer%
\ \protect \BOthers {.}}{%
{\protect \APACyear {2016}}%
}]{%
mkaouer2016use}
\APACinsertmetastar {%
mkaouer2016use}%
\begin{APACrefauthors}%
Mkaouer, M.W.%
, Kessentini, M.%
, Bechikh, S.%
, {\'O}~Cinn{\'e}ide, M.%
\BCBL {} Deb, K.%
\end{APACrefauthors}%
\unskip\
\newblock
\APACrefYearMonthDay{2016}{}{}.
\newblock
{\BBOQ}\APACrefatitle {On the use of many quality attributes for software
  refactoring: a many-objective search-based software engineering approach} {On
  the use of many quality attributes for software refactoring: a many-objective
  search-based software engineering approach}.{\BBCQ}
\newblock
\APACjournalVolNumPages{Empirical Software Engineering}{21}{}{2503--2545,}
\newblock

\newblock

\PrintBackRefs{\CurrentBib}

\bibitem [\protect \citeauthoryear {%
Mongiovi%
\ \protect \BOthers {.}}{%
Mongiovi%
\ \protect \BOthers {.}}{%
{\protect \APACyear {2017}}%
}]{%
mongiovi2017detecting}
\APACinsertmetastar {%
mongiovi2017detecting}%
\begin{APACrefauthors}%
Mongiovi, M.%
, Gheyi, R.%
, Soares, G.%
, Ribeiro, M.%
, Borba, P.%
\BCBL {} Teixeira, L.%
\end{APACrefauthors}%
\unskip\
\newblock
\APACrefYearMonthDay{2017}{}{}.
\newblock
{\BBOQ}\APACrefatitle {Detecting overly strong preconditions in refactoring
  engines} {Detecting overly strong preconditions in refactoring
  engines}.{\BBCQ}
\newblock
\APACjournalVolNumPages{IEEE Transactions on Software
  Engineering}{44}{5}{429--452,}
\newblock

\newblock

\PrintBackRefs{\CurrentBib}

\bibitem [\protect \citeauthoryear {%
Orailoglu%
\ \BBA {} Gajski%
}{%
Orailoglu%
\ \BBA {} Gajski%
}{%
{\protect \APACyear {1986}}%
}]{%
orailoglu1986flow}
\APACinsertmetastar {%
orailoglu1986flow}%
\begin{APACrefauthors}%
Orailoglu, A.%
\BCBT {}\ \BBA {} Gajski, D.D.%
\end{APACrefauthors}%
\unskip\
\newblock
\APACrefYearMonthDay{1986}{}{}.
\newblock
{\BBOQ}\APACrefatitle {Flow graph representation} {Flow graph
  representation}.{\BBCQ}
\newblock
 \APACrefbtitle {Proceedings of the 23rd ACM/IEEE Design Automation Conference}
  {Proceedings of the 23rd acm/ieee design automation conference}\ (\BPGS\
  503--509).
\PrintBackRefs{\CurrentBib}

\bibitem [\protect \citeauthoryear {%
Ott%
\ \BBA {} Thuss%
}{%
Ott%
\ \BBA {} Thuss%
}{%
{\protect \APACyear {1993}}%
}]{%
ott1993slice}
\APACinsertmetastar {%
ott1993slice}%
\begin{APACrefauthors}%
Ott, L.M.%
\BCBT {}\ \BBA {} Thuss, J.J.%
\end{APACrefauthors}%
\unskip\
\newblock
\APACrefYearMonthDay{1993}{}{}.
\newblock
{\BBOQ}\APACrefatitle {Slice based metrics for estimating cohesion} {Slice
  based metrics for estimating cohesion}.{\BBCQ}
\newblock
 \APACrefbtitle {[1993] Proceedings First International Software Metrics
  Symposium} {[1993] proceedings first international software metrics
  symposium}\ (\BPGS\ 71--81).
\PrintBackRefs{\CurrentBib}

\bibitem [\protect \citeauthoryear {%
Ottenstein%
\ \BBA {} Ottenstein%
}{%
Ottenstein%
\ \BBA {} Ottenstein%
}{%
{\protect \APACyear {1984}}%
}]{%
ottenstein1984program}
\APACinsertmetastar {%
ottenstein1984program}%
\begin{APACrefauthors}%
Ottenstein, K.J.%
\BCBT {}\ \BBA {} Ottenstein, L.M.%
\end{APACrefauthors}%
\unskip\
\newblock
\APACrefYearMonthDay{1984}{}{}.
\newblock
{\BBOQ}\APACrefatitle {The program dependence graph in a software development
  environment} {The program dependence graph in a software development
  environment}.{\BBCQ}
\newblock
\APACjournalVolNumPages{ACM Sigplan Notices}{19}{5}{177--184,}
\newblock

\newblock

\PrintBackRefs{\CurrentBib}

\bibitem [\protect \citeauthoryear {%
Palomba%
, Panichella%
, De~Lucia%
, Oliveto%
\BCBL {}\ \BBA {} Zaidman%
}{%
Palomba%
\ \protect \BOthers {.}}{%
{\protect \APACyear {2016}}%
}]{%
palomba2016textual}
\APACinsertmetastar {%
palomba2016textual}%
\begin{APACrefauthors}%
Palomba, F.%
, Panichella, A.%
, De~Lucia, A.%
, Oliveto, R.%
\BCBL {} Zaidman, A.%
\end{APACrefauthors}%
\unskip\
\newblock
\APACrefYearMonthDay{2016}{}{}.
\newblock
{\BBOQ}\APACrefatitle {A textual-based technique for smell detection} {A
  textual-based technique for smell detection}.{\BBCQ}
\newblock
 \APACrefbtitle {2016 IEEE 24th international conference on program
  comprehension (ICPC)} {2016 ieee 24th international conference on program
  comprehension (icpc)}\ (\BPGS\ 1--10).
\PrintBackRefs{\CurrentBib}

\bibitem [\protect \citeauthoryear {%
Pecorelli%
, Di~Nucci%
, De~Roover%
\BCBL {}\ \BBA {} De~Lucia%
}{%
Pecorelli%
\ \protect \BOthers {.}}{%
{\protect \APACyear {2020}}%
}]{%
pecorelli2020large}
\APACinsertmetastar {%
pecorelli2020large}%
\begin{APACrefauthors}%
Pecorelli, F.%
, Di~Nucci, D.%
, De~Roover, C.%
\BCBL {} De~Lucia, A.%
\end{APACrefauthors}%
\unskip\
\newblock
\APACrefYearMonthDay{2020}{}{}.
\newblock
{\BBOQ}\APACrefatitle {A large empirical assessment of the role of data
  balancing in machine-learning-based code smell detection} {A large empirical
  assessment of the role of data balancing in machine-learning-based code smell
  detection}.{\BBCQ}
\newblock
\APACjournalVolNumPages{Journal of Systems and Software}{169}{}{110693,}
\newblock

\newblock

\PrintBackRefs{\CurrentBib}

\bibitem [\protect \citeauthoryear {%
Robert%
}{%
Robert%
}{%
{\protect \APACyear {2017}}%
}]{%
robert2017clean}
\APACinsertmetastar {%
robert2017clean}%
\begin{APACrefauthors}%
Robert, M.%
\end{APACrefauthors}%
\unskip\
\newblock
\APACrefYearMonthDay{2017}{}{}.
\newblock
\APACrefbtitle {Clean architecture: a craftsman's guide to software structure
  and design.} {Clean architecture: a craftsman's guide to software structure
  and design.}
\newblock
\APACaddressPublisher{}{Prentice Hall}.
\PrintBackRefs{\CurrentBib}

\bibitem [\protect \citeauthoryear {%
Shahidi%
, Ashtiani%
\BCBL {}\ \BBA {} Zakeri-Nasrabadi%
}{%
Shahidi%
\ \protect \BOthers {.}}{%
{\protect \APACyear {2022}}%
}]{%
shahidi2022automated}
\APACinsertmetastar {%
shahidi2022automated}%
\begin{APACrefauthors}%
Shahidi, M.%
, Ashtiani, M.%
\BCBL {} Zakeri-Nasrabadi, M.%
\end{APACrefauthors}%
\unskip\
\newblock
\APACrefYearMonthDay{2022}{}{}.
\newblock
{\BBOQ}\APACrefatitle {An automated extract method refactoring approach to
  correct the long method code smell} {An automated extract method refactoring
  approach to correct the long method code smell}.{\BBCQ}
\newblock
\APACjournalVolNumPages{Journal of Systems and Software}{187}{}{111221,}
\newblock

\newblock

\PrintBackRefs{\CurrentBib}

\bibitem [\protect \citeauthoryear {%
Silva%
, Terra%
\BCBL {}\ \BBA {} Valente%
}{%
Silva%
\ \protect \BOthers {.}}{%
{\protect \APACyear {2014}}%
}]{%
silva2014recommending}
\APACinsertmetastar {%
silva2014recommending}%
\begin{APACrefauthors}%
Silva, D.%
, Terra, R.%
\BCBL {} Valente, M.T.%
\end{APACrefauthors}%
\unskip\
\newblock
\APACrefYearMonthDay{2014}{}{}.
\newblock
{\BBOQ}\APACrefatitle {Recommending automated extract method refactorings}
  {Recommending automated extract method refactorings}.{\BBCQ}
\newblock
 \APACrefbtitle {Proceedings of the 22nd International Conference on Program
  Comprehension} {Proceedings of the 22nd international conference on program
  comprehension}\ (\BPGS\ 146--156).
\PrintBackRefs{\CurrentBib}

\bibitem [\protect \citeauthoryear {%
Tiwari%
\ \BBA {} Joshi%
}{%
Tiwari%
\ \BBA {} Joshi%
}{%
{\protect \APACyear {2022}}%
}]{%
tiwari2022identifying}
\APACinsertmetastar {%
tiwari2022identifying}%
\begin{APACrefauthors}%
Tiwari, O.%
\BCBT {}\ \BBA {} Joshi, R.%
\end{APACrefauthors}%
\unskip\
\newblock
\APACrefYearMonthDay{2022}{}{}.
\newblock
{\BBOQ}\APACrefatitle {Identifying extract method refactorings} {Identifying
  extract method refactorings}.{\BBCQ}
\newblock
 \APACrefbtitle {15th Innovations in Software Engineering Conference} {15th
  innovations in software engineering conference}\ (\BPGS\ 1--11).
\PrintBackRefs{\CurrentBib}

\bibitem [\protect \citeauthoryear {%
Tsantalis%
, Chaikalis%
\BCBL {}\ \BBA {} Chatzigeorgiou%
}{%
Tsantalis%
\ \protect \BOthers {.}}{%
{\protect \APACyear {2018}}%
}]{%
tsantalis2018ten}
\APACinsertmetastar {%
tsantalis2018ten}%
\begin{APACrefauthors}%
Tsantalis, N.%
, Chaikalis, T.%
\BCBL {} Chatzigeorgiou, A.%
\end{APACrefauthors}%
\unskip\
\newblock
\APACrefYearMonthDay{2018}{}{}.
\newblock
{\BBOQ}\APACrefatitle {Ten years of JDeodorant: Lessons learned from the hunt
  for smells} {Ten years of jdeodorant: Lessons learned from the hunt for
  smells}.{\BBCQ}
\newblock
 \APACrefbtitle {2018 IEEE 25th international conference on software analysis,
  evolution and reengineering (SANER)} {2018 ieee 25th international conference
  on software analysis, evolution and reengineering (saner)}\ (\BPGS\ 4--14).
\PrintBackRefs{\CurrentBib}

\bibitem [\protect \citeauthoryear {%
Tsantalis%
\ \BBA {} Chatzigeorgiou%
}{%
Tsantalis%
\ \BBA {} Chatzigeorgiou%
}{%
{\protect \APACyear {2011}}%
}]{%
tsantalis2011identification}
\APACinsertmetastar {%
tsantalis2011identification}%
\begin{APACrefauthors}%
Tsantalis, N.%
\BCBT {}\ \BBA {} Chatzigeorgiou, A.%
\end{APACrefauthors}%
\unskip\
\newblock
\APACrefYearMonthDay{2011}{}{}.
\newblock
{\BBOQ}\APACrefatitle {Identification of extract method refactoring
  opportunities for the decomposition of methods} {Identification of extract
  method refactoring opportunities for the decomposition of methods}.{\BBCQ}
\newblock
\APACjournalVolNumPages{Journal of Systems and Software}{84}{10}{1757--1782,}
\newblock

\newblock

\PrintBackRefs{\CurrentBib}

\bibitem [\protect \citeauthoryear {%
Wakaito%
}{%
Wakaito%
}{%
{\protect \APACyear {2011}}%
}]{%
Weka2011}
\APACinsertmetastar {%
Weka2011}%
\begin{APACrefauthors}%
Wakaito%
\end{APACrefauthors}%
\unskip\
\newblock
\APACrefYearMonthDay{2011}{}{}.
\newblock
\APACrefbtitle {Weka 3: Machine learning software in Java.} {Weka 3: Machine
  learning software in java.}
\newblock
\begin{APACrefURL} {http://old-www.cms.waikato.ac.nz/~ml/weka/}
  \end{APACrefURL}
\PrintBackRefs{\CurrentBib}

\bibitem [\protect \citeauthoryear {%
Weiser%
}{%
Weiser%
}{%
{\protect \APACyear {1984}}%
}]{%
weiser1984program}
\APACinsertmetastar {%
weiser1984program}%
\begin{APACrefauthors}%
Weiser, M.%
\end{APACrefauthors}%
\unskip\
\newblock
\APACrefYearMonthDay{1984}{}{}.
\newblock
{\BBOQ}\APACrefatitle {Program slicing} {Program slicing}.{\BBCQ}
\newblock
\APACjournalVolNumPages{IEEE Transactions on software
  engineering}{}{4}{352--357,}
\newblock

\newblock

\PrintBackRefs{\CurrentBib}

\bibitem [\protect \citeauthoryear {%
Xu%
, Sivaraman%
, Khoo%
\BCBL {}\ \BBA {} Xu%
}{%
Xu%
\ \protect \BOthers {.}}{%
{\protect \APACyear {2017}}%
}]{%
xu2017gems}
\APACinsertmetastar {%
xu2017gems}%
\begin{APACrefauthors}%
Xu, S.%
, Sivaraman, A.%
, Khoo, S\BHBI C.%
\BCBL {} Xu, J.%
\end{APACrefauthors}%
\unskip\
\newblock
\APACrefYearMonthDay{2017}{}{}.
\newblock
{\BBOQ}\APACrefatitle {Gems: An extract method refactoring recommender} {Gems:
  An extract method refactoring recommender}.{\BBCQ}
\newblock
 \APACrefbtitle {2017 IEEE 28th International Symposium on Software Reliability
  Engineering (ISSRE)} {2017 ieee 28th international symposium on software
  reliability engineering (issre)}\ (\BPGS\ 24--34).
\PrintBackRefs{\CurrentBib}

\end{thebibliography}

\end{document}